\documentclass[10pt,aps,prc,reprint,floatfix,superscriptaddress,nofootinbib,twocolumn]{revtex4-1} 
\usepackage{bm,physics}
\usepackage[dvipsnames]{xcolor}
\usepackage{amsfonts,amsmath,amssymb,bm}
\usepackage{mathtools}
\usepackage{aas_macros}
\newcommand{\longmid}{\rule[0ex]{0.4pt}{3ex}}
\usepackage{graphicx}
\usepackage[utf8]{inputenc}
\usepackage{xspace}
\usepackage{isotope}
\usepackage{xparse}
\usepackage{multirow}
\usepackage[pdfpagelabels, pdfencoding=auto, psdextra]{hyperref}
\hypersetup{%
  pdfsubject=Paper,
  pdfkeywords={nuclear physics} {Bayesian} {chiral EFT} {nuclear matter} {reduced basis method} {emulators} {nucleon deuteron scattering} {Greedy algorithm},
  unicode = true,
  breaklinks = true,
  colorlinks = true,
  linkcolor = blue,
  citecolor = blue,
  menucolor = blue,
  citecolor = blue,
  urlcolor = blue
}

\usepackage{orcidlink}

\graphicspath{{./}} 

\newcommand{\half}{\frac{1}{2}}

\newcommand{\abi}{\textit{ab~initio}\xspace}
\newcommand{\txvec}{\widetilde{\vec{x}}^a}
\newcommand{\Xmatzero}{\vb{X}^{0,a}}   
\newcommand{\Ymatzero}{\vb{Y}^{0,a}}

\newcommand{\matrixform}[1]{\begin{bmatrix} #1   \end{bmatrix}}
\renewcommand{\vec}{\mathbf}

\newcommand{\param}{\theta} 
\newcommand{\cE}{c_E}
\newcommand{\cD}{c_D}
\newcommand{\rmat}[2]{\mathcal{R}_{#2 \to #1 } }
\newcommand{\cemul}{c} 
\newcommand{\cemulwa}{c^a} 
\newcommand{\cHH}{b}
\newcommand{\dimL}{N_b}
\newcommand{\dimH}{N_h}
\newcommand{\dimxi}{N_s}
\newcommand{\dimparam}{N_\theta}
\newcommand{\dimC}{N_c}
\newcommand{\halfplus}{\frac{1}{2}^+}
\newcommand{\halfminus}{\frac{1}{2}^-}

\newcommand{\rfunc}[2]{{\cal F}_{#2 \to #1}}
\newcommand{\schro}{Schr\"{o}dinger\xspace}

\newcommand{\yvec}{\phi}

\newcommand{\wf}[1]{\Psi^{#1}}
\newcommand{\paramVec}{\vb*{\theta}} 
\newcommand{\thetavec}{\paramVec} 

\newcommand{\Amat}{\vb{A}}  
\newcommand{\tAmat}{\widetilde{\Amat}^a}
\newcommand{\svec}{\vb{s}^a}   
\newcommand{\tsvec}{\widetilde{\vb{s}}^a}

\newcommand{\xvec}{\vec{x}^a}
\newcommand{\xvecwoa}{\vec{x}}

\newcommand{\cvec}{\vb{c} }   
\newcommand{\cvecwa}{\vb{c}^a}

\newcommand{\Xmat}{\vb{X}^a}  

\newcommand{\Ymat}{\vb{Y}^a}  

\newcommand{\residual}{\vb{r}}
\newcommand{\rvec}{\residual}   
\newcommand{\rvecwa}{\vb{r}^a}

\newcommand{\nTwoLO}{N$^2$LO\xspace}

\newcommand{\chiEFT}{$\chi$EFT\xspace}

\begin{document}

\title{Emulation of Proton-Deuteron Scattering via the Reduced Basis Method \texorpdfstring{\\}{} and Active Learning: Detailed Description}

\author{Alex Gnech~\orcidlink{0000-0002-2077-3866}}
\email{agnech@odu.edu}
\affiliation{Department of Physics, \href{https://ror.org/04zjtrb98}{Old Dominion University}, Norfolk, Virginia 23529, USA}
\affiliation{Theory Center, \href{https://ror.org/02vwzrd76}{Jefferson Lab}, Newport News, Virginia 23610, USA}

\author{Xilin Zhang~\orcidlink{0000-0001-9278-5359}} 
 \email{zhangx@frib.msu.edu}
\affiliation{\href{https://ror.org/03r4g9w46}{Facility for Rare Isotope Beams}, \href{https://ror.org/05hs6h993}{Michigan State University}, East Lansing, MI~48824, USA}

\author{Christian Drischler~\orcidlink{0000-0003-1534-6285}}
\email{drischler@ohio.edu}
\affiliation{Department of Physics and Astronomy, \href{https://ror.org/01jr3y717}{Ohio University}, Athens, OH~45701, USA}
\affiliation{\href{https://ror.org/03r4g9w46}{Facility for Rare Isotope Beams}, \href{https://ror.org/05hs6h993}{Michigan State University}, East Lansing, MI~48824, USA}

\author{R.~J. Furnstahl~\orcidlink{0000-0002-3483-333X}}
\email{furnstahl.1@osu.edu}
\affiliation{Department of Physics, \href{https://ror.org/00rs6vg23}{The Ohio State University}, Columbus, OH 43210, USA}

\author{Alessandro Grassi~\orcidlink{0000-0002-5703-3183}}
\email{alessandro.grassi@df.unipi.it}
\affiliation{Department of Physics “E. Fermi”, \href{https://ror.org/03ad39j10}{University of Pisa}, I-56127, Pisa, Italy}
\affiliation{\href{https://ror.org/01vj6ck58}{Istituto Nazionale di Fisica Nucleare, Sezione di Pisa}, I-56127 Pisa, Italy}

\author{Alejandro Kievsky~\orcidlink{0000-0003-4855-6326}}
\email{alejandro.kievsky@pi.infn.it}
\affiliation{\href{https://ror.org/01vj6ck58}{Istituto Nazionale di Fisica Nucleare, Sezione di Pisa}, I-56127 Pisa, Italy}

\author{Laura E. Marcucci~\orcidlink{0000-0003-3387-0590}}
\email{laura.elisa.marcucci@unipi.it}
\affiliation{Department of Physics “E. Fermi”, \href{https://ror.org/03ad39j10}{University of Pisa}, I-56127, Pisa, Italy}
\affiliation{\href{https://ror.org/01vj6ck58}{Istituto Nazionale di Fisica Nucleare, Sezione di Pisa}, I-56127 Pisa, Italy}

\author{Michele Viviani~\orcidlink{0000-0002-4682-4924}}
\email{michele.viviani@pi.infn.it}
\affiliation{\href{https://ror.org/01vj6ck58}{Istituto Nazionale di Fisica Nucleare, Sezione di Pisa}, I-56127 Pisa, Italy}

\date{\today}

\begin{abstract}
Nucleon-deuteron ($Nd$) scattering can be used to constrain three-nucleon forces in chiral effective field theory ($\chi$EFT). However, high-fidelity calculations, such as the Hyperspherical Harmonic (HH) method, are computationally expensive, making it difficult or even prohibitive to explore the vast parameter space of $\chi$EFT\xspace. To address this challenge, specifically for proton-deuteron ($pd$) scattering below the deuteron breakup threshold, we developed model-driven emulators based on the Reduced Basis Method (RBM) and active learning techniques, as presented in \href{https://arxiv.org/abs/2511.01844}{arXiv:2511.01844}. The method exploits the similarities between solutions at different parameter points to significantly reduce computational costs. 
In this companion paper, we provide a comprehensive description of our HH-based high-fidelity calculations and implementation of both variational-method-based and Galerkin-projection-based scattering emulators. 
We demonstrate the effectiveness of active learning in the form of greedy algorithms for selecting optimal training points in the parameter space, and the high accuracy and speed of the emulators, for two different nucleon forces and two scattering channels (${1/2}^+$ and ${1/2}^-$). For example, in a two-dimensional parameter space, the relative emulation errors can be reduced to $10^{-7}$ with fewer than 10 training points. Our work paves the way for the efficient calibration of $\chi$EFT\xspace nucleon interactions using Bayesian statistics, and the methodology can be applied to other nuclear scattering processes (including neutron-deuteron scattering), as well as other finite quantum systems. 
\end{abstract}

\maketitle

\section{Introduction}%
Nucleon-deuteron ($Nd$) scattering is an important process for calibrating, testing, and validating implementations of chiral effective field theory ($\chi$EFT)~\cite{LENPIC:2018ewt,Epelbaum:2002vt,Epelbaum:2008ga}. 
Of particular interest is the calibration of three-nucleon (3N) forces~\cite{Weinberg:1992yk, Epelbaum:2002vt,Hammer:2012id, Epelbaum:2012zz, Hebeler:2020ocj,Epelbaum:2019kcf,Machleidt:2024bwl}, which enter the chiral expansion at next-to-next-to-leading order (\nTwoLO) in Weinberg power counting~\cite{Weinberg:1990rz,Weinberg:1991um,Weinberg:1992yk,Epelbaum:2008ga,Machleidt:2011zz}, with principled uncertainty quantification (UQ)~\cite{Phillips:2020dmw,Wesolowski:2021cni}.
To this end, fast and accurate predictions for scattering observables are crucial. 
Although highly accurate (high-fidelity) methods~\cite{Gloeckle:1995jg,DeltuvaCoulombReview2008,Kievsky2008,Deltuva:2012kt,Lazauskas:2019rfb,Marcucci2019} exist to solve the few-body scattering problem, their application for UQ, where high-fidelity calculations must be repeated many times to sample the model parameter spaces, poses a formidable computational challenge.
Emulators, or surrogate models~\cite{Duguet:2023wuh, Melendez:2022kid, Drischler:2022ipa}, have become promising tools to facilitate these otherwise impossible few-body calculations with high accuracy and speed.

In Ref.~\cite{Nd_emulator_2025_short}, we introduced fast and accurate scattering emulators for proton-deuteron ($pd$) scattering below the deuteron breakup threshold, based on the so-called reduced-basis method (RBM)~\cite{hesthaven2015certified,Quarteroni:218966,Duguet:2023wuh,Melendez:2022kid,Drischler:2022ipa}.
This paper provides the necessary technical details.
We present a comprehensive description of the high-fidelity Hyperspherical Harmonic (HH) calculations~\cite{Kievsky2008,Marcucci2019}\footnote{See also Refs.~\cite{Leidemann:2012hr, Greene:2017cik, Rittenhouse_2011} for discussions of different HH-based methods and some early developments.} and a detailed implementation of three distinct RBM emulators. Furthermore, we discuss the greedy algorithms~\cite{Quarteroni:218966,Sarkar:2021fpz,Maldonado:2025ftg} used as an active learning tool that reduces the computational costs of emulator training. A compilation of emulation results for various \chiEFT nucleon interactions and scattering channels is included here, providing a general understanding of the emulator performance and a baseline for further developments. 

Chiral interactions are the fundamental inputs for modern \abi calculations of static properties of nuclei and nuclear dynamics~\cite{Weinberg:1990rz,Weinberg:1991um,Weinberg:1992yk,Epelbaum:2008ga,Machleidt:2011zz,Epelbaum:2019kcf}. It has been recognized that the 3N force is a key component for reproducing emergent phenomena~\cite{Hammer:2012id,vanKolck:1994yi,Epelbaum:2002vt,LENPIC:2015qsz}, for example, enabling nuclear saturation~\cite{Drischler:2021kxf} and reproducing experimental results for charge-radii and binding energies~\cite{Ekstrom:2015rta}. While part of the 3N force can be calibrated by analyzing two-nucleon scattering experimental data within the EFT framework, some other low-energy constants (LECs) need to be fixed against higher-body systems~\cite{Hebeler:2020ocj}. $Nd$ scattering has been a crucial source of experimental data for this purpose~\cite{Epelbaum:2002vt}. In this work, we vary the so-called $\cE$ and $\cD$ parameters in the leading-order contribution of the 3N force (in Weinberg power counting), and thus work with a two-dimensional parameter space.   

A long-standing computational challenge in using $Nd$ scattering data is the high cost of high-fidelity \abi calculations~\cite{Gloeckle:1995jg, Kievsky2008, Lazauskas:2019rfb}, such as the HH method discussed in this paper.
The HH approach~\cite{Kievsky2008,Marcucci2019} first constructs large vectors and matrices and then solves related high-dimensional linear equations. Typically, these matrices are dense, with a dimensionality that can reach $10^3$ to $10^4$ to ensure reasonable convergence for calculations of $Nd$ below-breakup scattering amplitudes. Consequently, it is not feasible to simply repeat these calculations when exploring the parameter space of the \chiEFT 3N force, especially in the context of Bayesian statistics~\cite{Phillips:2020dmw}, where a large number of samples of such calculations are needed to gain a reliable understanding of the range of allowed parameter values.
  
The emulators developed here are designed to significantly speed up the scattering amplitude predictions when exploring the parameter space. Specifically, our RBM emulators are model-driven rather than data-driven (the latter includes artificial neural networks~\cite{Bogojeski2020,2020MNRAS.494.2465B} and Gaussian processes~\cite{Mackay:1998introduction, rasmussen2006gaussian}). The model-driven emulators, studied extensively in the field of model order reduction (MOR)~\cite{hesthaven2015certified, Quarteroni:218966, Benner_2017aa, Benner2017modelRedApprox, benner2015survey}, are intrusive by incorporating insights into the general structure of the underlying equation systems, such as the \schro or Faddeev equation in $Nd$ scattering. In contrast, data-driven emulators learn a low-dimensional representation of the data without these insights. Although more broadly applicable, data-driven emulators typically require a significantly larger amount of training data and achieve a lower emulation accuracy~\cite{Konig:2019adq,Drischler:2022ipa}.

Previously, the RBM method~\cite{hesthaven2015certified,Quarteroni:218966}, also known as eigenvector continuation (EC)~\cite{Frame:2017fah,Duguet:2023wuh, Melendez:2022kid, Drischler:2022ipa} in nuclear theory, has been used successfully to develop model-driven emulators to compute nuclear ground and resonance states~\cite{Frame:2017fah,Sarkar:2020mad,Sarkar:2021fpz,Konig:2019adq,Demol:2019yjt,Ekstrom:2019lss,Demol:2020mzd,Yoshida:2021jbl,Anderson:2022jhq,Giuliani:2022yna,Yapa:2023xyf,Yapa:2024lya}, as well as various two-body~\cite{Furnstahl:2020abp,Drischler:2021qoy, Melendez:2021lyq,Zhang:2021jmi,Bai:2021xok, Drischler:2022yfb, Melendez:2022kid, Drischler:2022ipa, Bai:2022hjg, Garcia:2023slj, Odell:2023cun, Maldonado:2025ftg} and simple three-body~\cite{Zhang:2021jmi} scattering states.   
It should also be noted that the RBM method can be modified to form a data-driven approach, which has been further extended to a new machine learning method, called Parameterized Matrix Models (PMM)~\cite{Cook:2024toj}. 

The main strategy shared by all our emulators, and also the core idea of the RBMs, is to approximate the solution ($\yvec$) of our $Nd$ scattering equations (at a given scattering energy $E$), as a function of the parameter $\thetavec$ of the \chiEFT Hamiltonian $H$, by a linear sum of the high-fidelity solutions at a small set of points ($\thetavec_\mu$),
\begin{align}
    \yvec(\thetavec) \approx \sum_{\mu=1}^{\dimL} \cemul_\mu(\thetavec) \yvec(\thetavec_\mu) \label{eq:RBM_ansatz} \ . 
\end{align}
The reader should not confuse the coefficients $\cemul_\mu$ with the LECs (e.g.,~$\cE$ and $\cD$) of the chiral forces. This approximation enables projecting the original high-dimensional linear equation system onto a $\dimL$-dimensional subspace and forming a lower-dimensional linear equation system. For problems having $\yvec(\thetavec)$ smoothly varying with $\thetavec$, this projection is extremely efficient. Essentially, the projection enables us to leverage the similarities in solutions at different $\thetavec$ values, thereby eliminating redundancies and reducing overall computing costs. Also note that the basic structure of the original equation system is preserved in this approximation. 

Equation~\eqref{eq:RBM_ansatz} also separates the task of the parameter exploration into an offline (or training) stage and an online (or emulation) stage. In the offline stage, we obtain the solutions $\yvec(\thetavec_\mu)$, called snapshots. During the online stage, $\cemul_\mu(\thetavec)$ is quickly inferred at a general $\thetavec$, by solving the low-dimensional projected equation system. The efficiency of the subspace projection ensures the smallest possible number of snapshots required, resulting in the lowest possible computing costs for emulator training. Moreover, the small dimension of the subspace is also a key reason for achieving extreme computing speed during the online stage. 

On the other hand, the smooth $\thetavec$-dependence of $\yvec$  also means that the snapshot basis becomes numerically ``over-determined'' and the low-dimensional equation turns ill-conditioned very quickly when we keep including new snapshots. Extra procedures, such as Proper Orthogonal Decomposition (POD)~\cite{hesthaven2015certified,Quarteroni:218966}, could be implemented at the offline stage to identify the effective dimension of the subspace and also gain numerical stability (see, e.g., Refs.~\cite{Bonilla:2022rph,Giuliani:2022yna}). 

In this work, we utilize greedy algorithms, as developed in Ref.~\cite{Maldonado:2025ftg} for two-body scattering emulations, to iteratively and strategically choose the locations of the high-fidelity training calculations in the parameter space, where the estimated emulation errors, based on numerically inexpensive error estimation, are the largest. In this way, we not only achieve numeric stability but also keep the training cost (i.e., $\dimL$) at its minimum. 

The emulators developed here can, in principle, be adapted to handle other types of scattering calculations, such as Faddeev approaches with real scattering energies~\cite{Glockle:1983,Gloeckle:1995jg,Nielsen:2001hbm,DeltuvaCoulombReview2008,Deltuva:2012kt,Lazauskas:2019hil}. Continuum calculations based on various Non-Hermitian quantum mechanics (NHQM) methods%
\footnote{
The NHQM category includes the well-known complex scaling~\cite{reinhardt1982complex,Moiseyev_2011,Myo:2014ypa}, and Berggren basis methods~\cite{Michel:2021jkx}. Recent works~\cite{Zhang:2024ril, Zhang:2024gac} argue that the so-called complex-energy~\cite{Schlessinger:1966zz,Carbonell:2013ywa} and Lorentz integral transform~\cite{Efros:2007nq} methods also fall into this group.}  can also be emulated using the RBMs. Early studies along this line can be found in Refs.~\cite{Zhang:2024ril, Zhang:2024gac, Liu:2024pqp}. 

The rest of the paper is organized as follows. We discuss the scope of this work, including the \chiEFT parameters and the $Nd$ scattering channels, in Sec.~\ref{sec:scope}. We introduce the employed HH method in Sec.~\ref{sec:HH}, with sufficient details to make the paper self-contained. Two main types of emulations are discussed in Secs.~\ref{sec:var_emulation} and~\ref{sec:G_LSPG_ROM}. In particular, the greedy algorithm is discussed in Sec.~\ref{sec:G_LSPG_ROM}. In Sec.~\ref{sec:results} we discuss the numerical results. We conclude this paper with a summary and outlook. The emulators developed here will be made public on the BUQEYE website~\cite{buqeye}.
 
\section{$Nd$ scattering based on \chiEFT nuclear force} \label{sec:scope}

The \chiEFT Hamiltonian has the following form:    
\begin{equation}
         H(\thetavec) = \sum_{i=0}^{\dimparam}  \theta_i \, H_i \ , \label{eq:Haffinedecomp}
\end{equation}
where $H_0$ collects all the parameter-independent parts of the Hamiltonian (with the corresponding $\param_{0} = 1$), such as the kinetic energy term, two-body, and some three-body interactions. The $H_{i\geq 1}$ terms are three-body interaction operators with LECs $\param_{i}$ controlling their overall strengths. 

The linear dependence of $H(\thetavec)$ on the LECs, which is carried into its representation in the emulation subspace, is crucial for achieving emulation efficiency~\cite{Drischler:2022ipa}. The computationally expensive, high-dimensional linear algebraic manipulations, such as projecting $H_i$ onto the subspace, only need to be performed at the training stage. During emulation, the weights of the different $H_i$ components can be adjusted instantly; see also the related discussions in Secs.~\ref{sec:var_emulation} and~\ref{sec:G_LSPG_ROM}. A more general affine parameter dependence, studied in MOR, offers the same advantage: the parameter dependence, albeit being nonlinear, is factorized out from the large computations. For problems that do not satisfy this property explicitly, existing strategies, such as incorporating data-driven approaches in the RBM framework~\cite{Zhang:2021jmi} or approximating the non-affine parametric dependence in terms of the affine dependence~\cite{Odell:2023cun} could be explored.
Examples of parameters that enter the chiral interactions with non-affine dependencies include the regulator cutoffs and pion masses.

For the remainder of the paper, it should be kept in mind that the index $i$ in $\param_{i}$ lists the $H_i$ component in $H(\thetavec)$, while the $\mu$ index in $\thetavec_\mu$ (e.g., in Eq.~\eqref{eq:RBM_ansatz}) labels the training solution (or snapshot). This distinction is always implicitly assumed. 

The particular Hamiltonian we choose to work with is the Norfolk \chiEFT interaction (NV) developed in Refs.~\cite{Piarulli2015,Piarulli2016}. The Coulomb interaction, which is crucial for low-energy $pd$ scattering, is also included.\footnote{Although this paper is focused on  $pd$ scattering, neutron-deuteron scattering emulators can be created by simply turning off the Coulomb interaction.}  To validate the emulators performance using different regularization schemes, two versions of the NV force,  NVIIa and NVIIb, are explored. NVIIa uses ultraviolet regulators with $(R_L, R_S)= (1.2, 0.8)$~fm, while NVIIb has harder cutoffs with $(R_L, R_S)= (1.0, 0.7)$~fm. Both forces were fitted to $NN$ scattering data up to $200$ MeV in the laboratory frame. 

We coupled the two-nucleon interaction with the 3N forces constructed in Ref.~\cite{Piarulli:2017dwd}. The contribution coming from the pion exchange diagrams inside the 3N forces is included as part of the $H_0$ term in Eq.~\eqref{eq:Haffinedecomp}. The terms that multiply $c_E$ and $c_D$ correspond to $H_1$ and $H_2$ in the same equation. That is, we identify $c_E$ and $c_D$ as $\theta_1$ and $\theta_2$ respectively, the only two varying parameters in this work. Since the two-nucleon interaction is fixed, so is the deuteron internal wave function, which simplifies the emulation process (see Eqs.~\eqref{eq:T00affine}--\eqref{eq:TXYaffine}). For more general cases, where the deuteron wave function is also varied, fast emulation can still be achieved, although it will be more involved~\cite{Zhang:2021jmi}. 

Finally, we study scattering in both $\halfplus$ and $\halfminus$ (in $J^{\pi}$ notation) scattering states. As will be seen later, the total scattering amplitudes below the breakup threshold are dominated by the $\halfplus$ component. The $\halfminus$ calculations, however, are used to further test our emulators in $p$-wave scatterings (note the $\halfplus$ is dominated by the $s$-waves). A thorough study of all the partial waves will be performed in future work. 

\section{Hyperspherical Harmonic method for computing three-body scattering} \label{sec:HH}

To describe $Nd$ scattering at low energies, below the breakup threshold in particular, we use partial-wave decomposition. 
For each partial wave, we write the  wave function (below the deuteron breakup threshold) by separating it into a ``core'' part, labeled $C$, and an ``asymptotic''  part, labeled  $A$~\cite{Kievsky2008,Marcucci2019}:\footnote{This type of decomposition was previously used in, e.g., Ref.~\cite{Kohn:1948col} and early HH developments~\cite{Efros1969HH, Efros:1971vff}.}
\begin{equation}
    \Psi^{LSJJ_z}=\Psi^{LSJJ_z}_{C}+\Psi^{LSJJ_z}_{A}\,. \label{eq:wf_decomp1}
\end{equation}
The superscript ${LSJJ_z}$ labels the incoming channel of the continuum state, where $L$, $S$ are the relative orbital and total spin of the system, and $J$ and $J_z$ are the total spin and its projection. The parity of the state ($\pi$) is given by $(-1)^L$.
For a given $J$ and $J_z$, states with different but allowed $L$ and $S$ values are generally coupled. The core part describes the $Nd$ system when the particles are close and the strong interaction is active. The asymptotic part describes the motion of the $Nd$ system when the strong interaction between $N$ and $d$ is negligible, leaving only the Coulomb interaction in the $pd$ case.  As briefly mentioned in Sec.~\ref{sec:scope}, we focus only on the channels with total angular momentum $J = \half$, which involve $L = 0, 1, 2$. Considering the conservation of parity, we can separate two scattering states $\halfplus$ and $\halfminus$, for which the $L$ and $S$ values are listed in Table~\ref{tab:channels}.

\begin{table}[tb]
\renewcommand{\arraystretch}{1.4}
\caption{Channels and corresponding labels used for the  $J^\pi=\halfplus$ and $\halfminus$ scattering states.}
    \centering
\begin{ruledtabular}
    \begin{tabular}{ccc} 
        \multirow{2}{*}{channel} & \multicolumn{2}{c}{scattering state} \\[2pt]
    \cline{2-3} 
     & $J^\pi=\halfplus$ & $J^\pi=\halfminus$ \\[2pt]
    \hline
    1 & $L=0\,, \ S=1/2$ & $L=1\,, \ S=1/2$\\
    2 & $L=2\,, \  S=3/2$ & $L=1\,, \ S=3/2$\\
    \end{tabular}
\end{ruledtabular}
    \label{tab:channels}
\end{table}

The core wave function is expanded in HH functions, exactly as it is done for $A=3$ bound states~\cite{Kievsky2008,Marcucci2019}:
\begin{equation}\label{eq:psicore}
    \Psi^{LSJJ_z}_C(\rho,\Omega_p)=\sum_{[K],m}\cHH^{LSJ}_{[K],m}f_m(\rho)\sum_{p=1}^3{\cal Y}_{[K]}(\Omega_p)\,,
\end{equation}
where $f_m(\rho)$ is the polynomial expansion on the hyperradius $\rho$, typically performed using regularized Laguerre polynomials, and $p$ represents the sum over the even permutations of the three particles. The term ${\cal{Y}}_{[K]}(\Omega_p)$ is the HH function of order $K$, where $\matrixform{K}$ represents the full set of quantum numbers needed to describe the HH function and $\Omega_p$ the hyperspherical angles. The linear coefficients $\cHH^{LSJ}_{[K],m}$ are determined by solving the scattering problem. The detailed form of the HH basis expansion can be found in Refs.~\cite{Kievsky2008,Marcucci2019}.

The asymptotic wave function is a linear combination of  the regular and irregular solutions of the \schro equation, written as
\begin{equation}
\Psi_A^{L S J J_z}=\sum_{L^{\prime} S^{\prime}}\left[\delta_{L L^{\prime}} \delta_{S S^{\prime}} \Omega_{L^{\prime} S^{\prime} J J_z}^R+\mathcal{R}_{L S\rightarrow L^{\prime} S^{\prime}}^J \Omega_{L^{\prime} S^{\prime} J J_z}^I\right]\,. \label{eq:HHwf_A}
\end{equation}
Here,  $R$ and $I$ indicate the regular and irregular part, respectively, and $\mathcal{R}^J_{L S, L^{\prime} S^{\prime}}$ is the $R$-matrix, which represents the relative weight between the regular and irregular component of the wave function at a given energy and can be determined by solving the scattering problem. The two asymptotic functions are given by the following expression:
\begin{align}
\Omega_{L S J J_z}^{R(I)}&=\frac{\mathcal{C}}{\sqrt{3}} \sum_{p=1}^{3}\left[\left[\chi_{1 / 2}(N) \phi_{S_d}(d)\right]_S Y_L\left({\hat{\bm y}}_p\right)\right]_{J J_z} \notag \\
 & \qquad\qquad\qquad\null\times
 R_L^{R(I)}\left(y_p\right)\,,
\end{align}
where again the sum over $p$ is over the three possible even permutations of the nucleons, $\chi_{1 / 2}(N)$ is the wave function of the nucleon (i.e., its spin and isospin state), $\phi_{S_d}(d)$ is the deuteron wave function, and $\bm {y}_p$ is the relative distance between the nucleon and the center of mass of the deuteron. The functions $R_L^{R(I)}\left(y_p\right)$
are the normalized solutions of the \schro equation of a two-body system (i.e., for the $N$$d$ relative motion) without nuclear interaction:

\begin{equation}
\begin{aligned}
& R_L^R\left(y_p\right)=\frac{1}{(2 L+1) ! ! q^L C_L(\eta)} \frac{F_L\left(\eta, q y_p\right)}{q y_p}\,, \\
& R_L^I\left(y_p\right)=(2 L+1) ! ! q^{L+1} C_L(\eta) f\left(b, y_p\right) \frac{G_L\left(\eta, q y_p\right)}{q y_p}\, ,
\end{aligned}
\end{equation}
where $\eta=Z_NZ_d\mu e^2/q$ is the Coulomb parameter, and $F_L\left(\eta, q y_p\right)$
and $G_L\left(\eta, q y_p\right)$ are the regular and irregular Coulomb functions with $q$ the modulus of the $Nd$ relative momentum, that can be obtained from the kinetic energy in the center-of-mass frame $E\equiv q^2/2\mu$. The factors $C_L(\eta)$ $f(b,y_p)$ are given by
\begin{equation}
C_L(\eta)=\frac{2^L \mathrm{e}^{-\frac{\pi \eta}{2}}|\Gamma(L+1+i \eta)|}{\Gamma(2 L+2)}\,,
\end{equation}
and 
\begin{equation}
    f(b,y_p)=(1-e^{-b y_p})^{2L+1}\,,  
\end{equation}
where the latter is needed to cure the divergent behavior of $G_L(\eta,qy_p)$ at small values of $y_p$. The tunable parameter $b$ controls how quickly the regularized $G_L(\eta,qy_p)$ approaches the desired asymptotic behavior when increasing $y_p$. The normalization coefficient $\cal C$ is selected in order to fulfill the condition:
\begin{equation}
\left\langle\Omega_{L S J J_z}^R \right|H-E \left| \Omega_{L S J J_z}^I\right\rangle-\left\langle\Omega_{L S J J_z}^I|H-E| \Omega_{L S J J_z}^R\right\rangle=1\,,
\end{equation}
which corresponds to the Wronskian relation among the regular and irregular parts of the asymptotic wave function.

Notice that the coefficients $\cHH^{LSJ}_{[K],m}$ and 
$\mathcal{R}^J_{L S, L^{\prime} S^{\prime}}$, as well as the asymptotic functions $\Omega_{L^{\prime} S^{\prime} J J_z}^{R(I)}$, are the only terms in the wave function that depend on $q$. A detailed discussion on the wave function construction can be found in Refs.~\cite{Kievsky2008,Marcucci2019}. Moreover, to deal with the neutron-deuteron scattering case, we can simply substitute the Coulomb functions with the corresponding spherical Bessel functions. 

To reduce the number of indices, we now define $\xi= \{ [K],m \}$ and $a={L,S}$ and work implicitly with a given $J$ and $J_z$. Using the new labels and the ket notation, Eq.~\eqref{eq:wf_decomp1} turns into

\begin{align}
    |\wf{a} \rangle  
    &= \sum_{\xi=1}^{\dimxi} \cHH^{a}_{\xi}|\xi \rangle + \sum_{a_0=1}^{\dimC} \left(\delta_{a,a_0}  |\Omega_{a_0}^R\rangle+\rmat{a_0}{a} | \Omega_{a_0}^I\rangle\right)  \ , \label{eq:HHwfv2}
\end{align}
where $|\xi\rangle$ is an abbreviated notation for the HH expansion given in Eq.~\eqref{eq:psicore}, $N_s$ is the total number of states used in the HH expansion, and $N_c$ is the number of coupled channels for a given $J$ and parity.

At this point, in order to fully determine our wave function, we use the  Kohn variational principle (KVP)~\cite{Kohn:1948col,JoachainQCT1975,newton2002scattering} to obtain the parameters $\cHH^{a}_{\xi}$ and $\rmat{a_0}{a}$. The KVP states that, for a given pair of $a$ and $a'$, the functional,
\begin{equation}\label{eq:kvp}
\rfunc{a'}{a}\big[\wf{a},\wf{a'}\big]
=
\rmat{a'}{a}(E)-\big\langle\wf{a'}\big|H-E\big| \wf{a}\big\rangle\,,
\end{equation}
is stationary with respect to the variation of $\wf{a}$ and $\wf{a'}$ about the exact solutions. Importantly, the functional, if evaluated with trial $\wf{a}$ and $\wf{a'}$ that are close to the exact ones, provides a second-order estimate of the corresponding $\rmat{a'}{a}$. Bear in mind that in this paper, $\rmat{a'}{a}$ are mostly understood as the first-order estimates, unless stated explicitly otherwise. In Eq.~\eqref{eq:kvp}, we have made explicit the dependence on the energy $E$ of the coefficients $\rmat{a'}{a}$.

Based on our constructions of the wave functions, the stationary condition for the diagonal functional $\rfunc{a}{a}$ means   

\begin{equation}\label{eq:stat}
    \frac{\partial \rfunc{a}{a}}{\partial \cHH^{a}_{\xi}}=0\,,\qquad
    \frac{\partial \rfunc{a}{a}}{\partial \rmat{a_0}{a}}=0\,. 
\end{equation}
Again, as in Eq.~\eqref{eq:HHwfv2}, $a_0$ labels all the coupled channels, and thus it can be the same as $a$ or different from it. For a given $a$, we obtain a system of coupled linear equations that constrains all the parameters, including both the diagonal and the off-diagonal $\rmat{a_0}{a}$, since they are all part of the variational parameters inside $|\wf{a}\rangle$.

Note that we separate $a_0$ from $a'$ in our discussions to emphasize that~(i) $\rmat{a_0}{a}$ is the parameter inside the trial wave function $|\wf{a}\rangle$; its solved value from Eq.~\eqref{eq:stat} is the first-order estimate of the $R$-matrix element, and~(ii) $\rmat{a'}{a}$ is used in general discussions of the off-diagonal $R$-matrix elements.

Performing the variational calculations, we encounter the following matrix elements (here, $a'$ is just a general index):

\begin{align}
& T^{00}_{\xi,\xi'}=\langle \xi |H-E|\xi'\rangle=\langle \xi' |H-E|\xi\rangle\,,\label{eq:T_in_xi_Omega_1}\\
& T^{R}_{\xi,a}=\langle \xi |H-E|\Omega^R_{a}\rangle=\langle \Omega^R_{a}|H-E| \xi\rangle\,,\\
& T^{I}_{\xi,a}=\langle \xi |H-E|\Omega^I_{a}\rangle=\langle \Omega^I_{a}|H-E| \xi\rangle\,,\\
& T^{RI}_{a,a'}=\langle \Omega^R_{a} |H-E|\Omega^I_{a'}\rangle\,,\\
& T^{IR}_{a,a'}=\langle \Omega^I_{a}|H-E| \Omega^R_{a'}\rangle\,,\\
& T^{RR}_{a,a'}=\langle \Omega^R_{a} |H-E|\Omega^R_{a'}\rangle\,,\\
& T^{II}_{a,a'}=\langle \Omega^I_{a}|H-E| \Omega^I_{a'}\rangle\,.\label{eq:T_in_xi_Omega_-1}
\end{align}

\begin{widetext}
By solving Eq.~(\ref{eq:stat}), we obtain the following linear system for a given $a$ (see also the details in Appendix~\ref{app:HH_details}) in the matrix form $\Amat \xvec = \svec$:

\begin{equation}
\renewcommand{\arraystretch}{1.5}
\left(\begin{array}{c c c   c c c}
T^{00}_{\xi=1,\xi'=1} &  \cdots  &  T^{00}_{\xi=1,\xi'=\dimxi}   &  T^{I}_{\xi=1,a_0'=1} & \cdots & T^{I}_{\xi =1 ,a_0'=\dimC} \\
  & \vdots &  &   & \vdots &    \\ 
T^{00}_{\xi=\dimxi,\xi'=1} &  \cdots  &  T^{00}_{\xi=\dimxi,\xi'=\dimxi}     &  T^{I}_{\xi=\dimxi,a_0'=1} & \cdots & T^{I}_{\xi =\dimxi ,a_0'=\dimC} \\[2ex] 
T^{I}_{\xi'=1, a_0=1 } & \cdots & T^{I}_{\xi'=\dimxi,a_0=1}  & \widetilde{T}^{II}_{a_0=1,a_0'=1} & \cdots & \widetilde{T}^{II}_{a_0=1,a_0'=\dimC} \\ 
  & \vdots &   &   & \vdots &    \\ 
T^{I}_{\xi'=1, a_0=\dimC } & \cdots & T^{I}_{\xi'=\dimxi,a_0=\dimC}  & \widetilde{T}^{II}_{a_0=\dimC, a_0'=1} & \cdots & \widetilde{T}^{II}_{a_0=\dimC,a_0'=\dimC} \\  
\end{array} \right)
\left(\begin{array}{c}
\cHH^a_{\xi'=1} \\
\vdots \\ 
\cHH^a_{\xi'=\dimxi} \\[2ex] 
\rmat{a_0'=1}{a} \\ 
\vdots \\ 
\rmat{a_0'=\dimC}{a}\\
\end{array} \right)
= 
\left(\begin{array}{c}
-T^{R}_{\xi=1,a} \\
\vdots \\ 
-T^{R}_{\xi=\dimxi,a} \\[2ex]  
 \half \delta_{a,a_0=1}  - \widetilde{T}^{IR}_{a_0=1, a}  \\ 
\vdots \\ 
\half \delta_{a,a_0=\dimC} - \widetilde{T}^{IR}_{a_0=\dimC, a}\\
\end{array} \right) \,. \label{eq:lineareq_highfidelity_matrix_form}
\end{equation}
\end{widetext}
Here, $\Amat$ is a $\dimH \times \dimH$ generally dense matrix, and $\svec$ is a $\dimH$ long vector, with $\dimH \equiv \dimxi+\dimC$ (the dimensionality of the high-fidelity calculation for a fixed $a$).  Moreover,

\begin{align}
 \widetilde{T}^{II}_{a_0, a_0'}  & \equiv    \frac{1}{2}(T^{II}_{a_0,a_0'} + T^{II}_{a_0',a_0})\,, \\
 \widetilde{T}^{IR}_{a_0, a}  & \equiv    \frac{1}{2}(T^{IR}_{a_0, a}+T^{RI}_{a, a_0})\,.
\end{align}
It is interesting to note that $\Amat$ does not have explicit dependence on a specific channel $a$, while $\svec$ does.

After solving the equation for a given $a$, we get the first-order approximation for the wave function $|\wf{a}\rangle$, including the associated $R$-matrix elements, $\rmat{a_0}{a}$. The second-order estimate of $\rmat{a}{a}$ is readily achieved by evaluating $\rfunc{a}{a}$ with the first-order estimate of $|\wf{a}\rangle$. This step is repeated for all the relevant $a$, which produces the second-order estimates of all the diagonal $R$-matrix elements.  
At the first order the off-diagonal elements are in general numerically different, i.e. $\rmat{a_0}{a} \neq \rmat{a}{a_0}$, as they are constrained by independent equation systems---although, the difference becomes negligible in fully converged calculations. The second order estimate of the off-diagonal elements
are obtained directly from Eq.~\eqref{eq:kvp} using the first order wave functions. This reduces drastically the numerical discrepancy compared the first-order estimates.

Finally, it is easy to see the linear (or affine) dependence of the matrix elements on $\thetavec$ in Eqs.~\eqref{eq:T_in_xi_Omega_1}--\eqref{eq:T_in_xi_Omega_-1}, as inherited from the same properties of the $H(\thetavec)$ in  Eq.~\eqref{eq:Haffinedecomp}. They can be decomposed into individual terms associated with each $\theta_i$:
\begin{align}
 T^{00}_{\xi,\xi'}(i)&=\langle \xi |H_i-E\delta_{i,0}|\xi'\rangle\,, \label{eq:T00(i)def} \\
 T^{X}_{\xi,a}(i) & =\langle \xi |H_i-E\delta_{i,0}|\Omega^X_{a}\rangle \notag \\
                  & =\langle \Omega^X_{a}|H_i-E\delta_{i,0}| \xi\rangle\,, \label{eq:Txia(i)def} \\
 T^{XY}_{a,a'}(i)&=\langle \Omega^X_{a} |H_i-E\delta_{i,0}|\Omega^Y_{a'}\rangle\,, \label{eq:Taap(i)def}
\end{align}
where $X,Y=R,I$ and $i=0,\ldots,\dimparam$. Since we only vary the $3N$ force, the deuteron internal wave function is fixed, and so are the $\Omega^{I(R)}$ components. Therefore, the $\thetavec$ dependence of the matrix elements is strictly linear:
\begin{align}
     T^{00}_{\xi,\xi'}(\thetavec) &  = \sum_{i=0}^{\dimparam} \theta_i T^{00}_{\xi,\xi'}(i) \ , \label{eq:T00affine}\\
     T^{X}_{\xi,a}(\thetavec) &  = \sum_{i=0}^{\dimparam} \theta_i T^{X}_{\xi,a}(i) \ , \label{eq:TXaffine} \\
    T^{XY}_{a,a'}(\thetavec) &  = \sum_{i=0}^{\dimparam} \theta_i T^{XY}_{a,a'}(i) \ . \label{eq:TXYaffine}
\end{align}
Such dependence will be used to achieve fast emulations, as discussed in the following sections. 

\section{Variational emulations}
\label{sec:var_emulation}

Here, we follow the basic strategy of utilizing the KVP for emulations, as developed in, e.g., Refs.~\cite{Furnstahl:2020abp, Drischler:2021qoy, Zhang:2021jmi,Garcia:2023slj}. We identify, for a fixed energy $E$ and a given $a\to a'$ transition, the full HH scattering wave function  $ |\wf{a}(\thetavec) \rangle$ (and $ |\wf{a'}(\thetavec)\rangle $ if $a' \neq a$) as the $\yvec(\thetavec)$  variable in Eq.~\eqref{eq:RBM_ansatz}. 
The coefficients $\cemul_\mu(\thetavec)$ appearing in Eq.~\eqref{eq:RBM_ansatz} are now the new variational parameters in the trial wave functions that have to be determined, instead of the parameters in Eq.~\eqref{eq:HHwfv2}. We can then insert these new trial wave functions into the $\rfunc{a'}{a}$ functional in Eq.~\eqref{eq:kvp} and derive a new set of linear equations that determine the  $\cemul_\mu(\thetavec)$ parameters. 

Note that we use greedy algorithms, with a specific error estimator discussed in Sec.~\ref{sec:greedy_alg}, to choose the training points $\thetavec_\mu$ adaptively. The details can be found there and in Sec.~\ref{sec:results}.

The aforementioned approximation of the wave functions, for a particular $a \to a'$ transition and a given energy $E$, can be written explicitly as 
\begin{align}
   \begin{pmatrix}
    |  \wf{a}(\thetavec) \rangle \\[0.2em] 
    |  \wf{a'}(\thetavec) \rangle
   \end{pmatrix}
     & = \sum_{\mu=1}^{\dimL} \cemul_\mu^{a'a}(\thetavec) \begin{pmatrix}
    |  \wf{a}(\thetavec_\mu) \rangle \\[0.2em] 
    |  \wf{a'}(\thetavec_\mu) \rangle
   \end{pmatrix}\,, \label{eq:var_emulator_wf_ansatz} \\
    \sum_{\mu=1}^{\dimL} \cemul_\mu^{a'a} & = 1 \,.\label{eq:var_emulator_c_constraint}
\end{align}
When $a = a'$, the equation can be simplified to keep a single wave function. This approximation was already explored in Ref.~\cite{Garcia:2023slj} to develop RBM emulators for two-nucleon coupled-channel scattering.

The ${a'a}$ index of $\cemul_\mu^{a'a}(\thetavec)$ in the equation emphasizes the dependence of these coefficients on the particular transition of interest. A separate emulator needs to be developed for each different transition. 

The constraint enforced on $\cemul_\mu^{a'a}$ ensures that the $R$-matrix coefficients inside $ | \wf{a}(\thetavec)\rangle$ (as in the form of Eq.~\eqref{eq:HHwf_A}) are 
\begin{equation}
    \rmat{a_0}{a}(\thetavec) = \sum_{\mu=1}^{\dimL} \cemul_\mu^{a'a}(\thetavec) \rmat{a_0}{a}(\thetavec_\mu)\ , \label{eq:R1st_emul}
\end{equation}
where $\rmat{a_0}{a}(\thetavec_\mu)$ with $a_0 = 1,\cdots,\dimC$ are the first-order estimates of the $R$-matrix elements obtained in the high-fidelity calculations with $H(\thetavec_\mu)$. 

We then determine $\cemul_\mu^{a'a}(\thetavec)$ by finding the stationary solution of the $\rfunc{a'}{a}$. The Lagrange multiplier technique is used to enforce the constraint in Eq.~\eqref{eq:var_emulator_c_constraint}, leading to the following system of equations:
\begin{align}
\frac{\partial }{\partial \cemul_\mu^{a'a}} \left(\rfunc{a'}{a} - \lambda (\sum_\mu \cemul_\mu^{a'a} -1 ) \right) & = 0  \, , \label{eq:var1}   \\ 
\frac{\partial }{\partial \lambda } \left(\rfunc{a'}{a} - \lambda (\sum_\mu \cemul_\mu^{a'a} -1 ) \right) & = 0 \, . \label{eq:var2}
\end{align}
Here, $\rfunc{a'}{a}$ is  a function of $\cvec^{a'a}$ (defined as $\{\cemul_\mu^{a'a}(\thetavec)\}_{\mu=1\dots \dimL}$), as $|\wf{a}\rangle$ and $|\wf{a'}\rangle$ are determined by  $\cvec^{a'a}$ via Eq.~\eqref{eq:var_emulator_wf_ansatz}. The explicit expression of the functional can be found in Eq.~\eqref{eq:varfunc_direct_emul}.

The resulting system of linear equations has the matrix form:
\begin{widetext}
\begin{equation}\label{eq:emul_var_large}
\renewcommand{\arraystretch}{1.2}
\left(\begin{array}{c c c  c } 
U^{a'a}_{\mu=1,\mu'=1} &  \cdots  &  U^{a'a}_{\mu=1,\mu'=\dimL}   &  1   \\
 & \vdots &   &  \vdots \\ 
U^{a'a}_{\mu=\dimL,\mu'=1} &  \cdots  &  U^{a'a}_{\mu=\dimL,\mu'=\dimL}   &  1   \\[1ex] 
1 & \cdots & 1 & 0
\end{array} \right)
\left(\begin{array}{c}
\cemul_{\mu'=1}^{a'a} \\
\vdots \\ 
\cemul_{\mu'=\dimL}^{a'a} \\[1ex] 
\lambda \\
\end{array} 
\right) 
= 
\left(\begin{array}{c}
\rmat{a'}{a}(\thetavec_{\mu=1}) \\
\vdots \\ 
\rmat{a'}{a}(\thetavec_{\mu = \dimL}) \\[1ex] 
1 \\
\end{array} \right) \ , 
\end{equation}
\end{widetext}
with, 

\begin{align}
     U^{a'a}_{\mu',\mu}(\thetavec) & \equiv T^{a'a}_{\mu',\mu}(\thetavec) + T^{a'a}_{\mu,\mu'}(\thetavec)\, , \\ 
     T_{\mu',\mu}^{a' a}(\thetavec)  & \equiv  \langle \wf{a'}(\thetavec_{\mu'}) | H(\thetavec)-E | \wf{a}(\thetavec_{\mu}) \rangle \,. \label{eq:Tinsubspace}
\end{align}

Note that the ordering of $a'$, $a$ and $\mu'$, $\mu$ in the definitions of $T^{a'a}_{\mu',\mu}$ and $U^{a'a}_{\mu',\mu}$ is associated with the $a\to a'$ transition. Correspondingly, $\rmat{a'}{a}(\thetavec_\mu)$ appear on the right side of Eq.~\eqref{eq:emul_var_large}. It is essential to have these indices ordered correctly, because, as pointed out toward the end of Sec.~\ref{sec:HH}, the first-order estimate of the $R$ matrix is not exactly symmetric in the high-fidelity calculations.
 
Equation~\eqref{eq:emul_var_large} can be re-casted in a compact form, as shown in the companion paper~\cite{Nd_emulator_2025_short}:  
\begin{equation}
    \begin{pmatrix}
       \quad \vb{U}^{a' a}(\thetavec)  \quad &  \vb{1}_{\dimL}\\
       \vb{1}_{\dimL}^T & 0
    \end{pmatrix} 
    \begin{pmatrix}
        \vb{\cemul}^{a'a}(\thetavec) \\
        \lambda
    \end{pmatrix}=    
    \begin{pmatrix}
        \vb*{\mathcal{R}}_{aa'}\\
       1
    \end{pmatrix} \ ,   \label{eq:emul_var_compact}
\end{equation}
with $\vb{U}^{a'a}(\thetavec)$ as the matrix form of $U^{a'a}_{\mu',\mu}(\thetavec)$, and $ \vb*{\mathcal{R}}_{aa'} = \{\rmat{a'}{a}{(\thetavec_\mu)}\}_{\mu=1\dots \dimL}$. 

Returning to Eq.~\eqref{eq:Tinsubspace}, in fact, $T_{\mu', \mu}^{a' a}(\thetavec)$ can be viewed as the $H(\thetavec)-E$ operator matrix element in the emulator subspace (with a dependence on the $(a', a)$ pair). We rely on the $\mu$ and $\mu'$ indices to separate this representation from  those in the $|\xi\rangle$ and $| \Omega^{I(R)}\rangle$ basis in Eqs.~\eqref{eq:T_in_xi_Omega_1}--\eqref{eq:T_in_xi_Omega_-1}.  

All the details about $T_{\mu', \mu}^{a' a}$  can be found in Eq.~\eqref{eq:emul_Var_kernel}. There, we can see explicitly
how the affine parameter dependence of $H(\thetavec)$ can be exploited to compute the $T_{\mu', \mu}^{a' a}(\thetavec)$ and $U_{\mu', \mu}^{a' a}(\thetavec)$. Once the $|\wf{a}(\thetavec_\mu)\rangle$ (and $|\wf{a'}(\thetavec_\mu)\rangle$ if $a \neq a'$) are available, the contractions of high-dimensional vectors and the tensors defined in Eqs.~\eqref{eq:T00(i)def}--\eqref{eq:Taap(i)def} can be performed in the training stage and stored. $T_{\mu', \mu}^{a' a}(\thetavec)$ is then quickly computed by adjusting $\thetavec$ in Eq.~\eqref{eq:emul_Var_kernel}.   
 
As a result, the low-dimensional system of linear equations is instantly defined at the online stage. 
Its solution, $\cvec^{a'a}(\thetavec)$, provide immediately the first-order estimate of the $R$-matrix from Eq.~\eqref{eq:R1st_emul}. The correction term inside $\rfunc{a'}{a}$ can be evaluated as well by performing contractions involving low-dimensional $T_{\mu', \mu}^{a' a}(\thetavec)$ tensors and $\cvec^{a'a}(\thetavec)$ (see Eq.~\eqref{eq:varfunc_direct_emul}), producing fast second-order estimate of the $R$ matrix element. 

Note, the reduced equation system could
become ill-conditioned if more than the
necessary training snapshots are included, as seen in previous scattering emulation studies~\cite{Furnstahl:2020abp, Drischler:2021qoy, Zhang:2021jmi,Garcia:2023slj}. Certain regularization techniques were employed to stabilize the solution of the equation system.  In this work, however, we did not encounter ill-conditioning issues, likely due to the greedy algorithm, which iteratively adds training points in relevant locations until a prescribed termination condition is achieved.

Evidently, when dealing with the diagonal $R$ matrix elements, the variational emulations share the same strategy as in the high-fidelity calculations. However, for the off-diagonal matrix element $\rmat{a'}{a}$, our emulator looks for the optimal trial wave functions by varying $\cvec^{a'a}$, in contrast to the high-fidelity treatment where the wave functions inferred from the diagonal $R$-matrix element calculations are directly inserted in $\rfunc{a'}{a}$ without further optimization.  We explored briefly the same approach in emulating the off-diagonal $\rmat{a'}{a}$ and found little difference from our main emulation approach in terms of emulator performance.

\section{Emulation of linear systems} \label{sec:G_LSPG_ROM}
We revisit Eq.~\eqref{eq:lineareq_highfidelity_matrix_form} for a given $a$, or its compact form: 
\begin{align}
    \Amat(\paramVec) \xvec(\paramVec) = \svec(\paramVec) \,. \label{eq:general_lineareq}
\end{align}
In both $\Amat(\thetavec)$ and $\svec(\thetavec)$, the dependence on $\thetavec$ is linear (affine). See the discussion at the end of Sec.~\ref{sec:HH}.

Our linear-equation emulators basically emulate the solution of this equation. Naturally, $\xvec$ is treated as the $\yvec$ variable in Eq.~\eqref{eq:RBM_ansatz}. This emulation strategy was previously studied in Ref.~\cite{Maldonado:2025ftg},
which dealt with two-body scatterings.%
\footnote{Recent works~\cite{Zhang:2024gac,Zhang:2024ril} also studied emulations for inhomogeneous linear equations related to quantum continuum physics. However, a variational approach was utilized to guide subspace projection. Moreover, the parameter space studied there is a combination of the complex energy plane and the space of other real input parameters.} 
Both the  Galerkin reduced-order model (G-ROM) and the Petrov-Galerkin least-square reduced-order model (LSPG-ROM) developed there are adopted here to create two different linear equation emulators.

The greedy algorithms developed in Ref.~\cite{Maldonado:2025ftg} are also implemented here to minimize the emulator training cost. 
Interestingly, the greedy algorithm could also be used to potentially eliminate Kohn anomalies~\cite{PhysRev.124.1468, nesbet1980variational, Drischler:2021qoy}, i.e., spurious singularities when the ROM is singular or near singular, as the new training point can be included to reduce the errors due to anomalies and thus mitigate the problem~\cite{Maldonado:2025ftg}.

In addition to the emulation of the solutions of the Eq.~\eqref{eq:lineareq_highfidelity_matrix_form}, we further evaluate the variational functional in Eq.~\eqref{eq:kvp} by inserting these solutions to get second-order estimates. This step, not needed in Ref.~\cite{Maldonado:2025ftg}, is necessary to achieve high emulation accuracy, given that solving Eq.~\eqref{eq:lineareq_highfidelity_matrix_form} only produces first-order estimates of the scattering wave functions.

\subsection{Implementation details} \label{sec:G_LSPG_ROM_details}

For both emulators in this class, with a given $a$, Eq.~\eqref{eq:RBM_ansatz} is used to approximate the solution vector $\xvec$ with  $\txvec$, and  
\begin{align}
    \txvec(\thetavec)  & = \sum_{\mu=1}^{\dimL} \cemulwa_\mu(\thetavec) \xvec(\thetavec_\mu) \equiv \Xmat \cvecwa(\thetavec) \,.
\end{align}
Here, $\cvecwa(\thetavec)$ is the vector form of the to-be-determined $\cemulwa_\mu(\thetavec)$ coefficients that vary with $\thetavec$; $\Xmat$ is the  snapshot matrix. To construct $\Xmat$, we start by constructing $\Xmatzero$, whose columns are the snapshot vectors spanning the emulator reduced space, 

\begin{equation}
\Xmatzero_{(\dimH+\dimC)\times\dimL} = 
 \begin{pmatrix}
      \longmid & \longmid &    & \longmid \\[0.5ex] 
      \xvec(\thetavec_1) &  \xvec(\thetavec_2) & \cdots &  \xvec(\thetavec_{\dimL})   \\[1ex]
      \longmid & \longmid &  & \longmid
 \end{pmatrix} \,.  
\end{equation}
For numerical stability, the columns of $\Xmatzero$ are orthonormalized via applying the QR 
factorization to $\Xmatzero$~\cite{Maldonado:2025ftg}: 
\begin{align}
\Xmatzero_{\dimH\times\dimL} & = \Xmat_{\dimH\times\dimL} \vec{R}_{\dimL\times\dimL} \, . \label{eq:QR}
\end{align}
such that ${\Xmat}^\dagger \Xmat = \mathbf{I}_{\dimL}$. Here $\vec{R}$ is a triangular matrix~\cite{Maldonado:2025ftg}.

We employ two different projection methods to determine the coefficient vector $\cvecwa(\thetavec)$:

{\bf I. G-ROM:}
This approach requires that the residual, 
\begin{equation}
\rvecwa(\thetavec) = \svec(\thetavec) - \Amat(\thetavec)\txvec(\thetavec)    \ , \label{eq:residualdef}
\end{equation}
be orthogonal to the reduced space spanned by the snapshot matrix, i.e.,
\begin{equation}
 {\Xmat}^\dagger \rvecwa(\thetavec) = 0    \,.
\end{equation}
This leads to the reduced system:
\begin{equation} 
    \tAmat(\thetavec)\cvecwa(\thetavec)   = \tsvec(\thetavec) \,, \label{eq:ROM_lineareq} 
\end{equation}
with 
\begin{align}
    \tAmat(\thetavec) &  = {\Xmat}^\dagger \Amat(\thetavec) \Xmat \,,  \\
    \tsvec(\thetavec) &  = {\Xmat}^\dagger \svec(\thetavec)\, .
\end{align}

{\bf II. LSPG-ROM:}
Here, $\cvecwa(\thetavec)$ is determined by minimizing the norm of the residual $\rvecwa(\thetavec)$~\cite{Quarteroni:218966}, which results in a least-squares problem. 

To derive the LSPG-ROM equations, we note that $\Amat(\thetavec) = \sum_{i=0}^{\dimparam} \theta_i \Amat_{(i)}$ and $\svec(\thetavec) = \sum_{i=0}^{\dimparam} \theta_i \svec_{(i)}$, where $\Amat_{(i)}$ and $\svec_{(i)}$ are parameter-independent matrices and vectors and $\theta_0 = 1$  (similar to the convention in Eq.~\eqref{eq:Haffinedecomp}). The residual $\rvecwa(\thetavec)$ can also be expressed in a linear form:
\begin{equation} \label{eq:residual_affine}
\rvecwa(\paramVec)
= \sum_{i=0}^{\dimparam} \left[  \svec_{(i)} - \Amat_{(i)} \Xmat \cvecwa(\paramVec) \right] \theta_i \,.   
\end{equation} 
This shows that $\rvecwa(\thetavec)$ for \emph{any} $\thetavec$ is in the subspace spanned by the column vectors of $\vb{B}_{(i)} = \Amat_{(i)} \Xmat$ ($\dimH \times \dimL$ matrices) and the $\svec_i$ (length-$\dimH$) vectors for all $i \in [0,1,\ldots,\dimparam]$.
We stack these vectors horizontally to obtain the $\dimH \times (\dimparam+1)(\dimL+1)$ matrix:
\begin{equation} \label{eq:def_Ymat}
    \Ymatzero = \begin{bmatrix} \vb{B}_{(0)} & \vb{B}_{(1)} & \cdots & \vb{B}_{(\dimparam)} & \vb{s}_{(0)} & \vb{s}_{(1)} & \cdots & \vb{s}_{(\dimparam)}\end{bmatrix} \,.
\end{equation}
Singular value decomposition (SVD) is then applied to truncate the dimension of $\Ymat$ to $\dimH\times N_{\Ymat}$, with $N_{\Ymat} \leqslant (\dimparam+1)(\dimL+1)$. Eventually, we get the  LSPG-ROM emulator, in which we need to solve a least-squares problem for the equation  with the same form as Eq.~\eqref{eq:ROM_lineareq} but with 
\begin{align}
    \tAmat(\paramVec) & = {\Ymat}^\dagger \Amat(\paramVec) \Xmat \,,\label{eq:atilde_lspg}  \\ 
    \tsvec(\paramVec) & = {\Ymat}^\dagger \svec(\paramVec) \,.\label{eq:stilde_lspg}
\end{align}
Bear in mind that this equation is over-constrained, since $N_{\Ymat}$ is (much) larger than $\dimL$, which is why this equation is solved in the least-squares sense. 

For both G-ROM and LSPG-ROM, the linear parameter dependence is preserved in $\tAmat$ and $\tsvec$, allowing efficient offline-online decompositions and thus rapid emulations.

\subsection{Greedy Algorithm for Snapshot Selection and Error Estimation} \label{sec:greedy_alg}

The accuracy of these emulators (including the variational ones) hinges on the choice of snapshot locations in the parameter space. We employ greedy algorithms~\cite{Maldonado:2025ftg, Sarkar:2021fpz} to iteratively refine the emulator accuracy and determine optimal snapshot locations, utilizing an active learning approach. 

The greedy process begins with an initial set of snapshots (e.g., randomly selected). In each iteration, it identifies the parameter space location where the emulator error is estimated to be largest. A new high-fidelity calculation is then performed at this location, and the resulting solution vector (i.e., snapshot) is added to the emulator basis. This iterative process, involving high-fidelity calculations, updating the emulator basis, and error estimation, continually improves the emulator accuracy by ``greedily'' placing snapshots where they contribute most to error reduction.

A crucial component of the greedy algorithm is robust and fast estimation of emulation error. Instead of directly computing the actual error $\mathbf{e}^a(\thetavec) = \xvec(\thetavec) - \txvec(\thetavec)$, which would defeat the purpose of an emulator, we use the norm of the residual vector $\rvecwa(\thetavec) = \svec(\thetavec) - \Amat(\thetavec)\txvec(\thetavec)$ defined in Eq.~\eqref{eq:residual_affine} as a proxy for the actual error.

Our empirical findings suggest that the magnitude of the residual is approximately proportional to the magnitude of the true error, making the residual magnitude a  highly effective error proxy for identifying regions in the parameter space where the error is the largest~\cite{Maldonado:2025ftg, Sarkar:2021fpz}.
The approximate proportionality can be further exploited to calibrate the residual (i.e., rescale it) to accurately approximate the magnitude of the true error, if it is necessary.   

Since $\Ymat$ column vectors are designed to fully project the $\rvecwa(\thetavec)$ with general $\thetavec$,  we have $||\rvecwa(\thetavec)|| = || {\Ymat}^\dagger \rvecwa(\thetavec)||$~\cite{Maldonado:2025ftg}.
Using this relation and taking advantage of the affine decomposition in Eq.~\eqref{eq:residual_affine} we can use the matrix elements in Eqs.~\eqref{eq:atilde_lspg} and ~\eqref{eq:stilde_lspg} to compute quickly $||\rvecwa(\thetavec)||$ working only in the reduced space. Note that this $\Ymat$-based means of computing $||\rvecwa(\thetavec)||$ is used in executing the greedy algorithms for training all of our emulators, including the variational, G-ROM, and LSPG-ROM emulators.

\subsection{Emulation of the 2nd order $R$ matrix} \label{sec:var_2nd_piece}
We first note that in both G-ROM and LSPG-ROM, for a specific $a$ channel,   

\begin{equation}\label{eq:GROM_emul}
     \xvec (\thetavec) =  
     \left(\begin{array}{c}
\cHH^a_{\xi'=1}(\thetavec) \\
\vdots \\ 
\cHH^a_{\xi'=\dimxi}(\thetavec) \\[2ex] 
\rmat{a_0=1}{a}(\thetavec) \\ 
\vdots \\ 
\rmat{a_0=\dimC}{a}(\thetavec)\\
\end{array} \right)     
     \approx 
    \sum_\mu \cemulwa_\mu(\thetavec)
    \left(\begin{array}{c}
 X^a_{\mu,\xi'=1} \\
\vdots \\ 
 X^a_{\mu,\xi'=\dimL} \\[2ex] 
X^{a}_{\mu,a_0=1} \\ 
\vdots \\ 
X^{a}_{\mu,a_0=\dimC} 
\end{array} \right)    \\\,,
 \end{equation}
with $X^a_{\mu, \xi'}$ and $X^a_{\mu, a_0}$ as the entries of the column vectors in $\Xmat$ or $\Ymat$. Note that these entries are different from $\cHH^a_{\xi'}(\thetavec_\mu)$ and $\rmat{a_0}{a}(\thetavec_\mu)$ due to the step in Eq.~\eqref{eq:QR}. 

To emulate the second-order estimate of the $\rmat{a'}{a}$, we construct the wave functions  from emulated $\xvec(\thetavec)$ (and $\xvecwoa^{a'}(\thetavec)$ if $a \neq a'$) and insert them into  Eq.~\eqref{eq:kvp}. The details for computing  $ \langle \Psi^{a'}(\thetavec) | H(\thetavec)-E | \Psi^a(\thetavec) \rangle$ in the functional are provided in Eq.~\eqref{eq:emul_GROM_kernel} in Appendix~\ref{app:emulation_details}. There, the calculations within square brackets, which deal with the full space, are performed during the offline training stage. The sum over $\mu$ and $\mu'$ is instead in the reduced space and can be done right after obtaining $\vec{c}^a$ and $\vec{c}^{a'}$ at the online stage. 

\section{Results} \label{sec:results}

\begin{figure*}[tbh!]
    \centering
    \includegraphics
    {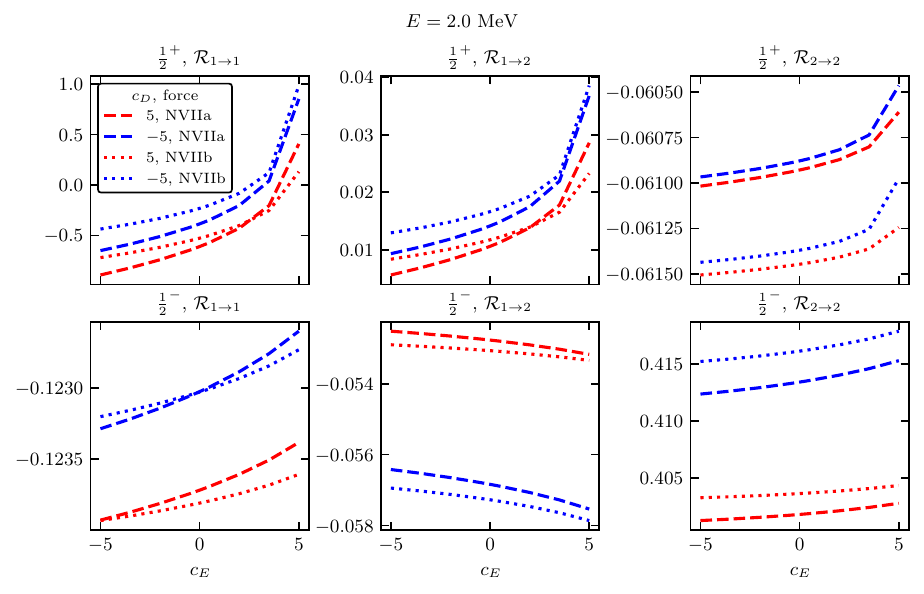}
    \caption{$R$-matrix elements vs.\ $\cE$ with fixed $\cD$ values (shown in the legend) and $E = 2 $ MeV. The results from the NVIIa and NVIIb forces are presented. }
    \label{fig:Rvsparam}
\end{figure*}

\begin{figure*}[tbh!]
    \centering
    \includegraphics[width=0.9\textwidth]{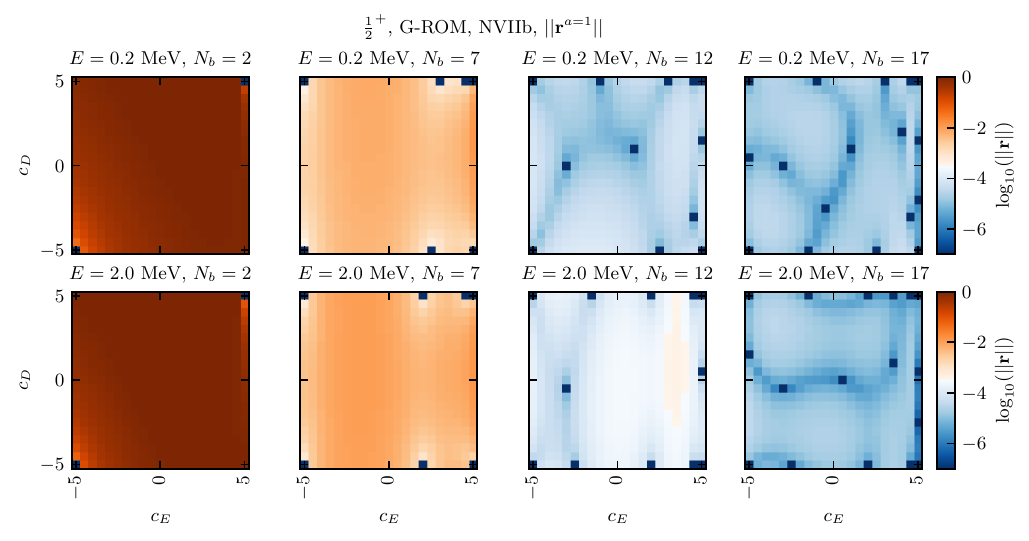}
     \caption{The progressive reductions (seen in the $(\cE,\cD)$ plane) of the residue $||\rvec^{a=1}(\thetavec)||$ (for the $\halfplus$ channel)  with increasing number of training points ($\dimL$). The residual is computed within the G-ROM emulations; the force is NVIIb. Two different energies are considered, one close to the elastic scattering threshold and the other close to the deuteron breakup threshold. 
     }
    \label{fig:res_halfplus_GROM_NVIIb}
\end{figure*}

\begin{figure*}
    \centering
    \includegraphics[width=0.9\textwidth]{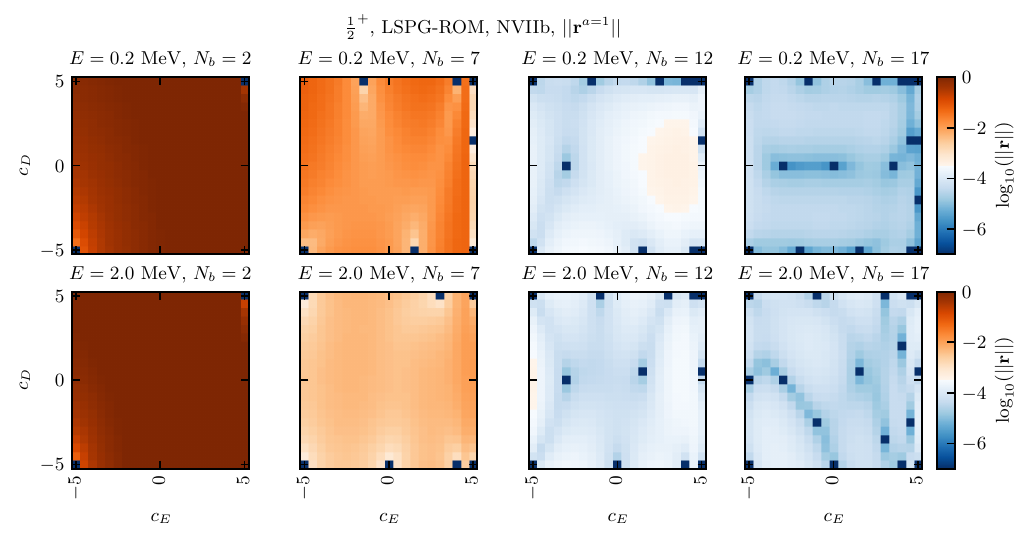}
    \caption{The same figure as Fig.~\ref{fig:res_halfplus_GROM_NVIIb} but now for the NVIIb force with LSPG-ROM emulator. 
    }
    \label{fig:res_halfplus_LSGROM_NVIIb}
\end{figure*}

Figure~\ref{fig:Rvsparam} provides a qualitative understanding of the diagonal and off-diagonal $R$-matrix elements at a scattering energy $E = 2$ MeV (close to deuteron breakup threshold) for both $J^\pi=\halfplus$ and $J^\pi=\halfminus$ channels and both NVII a and b forces. The ranges of the dimensionless $\cE$ and $\cD$ are chosen as between $-5$ and $5$, safely covering their typical variations in existing \chiEFT parameter fits.

It is immediately clear that $\rmat{1}{1}$ ($\halfplus$) and $\rmat{2}{2}$ ($\halfminus$) dominate in their contributions to the total scattering cross sections, based on their larger magnitudes compared to the other matrix elements in their respective channels. However, other observables, such as those related to polarizations, could be sensitive to the small matrix elements. Therefore, it is necessary to investigate emulator performance also for these small matrix elements.

On the parameter dependence, the matrix elements in the $\halfminus$ channel vary within no more than a percent, a range much smaller than the variations in $\halfplus$. Unsurprisingly, and as confirmed in the later plots, the emulation relative errors in $\halfminus$ are also much smaller, as compared to those in $\halfplus$. 

All the matrix elements have nonlinear dependence on $\cE$ with fixed $\cD$ values. In particular, the nonlinearities grow toward the large positive side of $\cE$ in $\halfplus$. The $\cD$ dependence is less nonlinear, and thus for a given force, the curves with two different $\cD$ values (with the same line style but different colors) are shifted relative to each other along the vertical direction by roughly the same amount. Changing regulators does not alter the qualitative parametric behaviors, as demonstrated by the comparisons between the results of the two forces, as shown here.  

We have observed qualitatively similar behaviors at other energies below the breakup threshold. The remainder of this section focuses on the performances of our emulators in reproducing these behaviors across different energies in the parameter space.  

\subsection{$J^\pi=\halfplus$ channel}

We begin by showing the residuals $||\residual||$, which are used as error proxies in training G-ROM emulators. Figure~\ref{fig:res_halfplus_GROM_NVIIb} shows, for the case of the NVIIb force and the $\halfplus$ channel, how the residuals $||\residual^{a=1}||$ decrease when the active learning processes progress, as guided by the greedy algorithm.  Two different energies, one (top row) close to the elastic scattering and the other (bottom row) to the breakup threshold, are demonstrated. The $a=1$ superscript in $||\residual^{a=1}||$ emphasizes that the residual vector is computed for emulating the solution of Eq.~\eqref{eq:general_lineareq} (or Eq.~\eqref{eq:lineareq_highfidelity_matrix_form}) with $a=1$. That is, $||\residual^{a=1}||$ comes out from the training of the G-ROM emulator for the $\rmat{1}{1}$.  

In Fig.~\ref{fig:res_halfplus_GROM_NVIIb}, from left to right, the greedy algorithm starts with two randomly chosen points as the initial points and then selects a new training point with the largest estimated $||\residual^{a=1}||$ at that step. The process eventually terminates when the average value of the residual decreases to a prefixed target. 

Initially, the new training points are located on the boundary of the parameter space, indicating that the greedy algorithm tries to minimize the errors of extrapolating to the boundaries. Later, the algorithm adds points inside the domain to reduce the residual step by step until it reaches approximately $10^{-5}$ in general--except at the training points, where the residuals are nearly zero (as indicated by the dark spots). 

Although the residuals at the two energies behave differently in detail, qualitatively they arrive at the targeted value at similar $\dimL$.

In principle, we could run the same greedy algorithm at the emulator training stage for the $\rmat{2}{2}$ matrix element. In our brief exploration, we observed similar choices of training points as in the case of $\rmat{1}{1}$, which is perhaps expected, considering that the $a=1$ and $a=2$ channels are coupled. To reduce computing cost for training, we always use the training points identified by $\residual^{a=1}$ as the error proxy to train the emulators for the other coupled channels.  

Moreover, the same active learning procedure can be performed when training the LSPG-ROM emulators. Since the $\cvec^a(\thetavec)$ coefficient vector in the G-ROM and LSPG-ROM are different, per Eq.~\eqref{eq:residualdef} (or~\eqref{eq:residual_affine}), we expect somewhat different residual behavior for the LSPG-ROM from the G-ROM, as shown in Fig.~\ref{fig:res_halfplus_LSGROM_NVIIb}. However, the general behaviors, such as concerning the number of training points needed, are similar. We also note that the residuals for the NVIIa force, as we have verified, do not exhibit significant differences from those presented here for the NVIIb case.

As a next step, we study how the true errors on the R-matrix  behave when running the greedy algorithms during emulator training. This investigation is relevant for understanding the quality of the error proxy used here and the emulator in general. However, this test is not expected to be feasible for extremely large computations, as complete checking requires a full sample of the high-fidelity calculations across the entire parameter space.

\begin{figure*}[tbh!]
    \centering\includegraphics{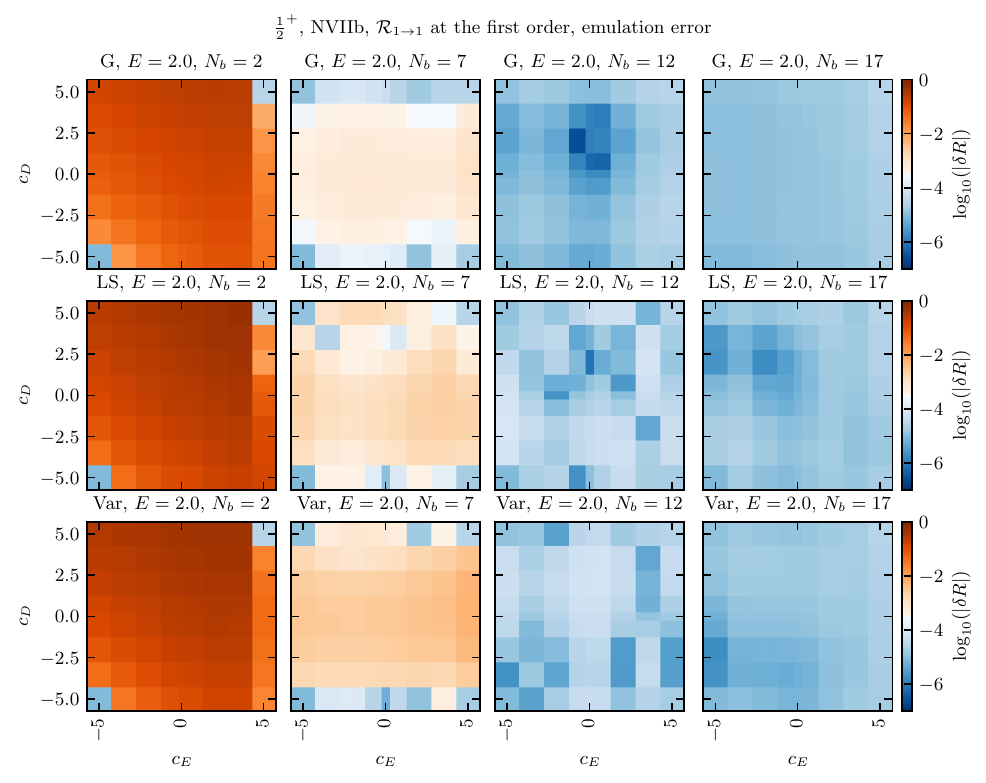}
    \caption{Emulation errors at $E=2$ MeV for $\rmat{1}{1}$ at its 1st order in the $\halfplus$ channel for the NVIIb force. From left to right, $\dimL$ increases, and the errors decrease. Three different emulators are compared from top to bottom:  ``G'' (G-ROM), ``LS'' (LSPG-ROM), and ``Var'' (variational).}
    \label{fig:emul_error_halfplus_R11_1st_NVIIb_multiple_emulators}
\end{figure*}

\begin{figure*}[tbh!]
    \centering
    \includegraphics{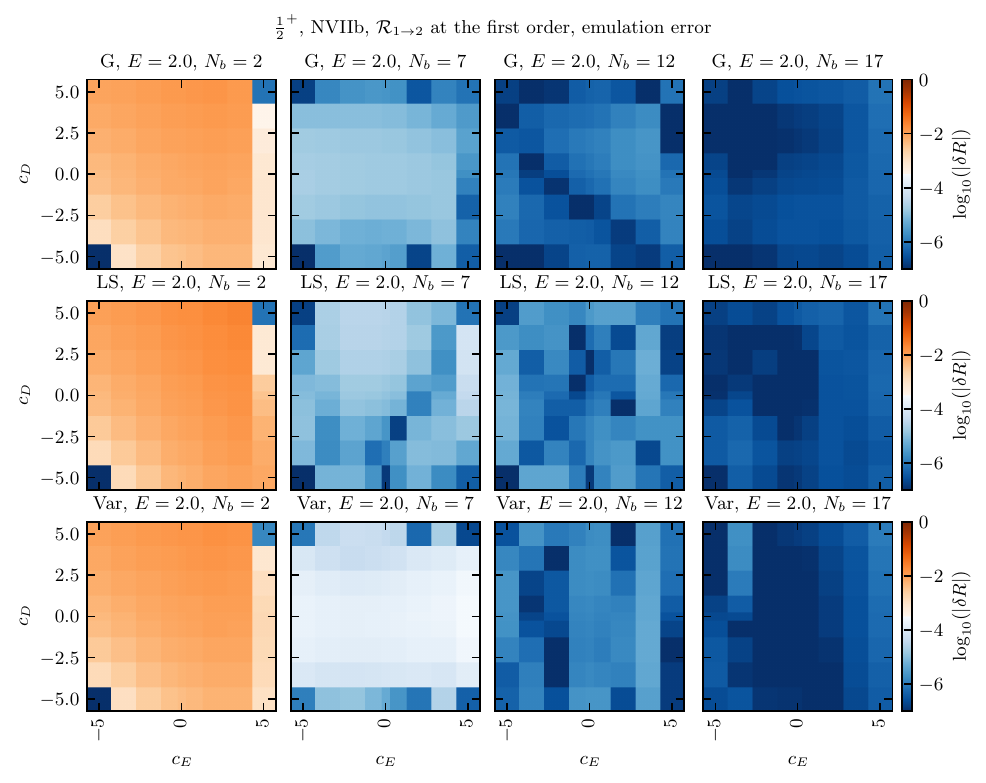}
    \caption{The same as Fig.~\ref{fig:emul_error_halfplus_R11_1st_NVIIb_multiple_emulators} but now for the $\rmat{2}{1}$ at its 1st order.}
    \label{fig:emul_error_halfplus_R12_1st_NVIIb_multiple_emulators}
\end{figure*}

Figures~\ref{fig:emul_error_halfplus_R11_1st_NVIIb_multiple_emulators} and~\ref{fig:emul_error_halfplus_R12_1st_NVIIb_multiple_emulators} show the emulation errors for the first-order estimates of $\rmat{1}{1}$ and $\rmat{2}{1}$ (for the NVIIb force and scattering energy $E=2$ MeV). All three emulators are compared. We see that for G-ROM (LSPG-ROM), the errors of $\rmat{1}{1}$ and $\rmat{2}{1}$ generally follow the pattern of the $||\rvecwa||$ shown in Fig.~\ref{fig:res_halfplus_GROM_NVIIb} (in Fig.~\ref{fig:res_halfplus_LSGROM_NVIIb}) (see the bottom row), in particular when $\dimL = 2$ and $7$, supporting the choice of the residual as the error proxy. Note that in the same Figs.~\ref{fig:emul_error_halfplus_R11_1st_NVIIb_multiple_emulators} and~\ref{fig:emul_error_halfplus_R12_1st_NVIIb_multiple_emulators}, when computing the $R$-matrix element errors for the variational emulations, the training points for the emulator are the same as those in the G-ROM. It will be interesting to develop the same error proxy for the variational emulator. Remarkably, the variational emulator errors for the two matrix elements are quite similar to those for the G-ROM. Note the emulation error for $\rmat{2}{2}$ can be found in Fig.~\ref{fig:emul_error_halfplus_R22_1st_NVIIb_multiple_emulators}. 

\begin{figure*}
    \centering
    \includegraphics{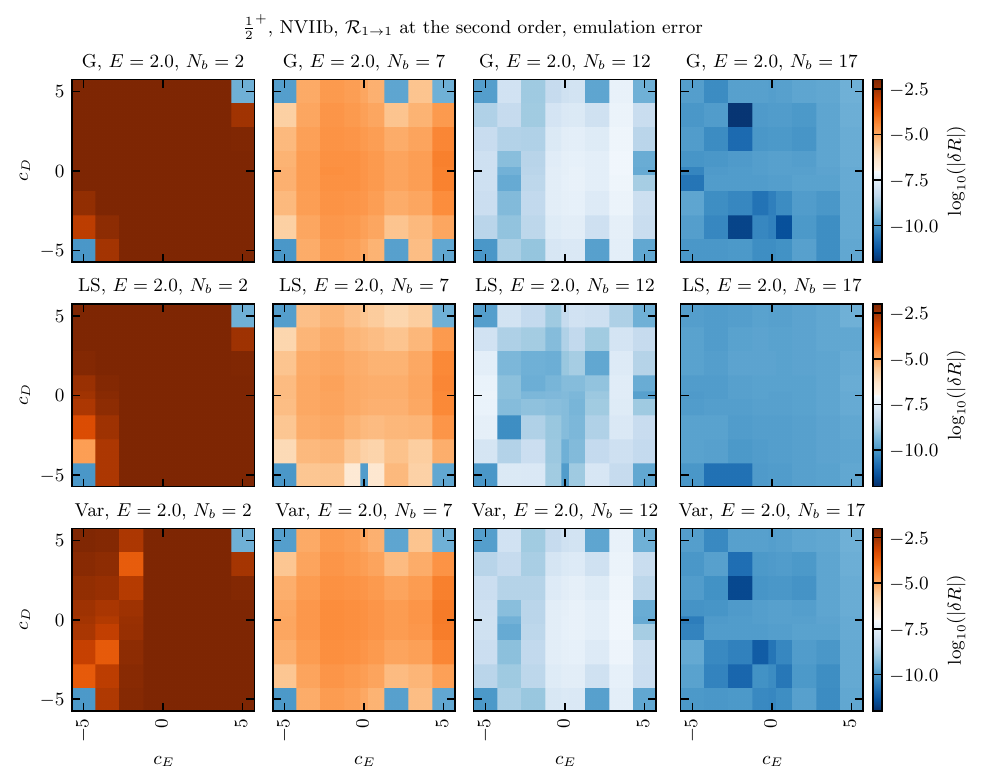}
    \caption{The same as Fig.~\ref{fig:emul_error_halfplus_R11_1st_NVIIb_multiple_emulators} but now for the $\rmat{1}{1}$ at its 2nd order.}
    \label{fig:emul_error_halfplus_R11_2nd_NVIIb_multiple_emulators}
\end{figure*}

\begin{figure*}[tbh!]    \centering
    \includegraphics{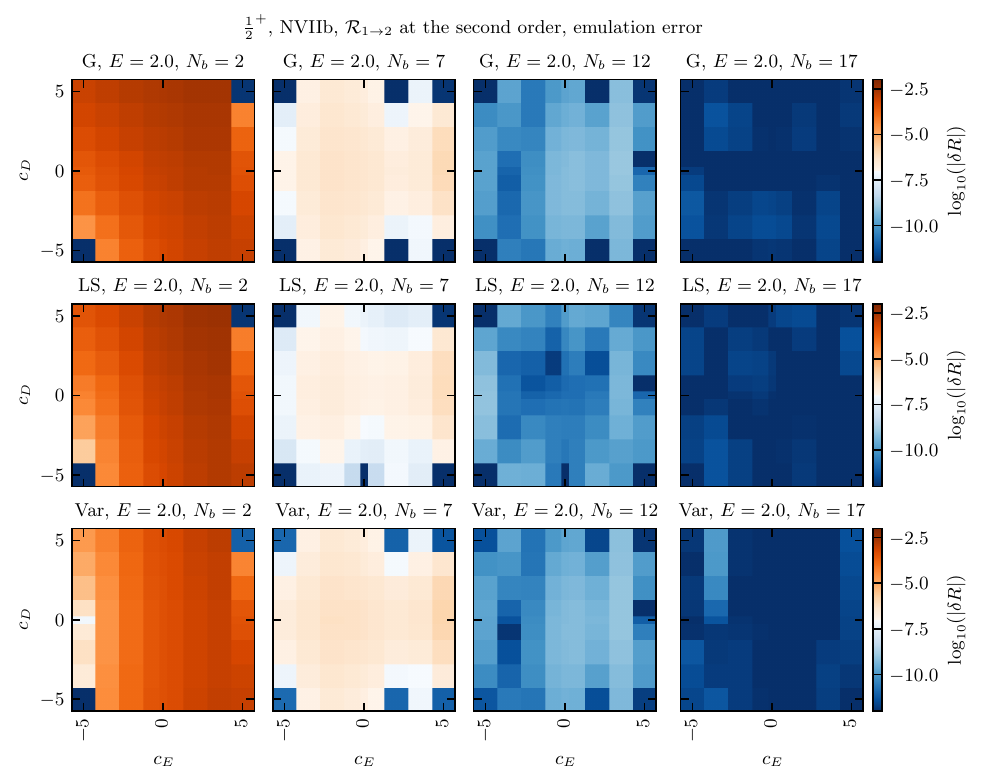}
    \caption{The same as Fig.~\ref{fig:emul_error_halfplus_R11_1st_NVIIb_multiple_emulators} but now for the $\rmat{2}{1}$ at its 2nd order.}
    \label{fig:emul_error_halfplus_R12_2nd_NVIIb_multiple_emulators}
\end{figure*}

Next, we investigate the emulations of the second-order estimates of the $R$-matrix elements, as shown in  Figs.~\ref{fig:emul_error_halfplus_R11_2nd_NVIIb_multiple_emulators}, \ref{fig:emul_error_halfplus_R12_2nd_NVIIb_multiple_emulators}, and~\ref{fig:emul_error_halfplus_R22_2nd_NVIIb_multiple_emulators}. We immediately notice significant reductions in the emulation error compared to those for the first-order estimates. This should not be surprising, as the second-order estimates bring all the estimates closer to the exact results. However, we do not observe a significant difference in pattern from the first-order estimates.

\begin{figure*}[tbh!]    \centering
    \includegraphics{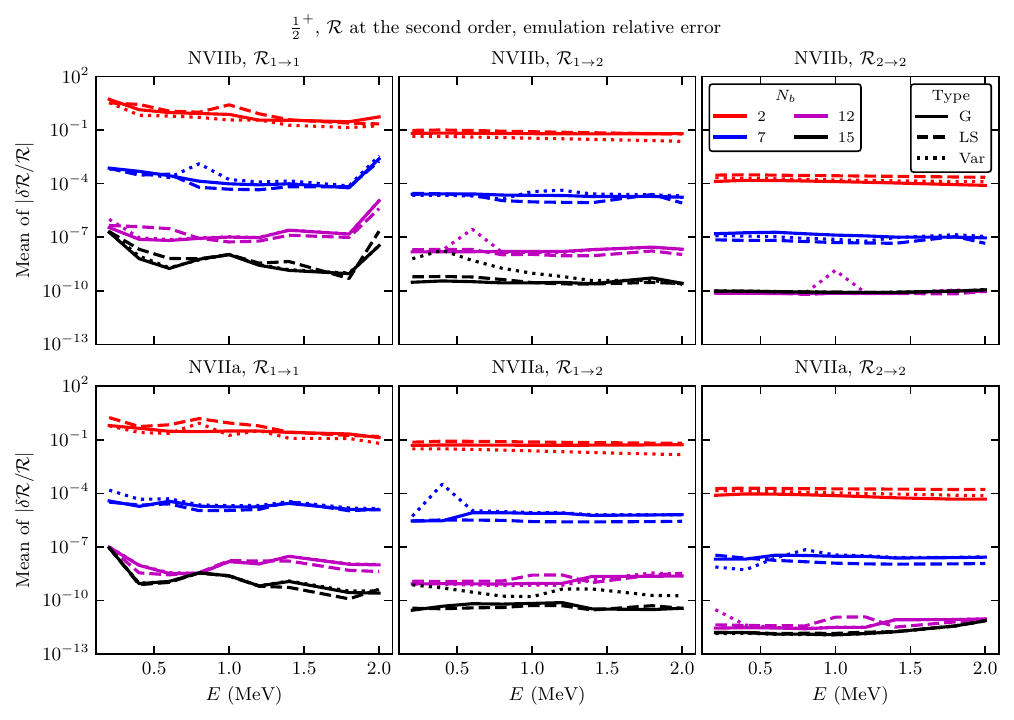}
    \caption{Emulator relative errors averaged over the parameter space vs.\ $E$ for $\rmat{a'}{a}$, with $a',a=1,2$, at its 2nd order in $\halfplus$ channel for the NVIIb (top row) and NVIIa (bottom row) force. Three different emulators are compared in each panel:  ``G'' (G-ROM, solid line), ``LS'' (LSPG-ROM, dashed line), and ``Var'' (variational, dotted line). Different colors correspond to emulators trained with different $\dimL$ values.}
    \label{fig:emul_rel_error_mean_halfplus_R_2nd_E_NVIIab_multiple_emulators}
\end{figure*}

To summarize the emulator performance, the \emph{relative} errors are averaged over the parameter space and plotted against energies in Fig.~\ref{fig:emul_rel_error_mean_halfplus_R_2nd_E_NVIIab_multiple_emulators}. The figure collects the information for all the second order $R$-matrix elements in the $\halfplus$ channel from all three emulators (solid, dashed, and dotted lines) with four different numbers of training points (red, blue, purple, and black). Each row corresponds to a particular NVII force. Results for $\rmat{1}{2}$ are not explicitly shown, as it is extremely close to $\rmat{2}{1}$ at the second-order estimates. 
We observe qualitatively the same behavior for all the emulators: when $\dimL$ increases beyond 12, the relative errors across energies and channels reduce to below $10^{-7}$ for the $\rmat{a'}{a}$ ($a',a=1,2$) matrix elements, which should be no worse than the error of the high-fidelity calculations. Moreover, the emulation performance is agnostic to the regulators employed in the NVII force. 
Importantly, the fact that~(i) the relative errors reach to $10^{-4}$ with $\dimL = 7$ for the dominant $\rmat{1}{1}$ and even smaller for the other two matrix elements, and~(ii) from $\dimL = 7$ to $12$, the errors reduce by four orders of magnitude, demonstrates an extreme emulation efficiency and accuracy. 

As a reference, we estimate that a generic interpolation methods, such as those based on splines or orthogonal polynomials, would require on the order of 10 or more training (interpolating) points in a one-dimensional parameter space to achieve a relative error of $10^{-4}$, based on our experience. For an $\dimparam$-dimensional space, the number of training points ($\dimL$) generally grows exponentially with $\dimparam$, which we estimate on the order of $10^2$ in our case of a two-dimensional parameter space. In contrast, our emulators require fewer than 10 training points to achieve the same accuracy, suggesting a much milder scaling of $\dimL$ in terms of $\dimparam$ compared to the generic methods. This scaling is key to our emulator's performance when the dimensionality $\dimparam$ of the parameter space increases, such as for full Bayesian calibration of nuclear forces.

\subsection{$J^\pi=\halfminus$ channel}

\begin{figure*}[tbh!]    \centering
    \includegraphics{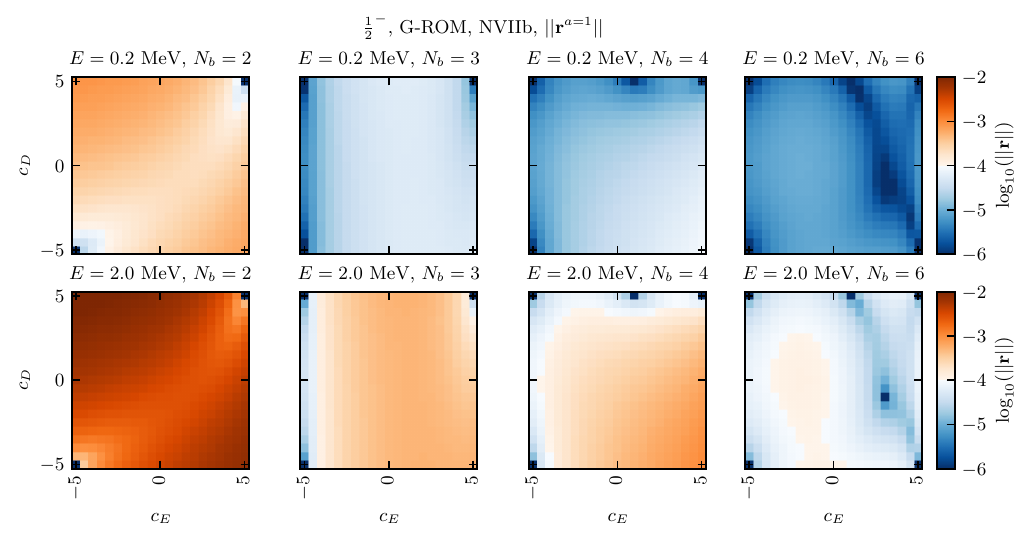}
     \caption{
     The progressive reductions (seen in the $(\cE,\cD)$ plane) of the residue $||\rvec^{a=1}(\thetavec)||$ (for the $\halfminus$ channel)  with increasing number of training points ($\dimL$). The residue is computed within the G-ROM emulations; the force is NVIIb. Two different energies are considered, one close to the elastic scattering threshold and the other close to the deuteron breakup threshold.   
     }
    \label{fig:res_halfminus_GROM_NVIIb}
\end{figure*}

\begin{figure*}[tbh!]
    \centering
    \includegraphics{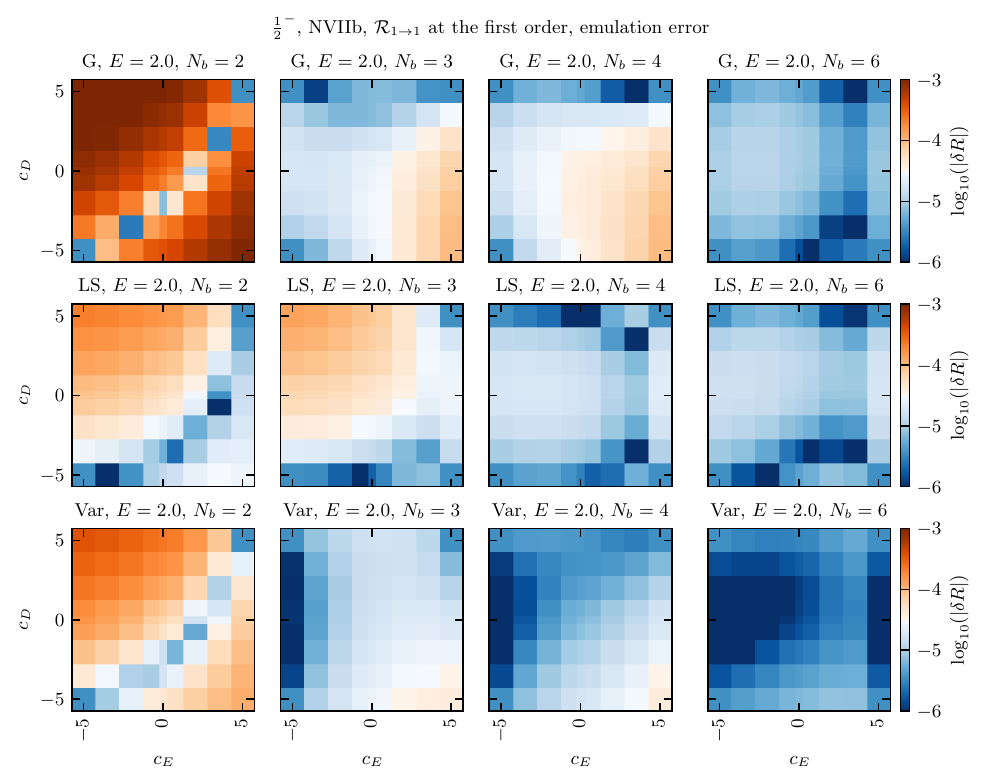}
    \caption{Emulation errors at $E=2$ MeV for $\rmat{1}{1}$ at its 1st order in $\halfminus$ channel for the NVIIb force. From left to right, $\dimL$ increases, and the errors decrease. Three different emulators are compared from top to bottom:  ``G'' (G-ROM), ``LS'' (LSPG-ROM), and ``Var'' (variational).}
    \label{fig:emul_error_halfminus_R11_1st_NVIIb_multiple_emulators}
\end{figure*}

\begin{figure*}[tbh!]
    \centering\includegraphics{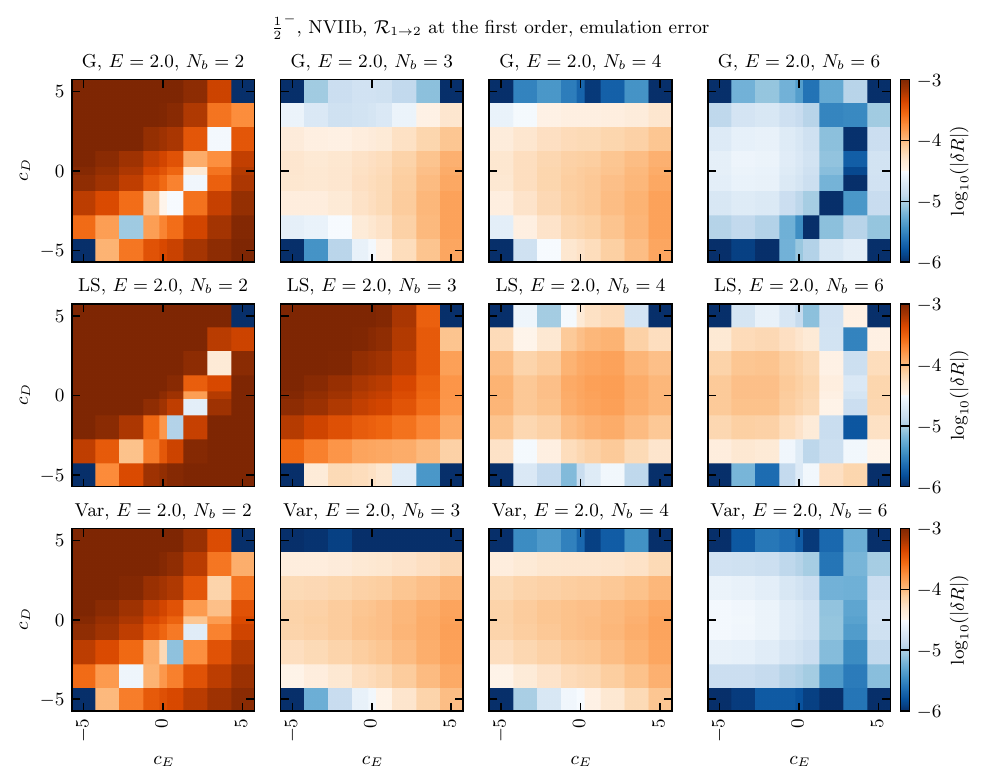}
    \caption{The same as Fig.~\ref{fig:emul_error_halfminus_R11_1st_NVIIb_multiple_emulators} but now for the $\rmat{2}{1}$ at its first order.}
    \label{fig:emul_error_halfminus_R12_1st_NVIIb_multiple_emulators}
\end{figure*}

\begin{figure*}[tbh!]    
    \centering \includegraphics{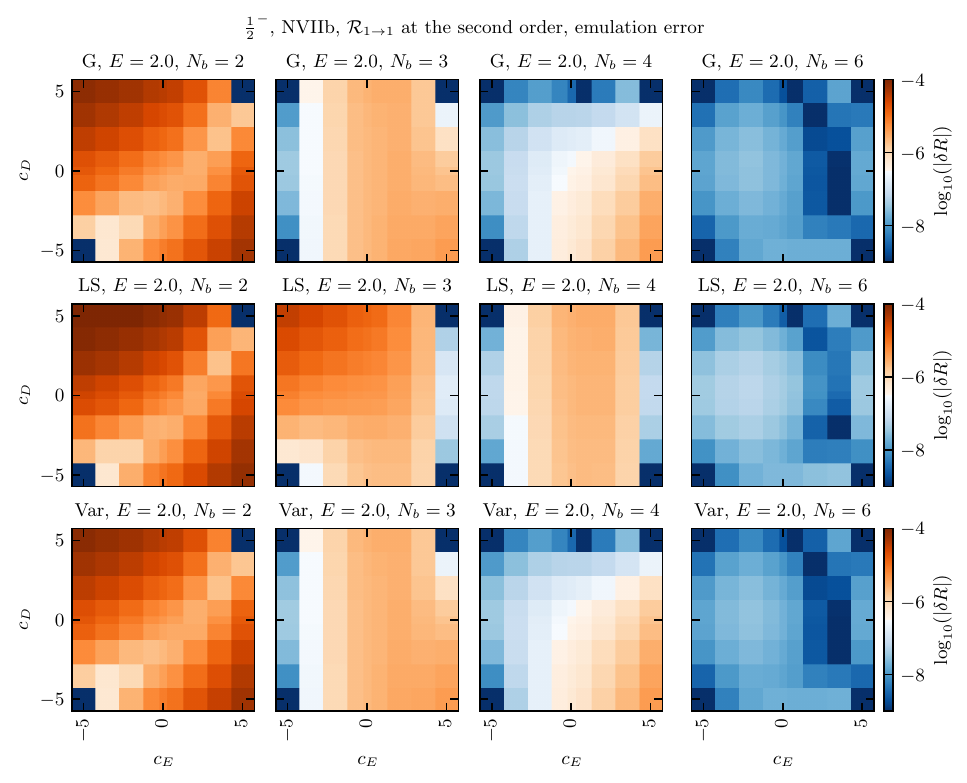}
    \caption{The same as Fig.~\ref{fig:emul_error_halfminus_R11_1st_NVIIb_multiple_emulators} but now for the $\rmat{1}{1}$ at its second order.}
    \label{fig:emul_error_halfminus_R11_2nd_NVIIb_multiple_emulators}
\end{figure*}

\begin{figure*}[tbh!]    
    \centering  \includegraphics{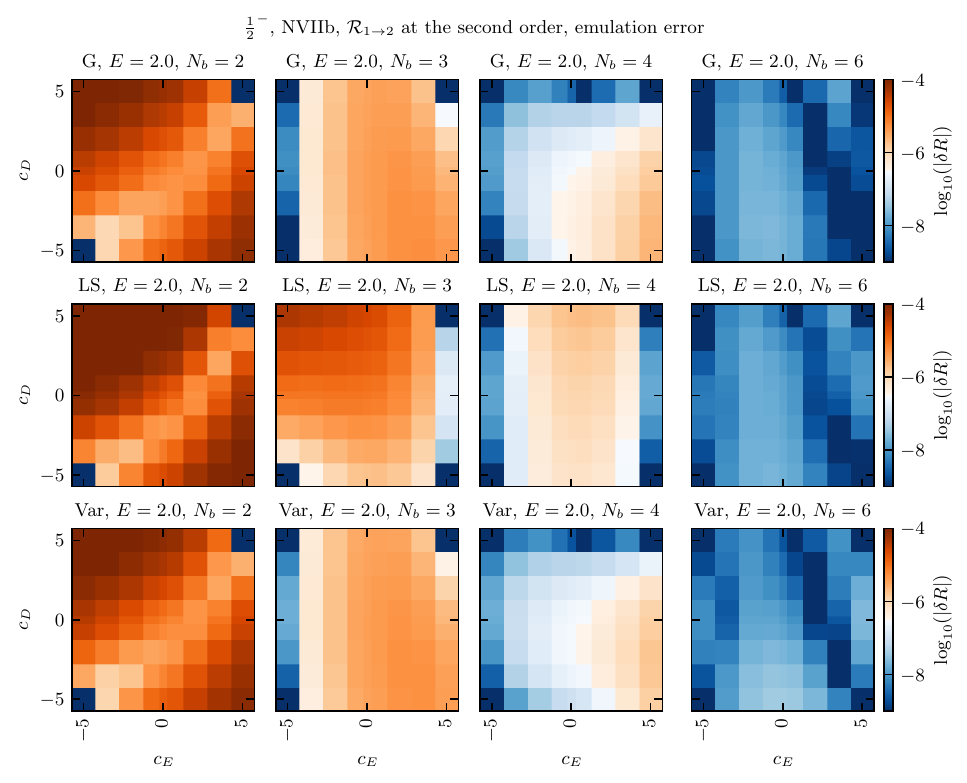}
    \caption{The same as Fig.~\ref{fig:emul_error_halfminus_R11_1st_NVIIb_multiple_emulators} but now for the $\rmat{2}{1}$ at its second order.}
    \label{fig:emul_error_halfminus_R12_2nd_NVIIb_multiple_emulators}
\end{figure*}

\begin{figure*}
    \centering
    \includegraphics{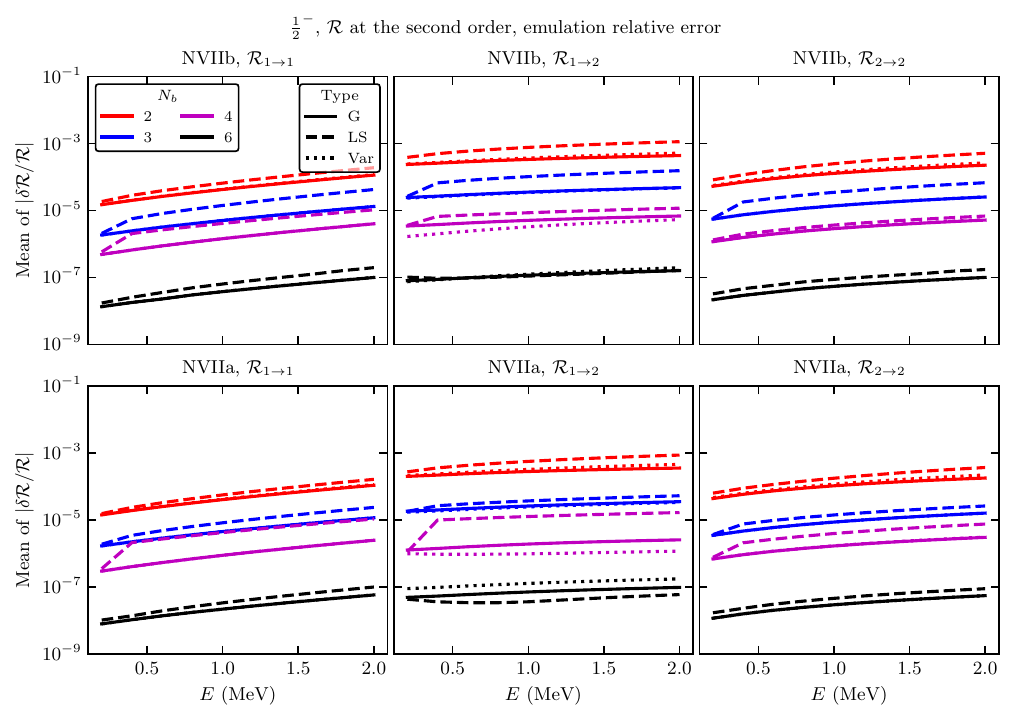}
    \caption{Emulator relative errors averaged over the parameter space vs.\ $E$ for $\rmat{a'}{a}$, with $a',a=1,2$, at its 2nd order in $\halfplus$ channel for the NVIIb (top row) and NVIIa (bottom row) force. From left to right, $\dimL$ increases, and the errors decrease. Three different emulators are compared in each panel:  ``G'' (G-ROM, solid line), ``LS'' (LSPG-ROM, dashed line), and ``Var'' (variational, dotted line). Different colors correspond to emulators trained with different $\dimL$ values.  }
    \label{fig:emul_rel_error_mean_halfminus_R_2nd_E_NVIIab_multiple_emulators}
\end{figure*}

Similar to the results presented for the $\halfplus$ channel, Figs.~\ref{fig:res_halfminus_GROM_NVIIb}, \ref{fig:emul_error_halfminus_R11_1st_NVIIb_multiple_emulators}, \ref{fig:emul_error_halfminus_R12_1st_NVIIb_multiple_emulators}, \ref{fig:emul_error_halfminus_R11_2nd_NVIIb_multiple_emulators}, \ref{fig:emul_error_halfminus_R12_2nd_NVIIb_multiple_emulators}, and~\ref{fig:emul_rel_error_mean_halfminus_R_2nd_E_NVIIab_multiple_emulators} depict the corresponding results for the $\halfminus$ channel. Note that the scales of the plots are different from those in the $\halfplus$ case. 

As already anticipated at the beginning of this section, the emulation relative errors are smaller than in $\halfplus$, given the same number of training points. At $\dimL = 6$, the emulation relative errors already reduce to below $10^{-7}$. 
Such a dramatic difference is, in fact, mostly due to the small variations of the matrix elements in this channel as function of $c_E$ and $c_D$. To appreciate the emulator performance, we should examine $\delta R/R_\mathrm{ref}$, where $R_\mathrm{ref}$ represents the typical size of a matrix element's variation. Going back to Fig.~\ref{fig:Rvsparam}, we estimate $R_\mathrm{ref}$ to be on the level of percent of the $R$ matrix elements themselves, while that ratio is on the order of $1$ for the $\rmat{1}{1}$ in the $\halfplus$ scattering state. Therefore, to better make comparisons between $\halfminus$ and $\halfplus$, we should multiply the relative errors in $\halfminus$ by a factor of $100$. By this metric, we observe similar performance, e.g., between the $\dimL = 6$ results in $\halfminus$ and the $\dimL=7$ results in $\halfplus$. 

\section{Summary and outlook}

In this work, we have developed and analyzed the RBM emulators for the $pd$ scattering below the deuteron breakup threshold. The high-fidelity calculations are performed using the well-established HH method. Our primary goal was to create efficient tools to eventually overcome the computational challenges associated with exploring the high-dimensional parameter space of \chiEFT for calibrating 3N forces.

We presented a detailed description of the HH method and the implementation of three different types of emulators: a variational-based emulator, G-ROM, and LSPG-ROM, where the ROMs are based on Petrov-Galerkin projections. The emulators exploit the affine dependence of the Hamiltonian on LECs in \chiEFT, allowing for computationally intensive parts to be performed once in an offline (or training) stage, while efficient predictions are made online.

A key aspect of our approach is the application of greedy algorithms for active learning. This algorithm enriches the emulator basis by iteratively adding training points where the emulation error is estimated to be the largest at each step. This process is locally optimal and facilitated by the residual vector, which serves as a reliable proxy for the true error. This method ensures that the emulator accuracy is efficiently and iteratively improved with a minimal number of training (high-fidelity) calculations.

Our results demonstrate that all three emulators perform well in reproducing the behavior of diagonal and off-diagonal $R$-matrix elements across different energies and parameter spaces for different chiral forces (NVIIa and NVIIb) and scattering channels ($\halfplus$ and $\halfminus$). While all emulators show a significant reduction in error with an increasing number of training points, the second-order-estimate emulations of the $R$-matrix elements consistently yield much better accuracy. With approximately $10$ training points in a two-dimensional parameter space, our emulators achieve relative errors of less than $10^{-7}$.  

The developed methodology provides a robust and efficient framework for future research, particularly for applications involving Bayesian statistics to constrain and calibrate nuclear chiral forces. The presented methodology, which includes the application of RBM and active learning, can be generalized to other complex scattering processes, providing a powerful numerical tool in this research area.
Another essential step for principled UQ will be propagating the emulator error to the predictions and considering it on equal footing with the other theoretical uncertainties of the scattering calculations, such as the EFT truncation error. 
The DOE STREAMLINE2 collaboration is currently pursuing several of these avenues.


\begin{acknowledgments}
We thank Petar Mlinarić, Stephan Rave, and our STREAMLINE collaborators for fruitful discussions.
We also thank ECT* for support at the 2025 workshop ``Next generation ab initio nuclear theory'' during which this work was presented and further developed.
This material is based upon work supported by the U.S. Department of Energy, Office of Science, Office of Nuclear Physics, under the FRIB Theory Alliance award DE-SC0013617, under the STREAMLINE collaboration awards DE-SC0024586 (Michigan State University),  DE-SC0024233 (Ohio University) and DE-SC0024509 (Ohio State University), the STREAMLINE 2 collaboration awards DE-SC0026198, and by the National Science Foundation under awards PHY-2209442 and PHY-2514765.
R.J.F.\ also acknowledges support from the ExtreMe Matter Institute EMMI at the GSI Helmholtzzentrum für Schwerionenforschung GmbH, Darmstadt, Germany.
The work of A.G. (A.~Gnech)  is supported by the U.S. Department of Energy through the Nuclear Theory for New Physics Topical Collaboration,  under contract DE-SC0023663. A.G. (A.~Gnech) acknowledges also the support of Jefferson Lab supported by the U.S. Department of Energy under contract DE-AC05-06OR23177.
A.G. (A.~Grassi), A.K., L.E.M., and M.V. acknowledge the financial support of the
European Union - Next Generation EU, Mission 4 Component 1, CUP
I53D23001060006.
The following open-source Python libraries were used to generate the results in this work:
\texttt{matplotlib}~\cite{Hunter:2007},
\texttt{numpy}~\cite{harris2020array}, and
\texttt{scipy}~\cite{2020SciPy-NMeth}.

\end{acknowledgments}

\appendix

\section{Derivation of Eq.~\eqref{eq:lineareq_highfidelity_matrix_form} in the HH calculation}\label{app:HH_details}

The variation with respect to $\cHH_\xi^a$ gives us the following result
\begin{equation}\label{eq:first_cond}
    \frac{\delta \rfunc{a}{a}}{\delta \cHH^{a}_{\xi}}=-2\left(\sum_{\xi'=1}^{\dimxi}\cHH_{\xi'}^aT^{00}_{\xi,\xi'}+T^R_{\xi,a}+\sum_{a'=1}^{\dimC}\rmat{a'}{a}T^I_{\xi,a'}\right)\,,
\end{equation}
where we have used the fact that $T^{00}_{\xi,\xi'}=T^{00}_{\xi',\xi}$, $T^{R}_{a,\xi}=T^{R}_{\xi,a}$,  and $T^{I}_{\xi,a}=T^{I}_{a,\xi}$.
Doing the same for the variation respect to $\rmat{a_0}{a}$ we obtain
\begin{widetext}    
\begin{equation}
    \frac{\delta \rfunc{a}{a} }{\delta \rmat{a_0}{a}}=\delta_{a,a_0}-\left[2\sum_{\xi'=1}^{\dimxi} \cHH_{\xi'}^aT^{I}_{\xi',a_0}+T^{RI}_{a,a_0}+T^{IR}_{a_0,a}+\sum_{a'=1}^{\dimC}\rmat{a'}{a}(T^{II}_{a',a_0}+T^{II}_{a_0,a'})\right]\,,
\end{equation}
where again we make use of $T^{I}_{\xi,a}=T^{I}_{a,\xi}$. We then get the main equation system:

\begin{align}
      \sum_{\xi' = 1}^{\dimxi} T^{00}_{\xi,\xi'} \cHH^a_{\xi'}  +  \sum_{a' = 1}^{\dimC} T^{I}_{\xi,a'} \rmat{a'}{a} 
      & = -T^{R}_{\xi,a}  \quad \text{for}\ \xi = 1,\dots\dimxi\,, \label{eq:lineareq_highfidelity_1}  \\
      \sum_{\xi' = 1}^{\dimxi} T^{I}_{\xi',a_0} \cHH^a_{\xi'}
       + \sum_{a' = 1}^{\dimC} \frac{1}{2}(T^{II}_{a_0,a'}+ T^{II}_{a',a_0}) \rmat{a'}{a} & = 
        \frac{1}{2}\left(\delta_{a,a_0}-T^{IR}_{a_0,a}-T^{RI}_{a,a_0}\right)\quad \text{for}\ a_0 = 1,\dots\dimC\,. \label{eq:lineareq_highfidelity_2}
\end{align}
The matrix form of this equation system is in Eq.~\eqref{eq:lineareq_highfidelity_matrix_form}.

\section{Important formulas for emulations}\label{app:emulation_details}

For variational emulation of the $R$-matrix elements, the functional takes the form of 

\begin{align}
      \rfunc{a'}{a}(\cvec)-\lambda(\sum_\mu\cemul_\mu^{a'a}-1)   
    =       \sum_\mu \cemul_\mu^{a'a} \rmat{a'}{a}(\thetavec_{\mu}) -
    \sum_{\mu,\mu'} \cemul_\mu^{a'a} \cemul_{\mu'}^{a'a} T^{a'a}_{\mu',\mu}-\lambda(\sum_\mu\cemul_\mu^{a'a}-1) \, . \label{eq:varfunc_direct_emul}
\end{align}

In these variational emulations, one key component is the $T^{a'a}_{\mu',\mu}$ tensor, defined as

\begin{align}
    T_{\mu',\mu}^{a' a}(\thetavec) &= \langle \Psi^{a'}(\thetavec_{\mu'}) | H(\thetavec)-E | \Psi^{a}(\thetavec_{\mu}) \rangle\ ,\nonumber\\
    &= \sum_{i=0}^{\dimparam} \param_{i} \sum_{\xi' ,\xi }     \cHH^{a'}_{\xi'}(\thetavec_{\mu'}) \cHH^a_\xi(\thetavec_\mu) 
T^{00}_{\xi', \xi}(i) + \sum_{i=0}^{\dimparam}  \param_{i}  \sum_{\xi }   \cHH^a_\xi(\thetavec_\mu) 
    \Bigl[T^{R}_{\xi,a'}(i)+\sum_{b'}\rmat{b'}{a'}(\thetavec_{\mu'})T^I_{\xi,b'}(i)\Bigr]\nonumber\\
    & \quad\null +   \sum_{i=0}^{\dimparam}  \param_{i}  \sum_{\xi'}\cHH^{a'}_{\xi'}(\thetavec_{\mu'}) \Bigl[T^{R}_{\xi',a}(i)+\sum_b \rmat{b}{a}(\thetavec_{\mu}) T^I_{\xi',b}(i)\Bigr]\nonumber\\
    & \quad\null +  \sum_{i=0}^{\dimparam}  \param_{i} \Bigl[ T^{RR}_{a',a}(i)+\sum_{b'} \rmat{b'}{a'}(\thetavec_{\mu'}) T^{IR}_{b',a}(i) + \sum_{b} \rmat{b}{a}(\thetavec_{\mu}) T^{RI}_{a',b}(i)+\sum_{b,b'}  \rmat{b'}{a'}(\thetavec_{\mu'})
    \rmat{b}{a}(\thetavec_{\mu}) T^{II}_{b',b}(i)\Bigr]\, . \label{eq:emul_Var_kernel}
\end{align}
The expression explicitly shows the online-offline separations when computing the tensor at general $\thetavec$.

It is worth noting that in the previous works (e.g., Ref.~\cite{Furnstahl:2020abp}), the following formulas were invoked to simplify the calculations of $T_{\mu'\mu}^{a'a}$ (note $\theta_0$ is always $1$):
\begin{align}
    T_{\mu'\mu}^{a'a}(\thetavec)    = \sum_{i = 1}^{\dimparam} (  \param_i - \param_{\mu,i} )  \langle \Psi^{a'}(\thetavec_{\mu'}) | H_i | \Psi^{a}(\thetavec_{\mu}) \rangle \,. \label{eq:simplification}
 \end{align}
Here, $\theta_{\mu,i}$ means the $i$-th component of the $\thetavec$ vector at the $\mu$-th training point. Since $\theta_0$ is held to be constant $1$, the sum over the $i$ index in the equation starts effectively at $i = 1$. The simplification is based on that  $H(\thetavec_\mu)|\Psi^a(\thetavec_\mu)\rangle = E|\Psi^a(\thetavec_\mu)\rangle$ is satisfied to a very high degree. In our high-fidelity calculations, however, the discrepancy of the wave function solutions (first-order estimate) is significant enough that the simplification from Eq.~\eqref{eq:simplification} is not applicable. 
 
When performing a second-order estimate in  G-ROM and LSPG-ROM emulators, the following matrix elements need to be computed quickly:

\begin{align}
    \langle \Psi^{a'}(\thetavec) | H-E | \Psi^a(\thetavec) \rangle\ 
   &  \approx  \sum_{\mu\mu'} \cemul^{a'}_{\mu'}(\thetavec) \cemul^a_\mu(\thetavec) \sum_{i=0}^N \param_{i}\Bigl[ \sum_{\xi',\xi} X^{a'}_{\mu',\xi'}X^{a}_{\mu,\xi}  T^{00}_{\xi',\xi}(i)\Bigr]\nonumber\\
   & \quad\null + \sum_\mu \cemul^a_\mu(\thetavec) \sum_i \param_{i}\Bigl[\sum_\xi X_{\mu,\xi}^a
     T^{R}_{\xi,a'}(i)\Bigr]+\sum_{\mu,\mu'} \cemul^a_\mu(\thetavec)  \cemul^{a'}_{\mu'}(\thetavec) \sum_i \param_{i}\Bigl[\sum_{\xi,b'} X_{\mu,\xi}^a X^{a'}_{\mu',b'}T^I_{\xi,b'}(i)\Bigr]\nonumber\\
     & \quad\null + \sum_\mu \cemul^{a'}_{\mu'}(\thetavec)\sum_i \param_{i}\Bigl[\sum_\xi X_{\mu',\xi}^{a'}
     T^{R}_{\xi,a}(i)\Bigr]+\sum_{\mu\mu'} \cemul^{a}_{\mu}(\thetavec) \cemul^{a'}_{\mu'}(\thetavec)\sum_i \param_{i}\Bigr[\sum_{\xi,b} X_{\mu',\xi}^{a'} X^{a}_{\mu,b}T^I_{\xi,b}(i)\Bigr]\nonumber\\
     & \quad\null +  \sum_i \param_{i} T^{RR}_{a'a}(i)+\sum_\mu \cemul^{a}_{\mu}(\thetavec)\sum_i \param_{i}\Bigl[\sum_b X_{\mu,b}^aT^{RI}_{a'b}(i)\Bigr]+
     \sum_{\mu'} \cemul^{a'}_{\mu'}(\thetavec)\sum_i \param_{i}\Bigl[\sum_{b'} X_{\mu',b'}^{a'}T^{IR}_{b'a}(i)\Bigr]\nonumber\\
     & \quad\null + \sum_{\mu\mu'} \cemul^{a'}_{\mu'}(\thetavec) \cemul^{a}_{\mu}(\thetavec)  \sum_i \param_{i}\Bigl[\sum_{b,b'}X_{\mu',b'}^{a'} X_{\mu,b}^{a} T_{b'b}^{II}(i)\Bigr]\,. \label{eq:emul_GROM_kernel}
\end{align}
The matrix elements within square brackets are the only ones that are computed in the full space and can be precomputed in the offline stage, then stored. The sum over $\mu$ and $\mu'$ is instead in the reduced space and can be done online.

\section{Extra Results}\label{app:more_results}
Figures~\ref{fig:emul_error_halfplus_R22_1st_NVIIb_multiple_emulators} and~\ref{fig:emul_error_halfplus_R22_2nd_NVIIb_multiple_emulators} provide additional results for $\rmat{2}{2}$, corresponding to Fig.~\ref{fig:emul_error_halfplus_R11_1st_NVIIb_multiple_emulators} in the main text.

\begin{figure*}
    \centering
    \includegraphics{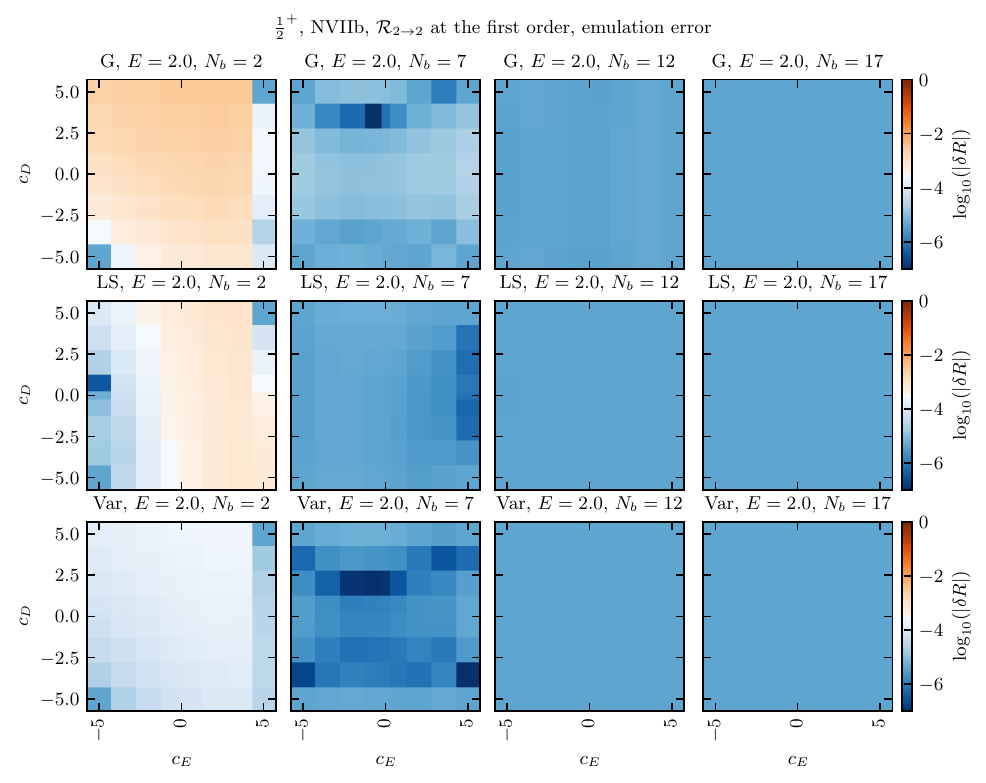}
    \caption{The same as Fig.~\ref{fig:emul_error_halfplus_R11_1st_NVIIb_multiple_emulators} but for $\rmat{2}{2}$ at its first order.}
    \label{fig:emul_error_halfplus_R22_1st_NVIIb_multiple_emulators}
\end{figure*}

\begin{figure*}
    \centering
    \includegraphics{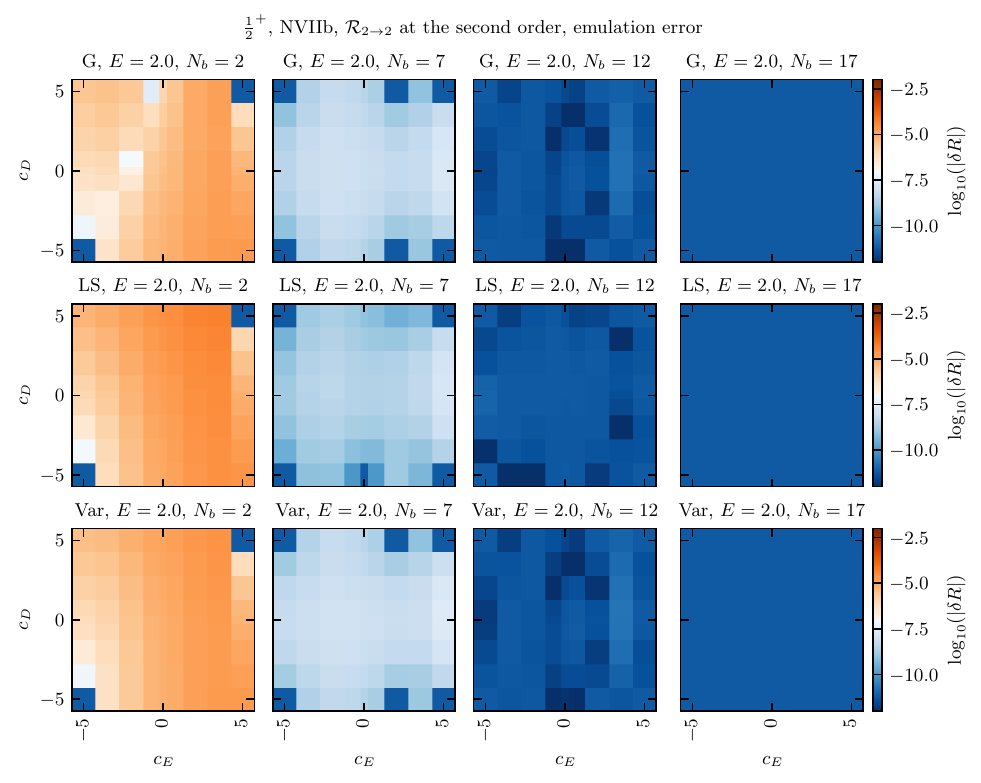}
    \caption{The same as Fig.~\ref{fig:emul_error_halfplus_R11_1st_NVIIb_multiple_emulators} but for $\rmat{2}{2}$ at its second order.}
    \label{fig:emul_error_halfplus_R22_2nd_NVIIb_multiple_emulators}
\end{figure*}

\end{widetext}


\begin{thebibliography}{91}%
\makeatletter
\providecommand \@ifxundefined [1]{%
 \@ifx{#1\undefined}
}%
\providecommand \@ifnum [1]{%
 \ifnum #1\expandafter \@firstoftwo
 \else \expandafter \@secondoftwo
 \fi
}%
\providecommand \@ifx [1]{%
 \ifx #1\expandafter \@firstoftwo
 \else \expandafter \@secondoftwo
 \fi
}%
\providecommand \natexlab [1]{#1}%
\providecommand \enquote  [1]{``#1''}%
\providecommand \bibnamefont  [1]{#1}%
\providecommand \bibfnamefont [1]{#1}%
\providecommand \citenamefont [1]{#1}%
\providecommand \href@noop [0]{\@secondoftwo}%
\providecommand \href [0]{\begingroup \@sanitize@url \@href}%
\providecommand \@href[1]{\@@startlink{#1}\@@href}%
\providecommand \@@href[1]{\endgroup#1\@@endlink}%
\providecommand \@sanitize@url [0]{\catcode `\\12\catcode `\$12\catcode `\&12\catcode `\#12\catcode `\^12\catcode `\_12\catcode `\%12\relax}%
\providecommand \@@startlink[1]{}%
\providecommand \@@endlink[0]{}%
\providecommand \url  [0]{\begingroup\@sanitize@url \@url }%
\providecommand \@url [1]{\endgroup\@href {#1}{\urlprefix }}%
\providecommand \urlprefix  [0]{URL }%
\providecommand \Eprint [0]{\href }%
\providecommand \doibase [0]{http://dx.doi.org/}%
\providecommand \selectlanguage [0]{\@gobble}%
\providecommand \bibinfo  [0]{\@secondoftwo}%
\providecommand \bibfield  [0]{\@secondoftwo}%
\providecommand \translation [1]{[#1]}%
\providecommand \BibitemOpen [0]{}%
\providecommand \bibitemStop [0]{}%
\providecommand \bibitemNoStop [0]{.\EOS\space}%
\providecommand \EOS [0]{\spacefactor3000\relax}%
\providecommand \BibitemShut  [1]{\csname bibitem#1\endcsname}%
\let\auto@bib@innerbib\@empty
\bibitem [{\citenamefont {Epelbaum}\ \emph {et~al.}(2019)\citenamefont {Epelbaum} \emph {et~al.}}]{LENPIC:2018ewt}%
  \BibitemOpen
  \bibfield  {author} {\bibinfo {author} {\bibfnamefont {E.}~\bibnamefont {Epelbaum}} \emph {et~al.} (\bibinfo {collaboration} {LENPIC}),\ }\href {\doibase 10.1103/PhysRevC.99.024313} {\bibfield  {journal} {\bibinfo  {journal} {Phys. Rev. C}\ }\textbf {\bibinfo {volume} {99}},\ \bibinfo {pages} {024313} (\bibinfo {year} {2019})},\ \Eprint {http://arxiv.org/abs/1807.02848} {arXiv:1807.02848 [nucl-th]} \BibitemShut {NoStop}%
\bibitem [{\citenamefont {Epelbaum}\ \emph {et~al.}(2002)\citenamefont {Epelbaum}, \citenamefont {Nogga}, \citenamefont {Gloeckle}, \citenamefont {Kamada}, \citenamefont {Mei{\ss}ner},\ and\ \citenamefont {Witala}}]{Epelbaum:2002vt}%
  \BibitemOpen
  \bibfield  {author} {\bibinfo {author} {\bibfnamefont {E.}~\bibnamefont {Epelbaum}}, \bibinfo {author} {\bibfnamefont {A.}~\bibnamefont {Nogga}}, \bibinfo {author} {\bibfnamefont {W.}~\bibnamefont {Gloeckle}}, \bibinfo {author} {\bibfnamefont {H.}~\bibnamefont {Kamada}}, \bibinfo {author} {\bibfnamefont {U.-G.}\ \bibnamefont {Mei{\ss}ner}}, \ and\ \bibinfo {author} {\bibfnamefont {H.}~\bibnamefont {Witala}},\ }\href {\doibase 10.1103/PhysRevC.66.064001} {\bibfield  {journal} {\bibinfo  {journal} {Phys. Rev. C}\ }\textbf {\bibinfo {volume} {66}},\ \bibinfo {pages} {064001} (\bibinfo {year} {2002})},\ \Eprint {http://arxiv.org/abs/nucl-th/0208023} {arXiv:nucl-th/0208023} \BibitemShut {NoStop}%
\bibitem [{\citenamefont {Epelbaum}\ \emph {et~al.}(2009)\citenamefont {Epelbaum}, \citenamefont {Hammer},\ and\ \citenamefont {Mei{\ss}ner}}]{Epelbaum:2008ga}%
  \BibitemOpen
  \bibfield  {author} {\bibinfo {author} {\bibfnamefont {E.}~\bibnamefont {Epelbaum}}, \bibinfo {author} {\bibfnamefont {H.-W.}\ \bibnamefont {Hammer}}, \ and\ \bibinfo {author} {\bibfnamefont {U.-G.}\ \bibnamefont {Mei{\ss}ner}},\ }\href {\doibase 10.1103/RevModPhys.81.1773} {\bibfield  {journal} {\bibinfo  {journal} {Rev. Mod. Phys.}\ }\textbf {\bibinfo {volume} {81}},\ \bibinfo {pages} {1773} (\bibinfo {year} {2009})},\ \Eprint {http://arxiv.org/abs/0811.1338} {arXiv:0811.1338} \BibitemShut {NoStop}%
\bibitem [{\citenamefont {Weinberg}(1992)}]{Weinberg:1992yk}%
  \BibitemOpen
  \bibfield  {author} {\bibinfo {author} {\bibfnamefont {S.}~\bibnamefont {Weinberg}},\ }\href@noop {} {\bibfield  {journal} {\bibinfo  {journal} {Phys. Lett. B}\ }\textbf {\bibinfo {volume} {295}},\ \bibinfo {pages} {114} (\bibinfo {year} {1992})},\ \Eprint {http://arxiv.org/abs/hep-ph/9209257} {hep-ph/9209257} \BibitemShut {NoStop}%
\bibitem [{\citenamefont {Hammer}\ \emph {et~al.}(2013)\citenamefont {Hammer}, \citenamefont {Nogga},\ and\ \citenamefont {Schwenk}}]{Hammer:2012id}%
  \BibitemOpen
  \bibfield  {author} {\bibinfo {author} {\bibfnamefont {H.-W.}\ \bibnamefont {Hammer}}, \bibinfo {author} {\bibfnamefont {A.}~\bibnamefont {Nogga}}, \ and\ \bibinfo {author} {\bibfnamefont {A.}~\bibnamefont {Schwenk}},\ }\href {\doibase 10.1103/RevModPhys.85.197} {\bibfield  {journal} {\bibinfo  {journal} {Rev. Mod. Phys.}\ }\textbf {\bibinfo {volume} {85}},\ \bibinfo {pages} {197} (\bibinfo {year} {2013})},\ \Eprint {http://arxiv.org/abs/1210.4273} {arXiv:1210.4273} \BibitemShut {NoStop}%
\bibitem [{\citenamefont {Epelbaum}(2012)}]{Epelbaum:2012zz}%
  \BibitemOpen
  \bibfield  {author} {\bibinfo {author} {\bibfnamefont {E.}~\bibnamefont {Epelbaum}},\ }\href {\doibase 10.1016/j.ppnp.2011.12.041} {\bibfield  {journal} {\bibinfo  {journal} {Prog. Part. Nucl. Phys.}\ }\textbf {\bibinfo {volume} {67}},\ \bibinfo {pages} {343} (\bibinfo {year} {2012})}\BibitemShut {NoStop}%
\bibitem [{\citenamefont {Hebeler}(2021)}]{Hebeler:2020ocj}%
  \BibitemOpen
  \bibfield  {author} {\bibinfo {author} {\bibfnamefont {K.}~\bibnamefont {Hebeler}},\ }\href {\doibase 10.1016/j.physrep.2020.08.009} {\bibfield  {journal} {\bibinfo  {journal} {Phys. Rept.}\ }\textbf {\bibinfo {volume} {890}},\ \bibinfo {pages} {1} (\bibinfo {year} {2021})},\ \Eprint {http://arxiv.org/abs/2002.09548} {arXiv:2002.09548 [nucl-th]} \BibitemShut {NoStop}%
\bibitem [{\citenamefont {Epelbaum}\ \emph {et~al.}(2020)\citenamefont {Epelbaum}, \citenamefont {Krebs},\ and\ \citenamefont {Reinert}}]{Epelbaum:2019kcf}%
  \BibitemOpen
  \bibfield  {author} {\bibinfo {author} {\bibfnamefont {E.}~\bibnamefont {Epelbaum}}, \bibinfo {author} {\bibfnamefont {H.}~\bibnamefont {Krebs}}, \ and\ \bibinfo {author} {\bibfnamefont {P.}~\bibnamefont {Reinert}},\ }\href {\doibase 10.3389/fphy.2020.00098} {\bibfield  {journal} {\bibinfo  {journal} {Front. Phys.}\ }\textbf {\bibinfo {volume} {8}},\ \bibinfo {pages} {98} (\bibinfo {year} {2020})},\ \Eprint {http://arxiv.org/abs/1911.11875} {arXiv:1911.11875} \BibitemShut {NoStop}%
\bibitem [{\citenamefont {Machleidt}\ and\ \citenamefont {Sammarruca}(2024)}]{Machleidt:2024bwl}%
  \BibitemOpen
  \bibfield  {author} {\bibinfo {author} {\bibfnamefont {R.}~\bibnamefont {Machleidt}}\ and\ \bibinfo {author} {\bibfnamefont {F.}~\bibnamefont {Sammarruca}},\ }\href {\doibase 10.1016/j.ppnp.2024.104117} {\bibfield  {journal} {\bibinfo  {journal} {Prog. Part. Nucl. Phys.}\ }\textbf {\bibinfo {volume} {137}},\ \bibinfo {pages} {104117} (\bibinfo {year} {2024})},\ \Eprint {http://arxiv.org/abs/2402.14032} {arXiv:2402.14032 [nucl-th]} \BibitemShut {NoStop}%
\bibitem [{\citenamefont {Weinberg}(1990)}]{Weinberg:1990rz}%
  \BibitemOpen
  \bibfield  {author} {\bibinfo {author} {\bibfnamefont {S.}~\bibnamefont {Weinberg}},\ }\href@noop {} {\bibfield  {journal} {\bibinfo  {journal} {Phys. Lett. B}\ }\textbf {\bibinfo {volume} {251}},\ \bibinfo {pages} {288} (\bibinfo {year} {1990})}\BibitemShut {NoStop}%
\bibitem [{\citenamefont {Weinberg}(1991)}]{Weinberg:1991um}%
  \BibitemOpen
  \bibfield  {author} {\bibinfo {author} {\bibfnamefont {S.}~\bibnamefont {Weinberg}},\ }\href@noop {} {\bibfield  {journal} {\bibinfo  {journal} {Nucl. Phys. B}\ }\textbf {\bibinfo {volume} {363}},\ \bibinfo {pages} {3} (\bibinfo {year} {1991})}\BibitemShut {NoStop}%
\bibitem [{\citenamefont {Machleidt}\ and\ \citenamefont {Entem}(2011)}]{Machleidt:2011zz}%
  \BibitemOpen
  \bibfield  {author} {\bibinfo {author} {\bibfnamefont {R.}~\bibnamefont {Machleidt}}\ and\ \bibinfo {author} {\bibfnamefont {D.~R.}\ \bibnamefont {Entem}},\ }\href {\doibase 10.1016/j.physrep.2011.02.001} {\bibfield  {journal} {\bibinfo  {journal} {Phys. Rept.}\ }\textbf {\bibinfo {volume} {503}},\ \bibinfo {pages} {1} (\bibinfo {year} {2011})},\ \Eprint {http://arxiv.org/abs/1105.2919} {arXiv:1105.2919} \BibitemShut {NoStop}%
\bibitem [{\citenamefont {Phillips}\ \emph {et~al.}(2021)\citenamefont {Phillips}, \citenamefont {Furnstahl}, \citenamefont {Heinz}, \citenamefont {Maiti}, \citenamefont {Nazarewicz}, \citenamefont {Nunes}, \citenamefont {Plumlee}, \citenamefont {Pratola}, \citenamefont {Pratt}, \citenamefont {Viens},\ and\ \citenamefont {Wild}}]{Phillips:2020dmw}%
  \BibitemOpen
  \bibfield  {author} {\bibinfo {author} {\bibfnamefont {D.~R.}\ \bibnamefont {Phillips}}, \bibinfo {author} {\bibfnamefont {R.~J.}\ \bibnamefont {Furnstahl}}, \bibinfo {author} {\bibfnamefont {U.}~\bibnamefont {Heinz}}, \bibinfo {author} {\bibfnamefont {T.}~\bibnamefont {Maiti}}, \bibinfo {author} {\bibfnamefont {W.}~\bibnamefont {Nazarewicz}}, \bibinfo {author} {\bibfnamefont {F.~M.}\ \bibnamefont {Nunes}}, \bibinfo {author} {\bibfnamefont {M.}~\bibnamefont {Plumlee}}, \bibinfo {author} {\bibfnamefont {M.~T.}\ \bibnamefont {Pratola}}, \bibinfo {author} {\bibfnamefont {S.}~\bibnamefont {Pratt}}, \bibinfo {author} {\bibfnamefont {F.~G.}\ \bibnamefont {Viens}}, \ and\ \bibinfo {author} {\bibfnamefont {S.~M.}\ \bibnamefont {Wild}},\ }\href {\doibase 10.1088/1361-6471/abf1df} {\bibfield  {journal} {\bibinfo  {journal} {J. Phys. G}\ }\textbf {\bibinfo {volume} {48}},\ \bibinfo {pages} {072001} (\bibinfo {year} {2021})},\ \Eprint {http://arxiv.org/abs/2012.07704} {arXiv:2012.07704 [nucl-th]} \BibitemShut {NoStop}%
\bibitem [{\citenamefont {Wesolowski}\ \emph {et~al.}(2021)\citenamefont {Wesolowski}, \citenamefont {Svensson}, \citenamefont {Ekstr\"om}, \citenamefont {Forss\'en}, \citenamefont {Furnstahl}, \citenamefont {Melendez},\ and\ \citenamefont {Phillips}}]{Wesolowski:2021cni}%
  \BibitemOpen
  \bibfield  {author} {\bibinfo {author} {\bibfnamefont {S.}~\bibnamefont {Wesolowski}}, \bibinfo {author} {\bibfnamefont {I.}~\bibnamefont {Svensson}}, \bibinfo {author} {\bibfnamefont {A.}~\bibnamefont {Ekstr\"om}}, \bibinfo {author} {\bibfnamefont {C.}~\bibnamefont {Forss\'en}}, \bibinfo {author} {\bibfnamefont {R.~J.}\ \bibnamefont {Furnstahl}}, \bibinfo {author} {\bibfnamefont {J.~A.}\ \bibnamefont {Melendez}}, \ and\ \bibinfo {author} {\bibfnamefont {D.~R.}\ \bibnamefont {Phillips}},\ }\href {\doibase 10.1103/PhysRevC.104.064001} {\bibfield  {journal} {\bibinfo  {journal} {Phys. Rev. C}\ }\textbf {\bibinfo {volume} {104}},\ \bibinfo {pages} {064001} (\bibinfo {year} {2021})},\ \Eprint {http://arxiv.org/abs/2104.04441} {arXiv:2104.04441 [nucl-th]} \BibitemShut {NoStop}%
\bibitem [{\citenamefont {Gloeckle}\ \emph {et~al.}(1996)\citenamefont {Gloeckle}, \citenamefont {Witala}, \citenamefont {Huber}, \citenamefont {Kamada},\ and\ \citenamefont {Golak}}]{Gloeckle:1995jg}%
  \BibitemOpen
  \bibfield  {author} {\bibinfo {author} {\bibfnamefont {W.}~\bibnamefont {Gloeckle}}, \bibinfo {author} {\bibfnamefont {H.}~\bibnamefont {Witala}}, \bibinfo {author} {\bibfnamefont {D.}~\bibnamefont {Huber}}, \bibinfo {author} {\bibfnamefont {H.}~\bibnamefont {Kamada}}, \ and\ \bibinfo {author} {\bibfnamefont {J.}~\bibnamefont {Golak}},\ }\href {\doibase 10.1016/0370-1573(95)00085-2} {\bibfield  {journal} {\bibinfo  {journal} {Phys. Rept.}\ }\textbf {\bibinfo {volume} {274}},\ \bibinfo {pages} {107} (\bibinfo {year} {1996})}\BibitemShut {NoStop}%
\bibitem [{\citenamefont {Deltuva}\ \emph {et~al.}(2008)\citenamefont {Deltuva}, \citenamefont {Fonseca},\ and\ \citenamefont {Sauer}}]{DeltuvaCoulombReview2008}%
  \BibitemOpen
  \bibfield  {author} {\bibinfo {author} {\bibfnamefont {A.}~\bibnamefont {Deltuva}}, \bibinfo {author} {\bibfnamefont {A.}~\bibnamefont {Fonseca}}, \ and\ \bibinfo {author} {\bibfnamefont {P.}~\bibnamefont {Sauer}},\ }\href {\doibase https://doi.org/10.1146/annurev.nucl.58.110707.171203} {\bibfield  {journal} {\bibinfo  {journal} {Annual Review of Nuclear and Particle Science}\ }\textbf {\bibinfo {volume} {58}},\ \bibinfo {pages} {27} (\bibinfo {year} {2008})}\BibitemShut {NoStop}%
\bibitem [{\citenamefont {Kievsky}\ \emph {et~al.}(2008)\citenamefont {Kievsky}, \citenamefont {Rosati}, \citenamefont {Viviani}, \citenamefont {Marcucci},\ and\ \citenamefont {Girlanda}}]{Kievsky2008}%
  \BibitemOpen
  \bibfield  {author} {\bibinfo {author} {\bibfnamefont {A.}~\bibnamefont {Kievsky}}, \bibinfo {author} {\bibfnamefont {S.}~\bibnamefont {Rosati}}, \bibinfo {author} {\bibfnamefont {M.}~\bibnamefont {Viviani}}, \bibinfo {author} {\bibfnamefont {L.}~\bibnamefont {Marcucci}}, \ and\ \bibinfo {author} {\bibfnamefont {L.}~\bibnamefont {Girlanda}},\ }\href {\doibase 10.1088/0954-3899/35/6/063101} {\bibfield  {journal} {\bibinfo  {journal} {J. Phys. G: Nucl. Part. Phys.}\ }\textbf {\bibinfo {volume} {35}},\ \bibinfo {pages} {063101} (\bibinfo {year} {2008})}\BibitemShut {NoStop}%
\bibitem [{\citenamefont {Deltuva}\ \emph {et~al.}(2014)\citenamefont {Deltuva}, \citenamefont {Fonseca},\ and\ \citenamefont {Lazauskas}}]{Deltuva:2012kt}%
  \BibitemOpen
  \bibfield  {author} {\bibinfo {author} {\bibfnamefont {A.}~\bibnamefont {Deltuva}}, \bibinfo {author} {\bibfnamefont {A.~C.}\ \bibnamefont {Fonseca}}, \ and\ \bibinfo {author} {\bibfnamefont {R.}~\bibnamefont {Lazauskas}},\ }\href {\doibase 10.1007/978-3-319-01077-9_1} {\bibfield  {journal} {\bibinfo  {journal} {Lect. Notes Phys.}\ }\textbf {\bibinfo {volume} {875}},\ \bibinfo {pages} {1} (\bibinfo {year} {2014})},\ \Eprint {http://arxiv.org/abs/1201.4979} {arXiv:1201.4979 [nucl-th]} \BibitemShut {NoStop}%
\bibitem [{\citenamefont {Lazauskas}\ and\ \citenamefont {Carbonell}(2019)}]{Lazauskas:2019rfb}%
  \BibitemOpen
  \bibfield  {author} {\bibinfo {author} {\bibfnamefont {R.}~\bibnamefont {Lazauskas}}\ and\ \bibinfo {author} {\bibfnamefont {J.}~\bibnamefont {Carbonell}},\ }\href {\doibase 10.1007/s00601-019-1529-5} {\bibfield  {journal} {\bibinfo  {journal} {Few Body Syst.}\ }\textbf {\bibinfo {volume} {60}},\ \bibinfo {pages} {62} (\bibinfo {year} {2019})},\ \Eprint {http://arxiv.org/abs/1908.04861} {arXiv:1908.04861 [quant-ph]} \BibitemShut {NoStop}%
\bibitem [{\citenamefont {Marcucci}\ \emph {et~al.}(2020)\citenamefont {Marcucci}, \citenamefont {Dohet-Eraly}, \citenamefont {Girlanda}, \citenamefont {Gnech}, \citenamefont {Kievsky},\ and\ \citenamefont {Viviani}}]{Marcucci2019}%
  \BibitemOpen
  \bibfield  {author} {\bibinfo {author} {\bibfnamefont {L.~E.}\ \bibnamefont {Marcucci}}, \bibinfo {author} {\bibfnamefont {J.}~\bibnamefont {Dohet-Eraly}}, \bibinfo {author} {\bibfnamefont {L.}~\bibnamefont {Girlanda}}, \bibinfo {author} {\bibfnamefont {A.}~\bibnamefont {Gnech}}, \bibinfo {author} {\bibfnamefont {A.}~\bibnamefont {Kievsky}}, \ and\ \bibinfo {author} {\bibfnamefont {M.}~\bibnamefont {Viviani}},\ }\href {\doibase 10.3389/fphy.2020.00069} {\bibfield  {journal} {\bibinfo  {journal} {Front. Phys.}\ }\textbf {\bibinfo {volume} {8}},\ \bibinfo {pages} {69} (\bibinfo {year} {2020})}\BibitemShut {NoStop}%
\bibitem [{\citenamefont {Duguet}\ \emph {et~al.}(2024)\citenamefont {Duguet}, \citenamefont {Ekstr\"om}, \citenamefont {Furnstahl}, \citenamefont {K\"onig},\ and\ \citenamefont {Lee}}]{Duguet:2023wuh}%
  \BibitemOpen
  \bibfield  {author} {\bibinfo {author} {\bibfnamefont {T.}~\bibnamefont {Duguet}}, \bibinfo {author} {\bibfnamefont {A.}~\bibnamefont {Ekstr\"om}}, \bibinfo {author} {\bibfnamefont {R.~J.}\ \bibnamefont {Furnstahl}}, \bibinfo {author} {\bibfnamefont {S.}~\bibnamefont {K\"onig}}, \ and\ \bibinfo {author} {\bibfnamefont {D.}~\bibnamefont {Lee}},\ }\href {\doibase 10.1103/RevModPhys.96.031002} {\bibfield  {journal} {\bibinfo  {journal} {Rev. Mod. Phys.}\ }\textbf {\bibinfo {volume} {96}},\ \bibinfo {pages} {031002} (\bibinfo {year} {2024})},\ \Eprint {http://arxiv.org/abs/2310.19419} {arXiv:2310.19419 [nucl-th]} \BibitemShut {NoStop}%
\bibitem [{\citenamefont {Melendez}\ \emph {et~al.}(2022)\citenamefont {Melendez}, \citenamefont {Drischler}, \citenamefont {Furnstahl}, \citenamefont {Garcia},\ and\ \citenamefont {Zhang}}]{Melendez:2022kid}%
  \BibitemOpen
  \bibfield  {author} {\bibinfo {author} {\bibfnamefont {J.~A.}\ \bibnamefont {Melendez}}, \bibinfo {author} {\bibfnamefont {C.}~\bibnamefont {Drischler}}, \bibinfo {author} {\bibfnamefont {R.~J.}\ \bibnamefont {Furnstahl}}, \bibinfo {author} {\bibfnamefont {A.~J.}\ \bibnamefont {Garcia}}, \ and\ \bibinfo {author} {\bibfnamefont {X.}~\bibnamefont {Zhang}},\ }\href {\doibase 10.1088/1361-6471/ac83dd} {\bibfield  {journal} {\bibinfo  {journal} {J. Phys. G}\ }\textbf {\bibinfo {volume} {49}},\ \bibinfo {pages} {102001} (\bibinfo {year} {2022})},\ \Eprint {http://arxiv.org/abs/2203.05528} {arXiv:2203.05528 [nucl-th]} \BibitemShut {NoStop}%
\bibitem [{\citenamefont {Drischler}\ \emph {et~al.}(2023)\citenamefont {Drischler}, \citenamefont {Melendez}, \citenamefont {Furnstahl}, \citenamefont {Garcia},\ and\ \citenamefont {Zhang}}]{Drischler:2022ipa}%
  \BibitemOpen
  \bibfield  {author} {\bibinfo {author} {\bibfnamefont {C.}~\bibnamefont {Drischler}}, \bibinfo {author} {\bibfnamefont {J.~A.}\ \bibnamefont {Melendez}}, \bibinfo {author} {\bibfnamefont {R.~J.}\ \bibnamefont {Furnstahl}}, \bibinfo {author} {\bibfnamefont {A.~J.}\ \bibnamefont {Garcia}}, \ and\ \bibinfo {author} {\bibfnamefont {X.}~\bibnamefont {Zhang}},\ }\href {\doibase 10.3389/fphy.2022.1092931} {\bibfield  {journal} {\bibinfo  {journal} {Front. Phys.}\ }\textbf {\bibinfo {volume} {10}},\ \bibinfo {pages} {92931} (\bibinfo {year} {2023})},\ \bibinfo {note} {supplemental, interactive Python code can be found on the companion website~\url{https://github.com/buqeye/frontiers-emulator-review}},\ \Eprint {http://arxiv.org/abs/2212.04912} {arXiv:2212.04912} \BibitemShut {NoStop}%
\bibitem [{\citenamefont {Gnech}\ \emph {et~al.}(2025)\citenamefont {Gnech}, \citenamefont {Zhang}, \citenamefont {Drischler}, \citenamefont {Furnstahl}, \citenamefont {Grassi}, \citenamefont {Kievsky}, \citenamefont {Marcucci},\ and\ \citenamefont {Viviani}}]{Nd_emulator_2025_short}%
  \BibitemOpen
  \bibfield  {author} {\bibinfo {author} {\bibfnamefont {A.}~\bibnamefont {Gnech}}, \bibinfo {author} {\bibfnamefont {X.}~\bibnamefont {Zhang}}, \bibinfo {author} {\bibfnamefont {C.}~\bibnamefont {Drischler}}, \bibinfo {author} {\bibfnamefont {R.~J.}\ \bibnamefont {Furnstahl}}, \bibinfo {author} {\bibfnamefont {A.}~\bibnamefont {Grassi}}, \bibinfo {author} {\bibfnamefont {A.}~\bibnamefont {Kievsky}}, \bibinfo {author} {\bibfnamefont {L.~E.}\ \bibnamefont {Marcucci}}, \ and\ \bibinfo {author} {\bibfnamefont {M.}~\bibnamefont {Viviani}},\ }\href@noop {} {\  (\bibinfo {year} {2025})},\ \Eprint {http://arxiv.org/abs/2511.01844} {arXiv:2511.01844 [nucl-th]} \BibitemShut {NoStop}%
\bibitem [{\citenamefont {Hesthaven}\ \emph {et~al.}(2015)\citenamefont {Hesthaven}, \citenamefont {Rozza},\ and\ \citenamefont {Stamm}}]{hesthaven2015certified}%
  \BibitemOpen
  \bibfield  {author} {\bibinfo {author} {\bibfnamefont {J.}~\bibnamefont {Hesthaven}}, \bibinfo {author} {\bibfnamefont {G.}~\bibnamefont {Rozza}}, \ and\ \bibinfo {author} {\bibfnamefont {B.}~\bibnamefont {Stamm}},\ }\href {\doibase 10.1007/978-3-319-22470-1} {\emph {\bibinfo {title} {Certified Reduced Basis Methods for Parametrized Partial Differential Equations}}},\ SpringerBriefs in Mathematics\ (\bibinfo  {publisher} {Springer International Publishing},\ \bibinfo {year} {2015})\BibitemShut {NoStop}%
\bibitem [{\citenamefont {Quarteroni}\ \emph {et~al.}(2016)\citenamefont {Quarteroni}, \citenamefont {Manzoni},\ and\ \citenamefont {Negri}}]{Quarteroni:218966}%
  \BibitemOpen
  \bibfield  {author} {\bibinfo {author} {\bibfnamefont {A.}~\bibnamefont {Quarteroni}}, \bibinfo {author} {\bibfnamefont {A.}~\bibnamefont {Manzoni}}, \ and\ \bibinfo {author} {\bibfnamefont {F.}~\bibnamefont {Negri}},\ }\href {\doibase 10.1007/978-3-319-15431-2} {\emph {\bibinfo {title} {Reduced Basis Methods for Partial Differential Equations. An Introduction}}},\ La Matematica per il 3+2. 92\ (\bibinfo  {publisher} {Springer International Publishing},\ \bibinfo {year} {2016})\BibitemShut {NoStop}%
\bibitem [{\citenamefont {Leidemann}\ and\ \citenamefont {Orlandini}(2013)}]{Leidemann:2012hr}%
  \BibitemOpen
  \bibfield  {author} {\bibinfo {author} {\bibfnamefont {W.}~\bibnamefont {Leidemann}}\ and\ \bibinfo {author} {\bibfnamefont {G.}~\bibnamefont {Orlandini}},\ }\href {\doibase 10.1016/j.ppnp.2012.09.001} {\bibfield  {journal} {\bibinfo  {journal} {Prog. Part. Nucl. Phys.}\ }\textbf {\bibinfo {volume} {68}},\ \bibinfo {pages} {158} (\bibinfo {year} {2013})},\ \Eprint {http://arxiv.org/abs/1204.4617} {arXiv:1204.4617 [nucl-th]} \BibitemShut {NoStop}%
\bibitem [{\citenamefont {Greene}\ \emph {et~al.}(2017)\citenamefont {Greene}, \citenamefont {Giannakeas},\ and\ \citenamefont {Perez-Rios}}]{Greene:2017cik}%
  \BibitemOpen
  \bibfield  {author} {\bibinfo {author} {\bibfnamefont {C.~H.}\ \bibnamefont {Greene}}, \bibinfo {author} {\bibfnamefont {P.}~\bibnamefont {Giannakeas}}, \ and\ \bibinfo {author} {\bibfnamefont {J.}~\bibnamefont {Perez-Rios}},\ }\href {\doibase 10.1103/RevModPhys.89.035006} {\bibfield  {journal} {\bibinfo  {journal} {Rev. Mod. Phys.}\ }\textbf {\bibinfo {volume} {89}},\ \bibinfo {pages} {035006} (\bibinfo {year} {2017})},\ \Eprint {http://arxiv.org/abs/1704.02029} {arXiv:1704.02029 [cond-mat.quant-gas]} \BibitemShut {NoStop}%
\bibitem [{\citenamefont {Rittenhouse}\ \emph {et~al.}(2011)\citenamefont {Rittenhouse}, \citenamefont {Stecher}, \citenamefont {D’Incao}, \citenamefont {Mehta},\ and\ \citenamefont {Greene}}]{Rittenhouse_2011}%
  \BibitemOpen
  \bibfield  {author} {\bibinfo {author} {\bibfnamefont {S.~T.}\ \bibnamefont {Rittenhouse}}, \bibinfo {author} {\bibfnamefont {J.~v.}\ \bibnamefont {Stecher}}, \bibinfo {author} {\bibfnamefont {J.~P.}\ \bibnamefont {D’Incao}}, \bibinfo {author} {\bibfnamefont {N.~P.}\ \bibnamefont {Mehta}}, \ and\ \bibinfo {author} {\bibfnamefont {C.~H.}\ \bibnamefont {Greene}},\ }\href {\doibase 10.1088/0953-4075/44/17/172001} {\bibfield  {journal} {\bibinfo  {journal} {Journal of Physics B: Atomic, Molecular and Optical Physics}\ }\textbf {\bibinfo {volume} {44}},\ \bibinfo {pages} {172001} (\bibinfo {year} {2011})}\BibitemShut {NoStop}%
\bibitem [{\citenamefont {Sarkar}\ and\ \citenamefont {Lee}(2022)}]{Sarkar:2021fpz}%
  \BibitemOpen
  \bibfield  {author} {\bibinfo {author} {\bibfnamefont {A.}~\bibnamefont {Sarkar}}\ and\ \bibinfo {author} {\bibfnamefont {D.}~\bibnamefont {Lee}},\ }\href {\doibase 10.1103/PhysRevResearch.4.023214} {\bibfield  {journal} {\bibinfo  {journal} {Phys. Rev. Res.}\ }\textbf {\bibinfo {volume} {4}},\ \bibinfo {pages} {023214} (\bibinfo {year} {2022})},\ \Eprint {http://arxiv.org/abs/2107.13449} {arXiv:2107.13449 [nucl-th]} \BibitemShut {NoStop}%
\bibitem [{\citenamefont {Maldonado}\ \emph {et~al.}(2025)\citenamefont {Maldonado}, \citenamefont {Drischler}, \citenamefont {Furnstahl},\ and\ \citenamefont {Mlinari{\'c}}}]{Maldonado:2025ftg}%
  \BibitemOpen
  \bibfield  {author} {\bibinfo {author} {\bibfnamefont {J.~M.}\ \bibnamefont {Maldonado}}, \bibinfo {author} {\bibfnamefont {C.}~\bibnamefont {Drischler}}, \bibinfo {author} {\bibfnamefont {R.~J.}\ \bibnamefont {Furnstahl}}, \ and\ \bibinfo {author} {\bibfnamefont {P.}~\bibnamefont {Mlinari{\'c}}},\ }\href {\doibase 10.1103/k77q-f82l} {\bibfield  {journal} {\bibinfo  {journal} {Phys. Rev. C}\ }\textbf {\bibinfo {volume} {112}},\ \bibinfo {pages} {024002} (\bibinfo {year} {2025})},\ \Eprint {http://arxiv.org/abs/2504.06092} {arXiv:2504.06092} \BibitemShut {NoStop}%
\bibitem [{\citenamefont {van Kolck}(1994)}]{vanKolck:1994yi}%
  \BibitemOpen
  \bibfield  {author} {\bibinfo {author} {\bibfnamefont {U.}~\bibnamefont {van Kolck}},\ }\href@noop {} {\bibfield  {journal} {\bibinfo  {journal} {Phys. Rev. C}\ }\textbf {\bibinfo {volume} {49}},\ \bibinfo {pages} {2932} (\bibinfo {year} {1994})}\BibitemShut {NoStop}%
\bibitem [{\citenamefont {Binder}\ \emph {et~al.}(2016)\citenamefont {Binder} \emph {et~al.}}]{LENPIC:2015qsz}%
  \BibitemOpen
  \bibfield  {author} {\bibinfo {author} {\bibfnamefont {S.}~\bibnamefont {Binder}} \emph {et~al.} (\bibinfo {collaboration} {LENPIC}),\ }\href {\doibase 10.1103/PhysRevC.93.044002} {\bibfield  {journal} {\bibinfo  {journal} {Phys. Rev. C}\ }\textbf {\bibinfo {volume} {93}},\ \bibinfo {pages} {044002} (\bibinfo {year} {2016})},\ \Eprint {http://arxiv.org/abs/1505.07218} {arXiv:1505.07218 [nucl-th]} \BibitemShut {NoStop}%
\bibitem [{\citenamefont {Drischler}\ \emph {et~al.}(2021{\natexlab{a}})\citenamefont {Drischler}, \citenamefont {Holt},\ and\ \citenamefont {Wellenhofer}}]{Drischler:2021kxf}%
  \BibitemOpen
  \bibfield  {author} {\bibinfo {author} {\bibfnamefont {C.}~\bibnamefont {Drischler}}, \bibinfo {author} {\bibfnamefont {J.~W.}\ \bibnamefont {Holt}}, \ and\ \bibinfo {author} {\bibfnamefont {C.}~\bibnamefont {Wellenhofer}},\ }\href {\doibase 10.1146/annurev-nucl-102419-041903} {\bibfield  {journal} {\bibinfo  {journal} {Annu. Rev. Nucl. Part. Sci.}\ }\textbf {\bibinfo {volume} {71}},\ \bibinfo {pages} {403} (\bibinfo {year} {2021}{\natexlab{a}})},\ \Eprint {http://arxiv.org/abs/2101.01709} {arXiv:2101.01709} \BibitemShut {NoStop}%
\bibitem [{\citenamefont {Ekström}\ \emph {et~al.}(2015)\citenamefont {Ekström}, \citenamefont {Jansen}, \citenamefont {Wendt}, \citenamefont {Hagen}, \citenamefont {Papenbrock}, \citenamefont {Carlsson}, \citenamefont {Forssén}, \citenamefont {Hjorth-Jensen}, \citenamefont {Navrátil},\ and\ \citenamefont {Nazarewicz}}]{Ekstrom:2015rta}%
  \BibitemOpen
  \bibfield  {author} {\bibinfo {author} {\bibfnamefont {A.}~\bibnamefont {Ekström}}, \bibinfo {author} {\bibfnamefont {G.~R.}\ \bibnamefont {Jansen}}, \bibinfo {author} {\bibfnamefont {K.~A.}\ \bibnamefont {Wendt}}, \bibinfo {author} {\bibfnamefont {G.}~\bibnamefont {Hagen}}, \bibinfo {author} {\bibfnamefont {T.}~\bibnamefont {Papenbrock}}, \bibinfo {author} {\bibfnamefont {B.~D.}\ \bibnamefont {Carlsson}}, \bibinfo {author} {\bibfnamefont {C.}~\bibnamefont {Forssén}}, \bibinfo {author} {\bibfnamefont {M.}~\bibnamefont {Hjorth-Jensen}}, \bibinfo {author} {\bibfnamefont {P.}~\bibnamefont {Navrátil}}, \ and\ \bibinfo {author} {\bibfnamefont {W.}~\bibnamefont {Nazarewicz}},\ }\href {\doibase 10.1103/PhysRevC.91.051301} {\bibfield  {journal} {\bibinfo  {journal} {Phys. Rev. C}\ }\textbf {\bibinfo {volume} {91}},\ \bibinfo {pages} {051301(R)} (\bibinfo {year} {2015})},\ \Eprint {http://arxiv.org/abs/1502.04682} {arXiv:1502.04682} \BibitemShut {NoStop}%
\bibitem [{\citenamefont {Bogojeski}\ \emph {et~al.}(2020)\citenamefont {Bogojeski}, \citenamefont {Vogt-Maranto}, \citenamefont {Tuckerman}, \citenamefont {M{\"u}ller},\ and\ \citenamefont {Burke}}]{Bogojeski2020}%
  \BibitemOpen
  \bibfield  {author} {\bibinfo {author} {\bibfnamefont {M.}~\bibnamefont {Bogojeski}}, \bibinfo {author} {\bibfnamefont {L.}~\bibnamefont {Vogt-Maranto}}, \bibinfo {author} {\bibfnamefont {M.~E.}\ \bibnamefont {Tuckerman}}, \bibinfo {author} {\bibfnamefont {K.-R.}\ \bibnamefont {M{\"u}ller}}, \ and\ \bibinfo {author} {\bibfnamefont {K.}~\bibnamefont {Burke}},\ }\href {\doibase 10.1038/s41467-020-19093-1} {\bibfield  {journal} {\bibinfo  {journal} {Nat. Commun.}\ }\textbf {\bibinfo {volume} {11}},\ \bibinfo {pages} {5223} (\bibinfo {year} {2020})}\BibitemShut {NoStop}%
\bibitem [{\citenamefont {{Breen}}\ \emph {et~al.}(2020)\citenamefont {{Breen}}, \citenamefont {{Foley}}, \citenamefont {{Boekholt}},\ and\ \citenamefont {{Zwart}}}]{2020MNRAS.494.2465B}%
  \BibitemOpen
  \bibfield  {author} {\bibinfo {author} {\bibfnamefont {P.~G.}\ \bibnamefont {{Breen}}}, \bibinfo {author} {\bibfnamefont {C.~N.}\ \bibnamefont {{Foley}}}, \bibinfo {author} {\bibfnamefont {T.}~\bibnamefont {{Boekholt}}}, \ and\ \bibinfo {author} {\bibfnamefont {S.~P.}\ \bibnamefont {{Zwart}}},\ }\href {\doibase 10.1093/mnras/staa713} {\bibfield  {journal} {\bibinfo  {journal} {\mnras}\ }\textbf {\bibinfo {volume} {494}},\ \bibinfo {pages} {2465} (\bibinfo {year} {2020})},\ \Eprint {http://arxiv.org/abs/1910.07291} {arXiv:1910.07291 [astro-ph.GA]} \BibitemShut {NoStop}%
\bibitem [{\citenamefont {MacKay}(1998)}]{Mackay:1998introduction}%
  \BibitemOpen
  \bibfield  {author} {\bibinfo {author} {\bibfnamefont {D.~J.~C.}\ \bibnamefont {MacKay}},\ }in\ \href@noop {} {\emph {\bibinfo {booktitle} {Neural {{Networks}} and {{Machine Learning}}}}},\ \bibinfo {series} {{{NATO ASI Series}}}, Vol.\ \bibinfo {volume} {168},\ \bibinfo {editor} {edited by\ \bibinfo {editor} {\bibfnamefont {C.~M.}\ \bibnamefont {Bishop}}}\ (\bibinfo  {publisher} {{Springer, Berlin}},\ \bibinfo {year} {1998})\ pp.\ \bibinfo {pages} {133--166}\BibitemShut {NoStop}%
\bibitem [{\citenamefont {Rasmussen}\ and\ \citenamefont {Williams}(2006)}]{rasmussen2006gaussian}%
  \BibitemOpen
  \bibfield  {author} {\bibinfo {author} {\bibfnamefont {C.~E.}\ \bibnamefont {Rasmussen}}\ and\ \bibinfo {author} {\bibfnamefont {C.~K.~I.}\ \bibnamefont {Williams}},\ }\href {https://books.google.com/books?id=vWtwQgAACAAJ} {\emph {\bibinfo {title} {Gaussian Processes for Machine Learning}}},\ Adaptive computation and machine learning series\ (\bibinfo  {publisher} {University Press Group Limited},\ \bibinfo {address} {Cambridge, MA},\ \bibinfo {year} {2006})\BibitemShut {NoStop}%
\bibitem [{\citenamefont {Benner}\ \emph {et~al.}(2017{\natexlab{a}})\citenamefont {Benner}, \citenamefont {Ohlberger}, \citenamefont {Patera}, \citenamefont {Rozza},\ and\ \citenamefont {Urban}}]{Benner_2017aa}%
  \BibitemOpen
  \bibinfo {editor} {\bibfnamefont {P.}~\bibnamefont {Benner}}, \bibinfo {editor} {\bibfnamefont {M.}~\bibnamefont {Ohlberger}}, \bibinfo {editor} {\bibfnamefont {A.}~\bibnamefont {Patera}}, \bibinfo {editor} {\bibfnamefont {G.}~\bibnamefont {Rozza}}, \ and\ \bibinfo {editor} {\bibfnamefont {K.}~\bibnamefont {Urban}},\ eds.,\ \href {\doibase 10.1007/978-3-319-58786-8} {\emph {\bibinfo {title} {Model Reduction of Parametrized Systems}}}\ (\bibinfo  {publisher} {Springer},\ \bibinfo {year} {2017})\BibitemShut {NoStop}%
\bibitem [{\citenamefont {Benner}\ \emph {et~al.}(2017{\natexlab{b}})\citenamefont {Benner}, \citenamefont {Cohen}, \citenamefont {Ohlberger},\ and\ \citenamefont {Willcox}}]{Benner2017modelRedApprox}%
  \BibitemOpen
  \bibfield  {author} {\bibinfo {author} {\bibfnamefont {P.}~\bibnamefont {Benner}}, \bibinfo {author} {\bibfnamefont {A.}~\bibnamefont {Cohen}}, \bibinfo {author} {\bibfnamefont {M.}~\bibnamefont {Ohlberger}}, \ and\ \bibinfo {author} {\bibfnamefont {K.}~\bibnamefont {Willcox}},\ }\href {\doibase doi:10.1137/1.9781611974829} {\emph {\bibinfo {title} {Model Reduction and Approximation}}}\ (\bibinfo  {publisher} {Society for Industrial and Applied Mathematics: Computational Science \& Engineering},\ \bibinfo {year} {2017})\BibitemShut {NoStop}%
\bibitem [{\citenamefont {Benner}\ \emph {et~al.}(2015)\citenamefont {Benner}, \citenamefont {Gugercin},\ and\ \citenamefont {Willcox}}]{benner2015survey}%
  \BibitemOpen
  \bibfield  {author} {\bibinfo {author} {\bibfnamefont {P.}~\bibnamefont {Benner}}, \bibinfo {author} {\bibfnamefont {S.}~\bibnamefont {Gugercin}}, \ and\ \bibinfo {author} {\bibfnamefont {K.}~\bibnamefont {Willcox}},\ }\href {\doibase 10.1137/130932715} {\bibfield  {journal} {\bibinfo  {journal} {SIAM Review}\ }\textbf {\bibinfo {volume} {57}},\ \bibinfo {pages} {483} (\bibinfo {year} {2015})}\BibitemShut {NoStop}%
\bibitem [{\citenamefont {K\"onig}\ \emph {et~al.}(2020)\citenamefont {K\"onig}, \citenamefont {Ekstr\"om}, \citenamefont {Hebeler}, \citenamefont {Lee},\ and\ \citenamefont {Schwenk}}]{Konig:2019adq}%
  \BibitemOpen
  \bibfield  {author} {\bibinfo {author} {\bibfnamefont {S.}~\bibnamefont {K\"onig}}, \bibinfo {author} {\bibfnamefont {A.}~\bibnamefont {Ekstr\"om}}, \bibinfo {author} {\bibfnamefont {K.}~\bibnamefont {Hebeler}}, \bibinfo {author} {\bibfnamefont {D.}~\bibnamefont {Lee}}, \ and\ \bibinfo {author} {\bibfnamefont {A.}~\bibnamefont {Schwenk}},\ }\href {\doibase 10.1016/j.physletb.2020.135814} {\bibfield  {journal} {\bibinfo  {journal} {Phys. Lett. B}\ }\textbf {\bibinfo {volume} {810}},\ \bibinfo {pages} {135814} (\bibinfo {year} {2020})},\ \Eprint {http://arxiv.org/abs/1909.08446} {arXiv:1909.08446 [nucl-th]} \BibitemShut {NoStop}%
\bibitem [{\citenamefont {Frame}\ \emph {et~al.}(2018)\citenamefont {Frame}, \citenamefont {He}, \citenamefont {Ipsen}, \citenamefont {Lee}, \citenamefont {Lee},\ and\ \citenamefont {Rrapaj}}]{Frame:2017fah}%
  \BibitemOpen
  \bibfield  {author} {\bibinfo {author} {\bibfnamefont {D.}~\bibnamefont {Frame}}, \bibinfo {author} {\bibfnamefont {R.}~\bibnamefont {He}}, \bibinfo {author} {\bibfnamefont {I.}~\bibnamefont {Ipsen}}, \bibinfo {author} {\bibfnamefont {D.}~\bibnamefont {Lee}}, \bibinfo {author} {\bibfnamefont {D.}~\bibnamefont {Lee}}, \ and\ \bibinfo {author} {\bibfnamefont {E.}~\bibnamefont {Rrapaj}},\ }\href {\doibase 10.1103/PhysRevLett.121.032501} {\bibfield  {journal} {\bibinfo  {journal} {Phys. Rev. Lett.}\ }\textbf {\bibinfo {volume} {121}},\ \bibinfo {pages} {032501} (\bibinfo {year} {2018})},\ \Eprint {http://arxiv.org/abs/1711.07090} {arXiv:1711.07090} \BibitemShut {NoStop}%
\bibitem [{\citenamefont {Sarkar}\ and\ \citenamefont {Lee}(2021)}]{Sarkar:2020mad}%
  \BibitemOpen
  \bibfield  {author} {\bibinfo {author} {\bibfnamefont {A.}~\bibnamefont {Sarkar}}\ and\ \bibinfo {author} {\bibfnamefont {D.}~\bibnamefont {Lee}},\ }\href {\doibase 10.1103/PhysRevLett.126.032501} {\bibfield  {journal} {\bibinfo  {journal} {Phys. Rev. Lett.}\ }\textbf {\bibinfo {volume} {126}},\ \bibinfo {pages} {032501} (\bibinfo {year} {2021})},\ \Eprint {http://arxiv.org/abs/2004.07651} {arXiv:2004.07651 [nucl-th]} \BibitemShut {NoStop}%
\bibitem [{\citenamefont {Demol}\ \emph {et~al.}(2020)\citenamefont {Demol}, \citenamefont {Duguet}, \citenamefont {Ekstr\"om}, \citenamefont {Frosini}, \citenamefont {Hebeler}, \citenamefont {K\"onig}, \citenamefont {Lee}, \citenamefont {Schwenk}, \citenamefont {Som\`a},\ and\ \citenamefont {Tichai}}]{Demol:2019yjt}%
  \BibitemOpen
  \bibfield  {author} {\bibinfo {author} {\bibfnamefont {P.}~\bibnamefont {Demol}}, \bibinfo {author} {\bibfnamefont {T.}~\bibnamefont {Duguet}}, \bibinfo {author} {\bibfnamefont {A.}~\bibnamefont {Ekstr\"om}}, \bibinfo {author} {\bibfnamefont {M.}~\bibnamefont {Frosini}}, \bibinfo {author} {\bibfnamefont {K.}~\bibnamefont {Hebeler}}, \bibinfo {author} {\bibfnamefont {S.}~\bibnamefont {K\"onig}}, \bibinfo {author} {\bibfnamefont {D.}~\bibnamefont {Lee}}, \bibinfo {author} {\bibfnamefont {A.}~\bibnamefont {Schwenk}}, \bibinfo {author} {\bibfnamefont {V.}~\bibnamefont {Som\`a}}, \ and\ \bibinfo {author} {\bibfnamefont {A.}~\bibnamefont {Tichai}},\ }\href {\doibase 10.1103/PhysRevC.101.041302} {\bibfield  {journal} {\bibinfo  {journal} {Phys. Rev. C}\ }\textbf {\bibinfo {volume} {101}},\ \bibinfo {pages} {041302} (\bibinfo {year} {2020})},\ \Eprint {http://arxiv.org/abs/1911.12578} {arXiv:1911.12578} \BibitemShut {NoStop}%
\bibitem [{\citenamefont {Ekström}\ and\ \citenamefont {Hagen}(2019)}]{Ekstrom:2019lss}%
  \BibitemOpen
  \bibfield  {author} {\bibinfo {author} {\bibfnamefont {A.}~\bibnamefont {Ekström}}\ and\ \bibinfo {author} {\bibfnamefont {G.}~\bibnamefont {Hagen}},\ }\href {\doibase 10.1103/PhysRevLett.123.252501} {\bibfield  {journal} {\bibinfo  {journal} {Phys. Rev. Lett.}\ }\textbf {\bibinfo {volume} {123}},\ \bibinfo {pages} {252501} (\bibinfo {year} {2019})},\ \Eprint {http://arxiv.org/abs/1910.02922} {arXiv:1910.02922 [nucl-th]} \BibitemShut {NoStop}%
\bibitem [{\citenamefont {Demol}\ \emph {et~al.}(2021)\citenamefont {Demol}, \citenamefont {Frosini}, \citenamefont {Tichai}, \citenamefont {Som\`a},\ and\ \citenamefont {Duguet}}]{Demol:2020mzd}%
  \BibitemOpen
  \bibfield  {author} {\bibinfo {author} {\bibfnamefont {P.}~\bibnamefont {Demol}}, \bibinfo {author} {\bibfnamefont {M.}~\bibnamefont {Frosini}}, \bibinfo {author} {\bibfnamefont {A.}~\bibnamefont {Tichai}}, \bibinfo {author} {\bibfnamefont {V.}~\bibnamefont {Som\`a}}, \ and\ \bibinfo {author} {\bibfnamefont {T.}~\bibnamefont {Duguet}},\ }\href {\doibase 10.1016/j.aop.2020.168358} {\bibfield  {journal} {\bibinfo  {journal} {Annals Phys.}\ }\textbf {\bibinfo {volume} {424}},\ \bibinfo {pages} {168358} (\bibinfo {year} {2021})},\ \Eprint {http://arxiv.org/abs/2002.02724} {arXiv:2002.02724 [nucl-th]} \BibitemShut {NoStop}%
\bibitem [{\citenamefont {Yoshida}\ and\ \citenamefont {Shimizu}(2022)}]{Yoshida:2021jbl}%
  \BibitemOpen
  \bibfield  {author} {\bibinfo {author} {\bibfnamefont {S.}~\bibnamefont {Yoshida}}\ and\ \bibinfo {author} {\bibfnamefont {N.}~\bibnamefont {Shimizu}},\ }\href {\doibase 10.1093/ptep/ptac057} {\bibfield  {journal} {\bibinfo  {journal} {PTEP}\ }\textbf {\bibinfo {volume} {2022}},\ \bibinfo {pages} {053D02} (\bibinfo {year} {2022})},\ \Eprint {http://arxiv.org/abs/2105.08256} {arXiv:2105.08256} \BibitemShut {NoStop}%
\bibitem [{\citenamefont {Anderson}\ \emph {et~al.}(2022)\citenamefont {Anderson}, \citenamefont {O'Donnell},\ and\ \citenamefont {Piekarewicz}}]{Anderson:2022jhq}%
  \BibitemOpen
  \bibfield  {author} {\bibinfo {author} {\bibfnamefont {A.~L.}\ \bibnamefont {Anderson}}, \bibinfo {author} {\bibfnamefont {G.~L.}\ \bibnamefont {O'Donnell}}, \ and\ \bibinfo {author} {\bibfnamefont {J.}~\bibnamefont {Piekarewicz}},\ }\href {\doibase 10.1103/PhysRevC.106.L031302} {\bibfield  {journal} {\bibinfo  {journal} {Phys. Rev. C}\ }\textbf {\bibinfo {volume} {106}},\ \bibinfo {pages} {L031302} (\bibinfo {year} {2022})},\ \Eprint {http://arxiv.org/abs/2206.14889} {arXiv:2206.14889 [nucl-th]} \BibitemShut {NoStop}%
\bibitem [{\citenamefont {Giuliani}\ \emph {et~al.}(2023)\citenamefont {Giuliani}, \citenamefont {Godbey}, \citenamefont {Bonilla}, \citenamefont {Viens},\ and\ \citenamefont {Piekarewicz}}]{Giuliani:2022yna}%
  \BibitemOpen
  \bibfield  {author} {\bibinfo {author} {\bibfnamefont {P.}~\bibnamefont {Giuliani}}, \bibinfo {author} {\bibfnamefont {K.}~\bibnamefont {Godbey}}, \bibinfo {author} {\bibfnamefont {E.}~\bibnamefont {Bonilla}}, \bibinfo {author} {\bibfnamefont {F.}~\bibnamefont {Viens}}, \ and\ \bibinfo {author} {\bibfnamefont {J.}~\bibnamefont {Piekarewicz}},\ }\href {\doibase 10.3389/fphy.2022.1054524} {\bibfield  {journal} {\bibinfo  {journal} {Front. Phys.}\ }\textbf {\bibinfo {volume} {10}} (\bibinfo {year} {2023}),\ 10.3389/fphy.2022.1054524},\ \Eprint {http://arxiv.org/abs/2209.13039} {arXiv:2209.13039} \BibitemShut {NoStop}%
\bibitem [{\citenamefont {Yapa}\ \emph {et~al.}(2023)\citenamefont {Yapa}, \citenamefont {Fossez},\ and\ \citenamefont {K\"onig}}]{Yapa:2023xyf}%
  \BibitemOpen
  \bibfield  {author} {\bibinfo {author} {\bibfnamefont {N.}~\bibnamefont {Yapa}}, \bibinfo {author} {\bibfnamefont {K.}~\bibnamefont {Fossez}}, \ and\ \bibinfo {author} {\bibfnamefont {S.}~\bibnamefont {K\"onig}},\ }\href {\doibase 10.1103/PhysRevC.107.064316} {\bibfield  {journal} {\bibinfo  {journal} {Phys. Rev. C}\ }\textbf {\bibinfo {volume} {107}},\ \bibinfo {pages} {064316} (\bibinfo {year} {2023})},\ \Eprint {http://arxiv.org/abs/2303.06139} {arXiv:2303.06139 [nucl-th]} \BibitemShut {NoStop}%
\bibitem [{\citenamefont {Yapa}\ \emph {et~al.}(2025)\citenamefont {Yapa}, \citenamefont {K{\"o}nig},\ and\ \citenamefont {Fossez}}]{Yapa:2024lya}%
  \BibitemOpen
  \bibfield  {author} {\bibinfo {author} {\bibfnamefont {N.}~\bibnamefont {Yapa}}, \bibinfo {author} {\bibfnamefont {S.}~\bibnamefont {K{\"o}nig}}, \ and\ \bibinfo {author} {\bibfnamefont {K.}~\bibnamefont {Fossez}},\ }\href {\doibase 10.1103/PhysRevC.111.064318} {\bibfield  {journal} {\bibinfo  {journal} {Phys. Rev. C}\ }\textbf {\bibinfo {volume} {111}},\ \bibinfo {pages} {064318} (\bibinfo {year} {2025})},\ \Eprint {http://arxiv.org/abs/2409.03116} {arXiv:2409.03116 [nucl-th]} \BibitemShut {NoStop}%
\bibitem [{\citenamefont {Furnstahl}\ \emph {et~al.}(2020)\citenamefont {Furnstahl}, \citenamefont {Garcia}, \citenamefont {Millican},\ and\ \citenamefont {Zhang}}]{Furnstahl:2020abp}%
  \BibitemOpen
  \bibfield  {author} {\bibinfo {author} {\bibfnamefont {R.~J.}\ \bibnamefont {Furnstahl}}, \bibinfo {author} {\bibfnamefont {A.~J.}\ \bibnamefont {Garcia}}, \bibinfo {author} {\bibfnamefont {P.~J.}\ \bibnamefont {Millican}}, \ and\ \bibinfo {author} {\bibfnamefont {X.}~\bibnamefont {Zhang}},\ }\href {\doibase 10.1016/j.physletb.2020.135719} {\bibfield  {journal} {\bibinfo  {journal} {Phys. Lett. B}\ }\textbf {\bibinfo {volume} {809}},\ \bibinfo {pages} {135719} (\bibinfo {year} {2020})},\ \Eprint {http://arxiv.org/abs/2007.03635} {arXiv:2007.03635 [nucl-th]} \BibitemShut {NoStop}%
\bibitem [{\citenamefont {Drischler}\ \emph {et~al.}(2021{\natexlab{b}})\citenamefont {Drischler}, \citenamefont {Quinonez}, \citenamefont {Giuliani}, \citenamefont {Lovell},\ and\ \citenamefont {Nunes}}]{Drischler:2021qoy}%
  \BibitemOpen
  \bibfield  {author} {\bibinfo {author} {\bibfnamefont {C.}~\bibnamefont {Drischler}}, \bibinfo {author} {\bibfnamefont {M.}~\bibnamefont {Quinonez}}, \bibinfo {author} {\bibfnamefont {P.~G.}\ \bibnamefont {Giuliani}}, \bibinfo {author} {\bibfnamefont {A.~E.}\ \bibnamefont {Lovell}}, \ and\ \bibinfo {author} {\bibfnamefont {F.~M.}\ \bibnamefont {Nunes}},\ }\href {\doibase 10.1016/j.physletb.2021.136777} {\bibfield  {journal} {\bibinfo  {journal} {Phys. Lett. B}\ }\textbf {\bibinfo {volume} {823}},\ \bibinfo {pages} {136777} (\bibinfo {year} {2021}{\natexlab{b}})},\ \Eprint {http://arxiv.org/abs/2108.08269} {arXiv:2108.08269 [nucl-th]} \BibitemShut {NoStop}%
\bibitem [{\citenamefont {Melendez}\ \emph {et~al.}(2021)\citenamefont {Melendez}, \citenamefont {Drischler}, \citenamefont {Garcia}, \citenamefont {Furnstahl},\ and\ \citenamefont {Zhang}}]{Melendez:2021lyq}%
  \BibitemOpen
  \bibfield  {author} {\bibinfo {author} {\bibfnamefont {J.~A.}\ \bibnamefont {Melendez}}, \bibinfo {author} {\bibfnamefont {C.}~\bibnamefont {Drischler}}, \bibinfo {author} {\bibfnamefont {A.~J.}\ \bibnamefont {Garcia}}, \bibinfo {author} {\bibfnamefont {R.~J.}\ \bibnamefont {Furnstahl}}, \ and\ \bibinfo {author} {\bibfnamefont {X.}~\bibnamefont {Zhang}},\ }\href {\doibase 10.1016/j.physletb.2021.136608} {\bibfield  {journal} {\bibinfo  {journal} {Phys. Lett. B}\ }\textbf {\bibinfo {volume} {821}},\ \bibinfo {pages} {136608} (\bibinfo {year} {2021})},\ \Eprint {http://arxiv.org/abs/2106.15608} {arXiv:2106.15608 [nucl-th]} \BibitemShut {NoStop}%
\bibitem [{\citenamefont {Zhang}\ and\ \citenamefont {Furnstahl}(2022)}]{Zhang:2021jmi}%
  \BibitemOpen
  \bibfield  {author} {\bibinfo {author} {\bibfnamefont {X.}~\bibnamefont {Zhang}}\ and\ \bibinfo {author} {\bibfnamefont {R.~J.}\ \bibnamefont {Furnstahl}},\ }\href {\doibase 10.1103/PhysRevC.105.064004} {\bibfield  {journal} {\bibinfo  {journal} {Phys. Rev. C}\ }\textbf {\bibinfo {volume} {105}},\ \bibinfo {pages} {064004} (\bibinfo {year} {2022})},\ \Eprint {http://arxiv.org/abs/2110.04269} {arXiv:2110.04269 [nucl-th]} \BibitemShut {NoStop}%
\bibitem [{\citenamefont {Bai}\ and\ \citenamefont {Ren}(2021)}]{Bai:2021xok}%
  \BibitemOpen
  \bibfield  {author} {\bibinfo {author} {\bibfnamefont {D.}~\bibnamefont {Bai}}\ and\ \bibinfo {author} {\bibfnamefont {Z.}~\bibnamefont {Ren}},\ }\href {\doibase 10.1103/PhysRevC.103.014612} {\bibfield  {journal} {\bibinfo  {journal} {Phys. Rev. C}\ }\textbf {\bibinfo {volume} {103}},\ \bibinfo {pages} {014612} (\bibinfo {year} {2021})},\ \Eprint {http://arxiv.org/abs/2101.06336} {arXiv:2101.06336 [nucl-th]} \BibitemShut {NoStop}%
\bibitem [{\citenamefont {Drischler}\ and\ \citenamefont {Zhang}(2022)}]{Drischler:2022yfb}%
  \BibitemOpen
  \bibfield  {author} {\bibinfo {author} {\bibfnamefont {C.}~\bibnamefont {Drischler}}\ and\ \bibinfo {author} {\bibfnamefont {X.}~\bibnamefont {Zhang}},\ }in\ \href {\doibase 10.1007/s00601-022-01749-x} {\emph {\bibinfo {booktitle} {{Nuclear Forces for Precision Nuclear Physics: A Collection of Perspectives}}}},\ Vol.~\bibinfo {volume} {63},\ \bibinfo {editor} {edited by\ \bibinfo {editor} {\bibfnamefont {I.}~\bibnamefont {Tews}}, \bibinfo {editor} {\bibfnamefont {Z.}~\bibnamefont {Davoudi}}, \bibinfo {editor} {\bibfnamefont {A.}~\bibnamefont {Ekstr{\"o}m}}, \ and\ \bibinfo {editor} {\bibfnamefont {J.~D.}\ \bibnamefont {Holt}}}\ (\bibinfo  {publisher} {Springer-Verlag GmbH Austria},\ \bibinfo {year} {2022})\ Chap.~\bibinfo {chapter} {8}, p.~\bibinfo {pages} {67},\ \Eprint {http://arxiv.org/abs/2202.01105} {arXiv:2202.01105} \BibitemShut {NoStop}%
\bibitem [{\citenamefont {Bai}(2022)}]{Bai:2022hjg}%
  \BibitemOpen
  \bibfield  {author} {\bibinfo {author} {\bibfnamefont {D.}~\bibnamefont {Bai}},\ }\href {\doibase 10.1103/PhysRevC.106.024611} {\bibfield  {journal} {\bibinfo  {journal} {Phys. Rev. C}\ }\textbf {\bibinfo {volume} {106}},\ \bibinfo {pages} {024611} (\bibinfo {year} {2022})}\BibitemShut {NoStop}%
\bibitem [{\citenamefont {Garcia}\ \emph {et~al.}(2023)\citenamefont {Garcia}, \citenamefont {Drischler}, \citenamefont {Furnstahl}, \citenamefont {Melendez},\ and\ \citenamefont {Zhang}}]{Garcia:2023slj}%
  \BibitemOpen
  \bibfield  {author} {\bibinfo {author} {\bibfnamefont {A.~J.}\ \bibnamefont {Garcia}}, \bibinfo {author} {\bibfnamefont {C.}~\bibnamefont {Drischler}}, \bibinfo {author} {\bibfnamefont {R.~J.}\ \bibnamefont {Furnstahl}}, \bibinfo {author} {\bibfnamefont {J.~A.}\ \bibnamefont {Melendez}}, \ and\ \bibinfo {author} {\bibfnamefont {X.}~\bibnamefont {Zhang}},\ }\href {\doibase 10.1103/PhysRevC.107.054001} {\bibfield  {journal} {\bibinfo  {journal} {Phys. Rev. C}\ }\textbf {\bibinfo {volume} {107}},\ \bibinfo {pages} {054001} (\bibinfo {year} {2023})},\ \Eprint {http://arxiv.org/abs/2301.05093} {arXiv:2301.05093 [nucl-th]} \BibitemShut {NoStop}%
\bibitem [{\citenamefont {Odell}\ \emph {et~al.}(2024)\citenamefont {Odell}, \citenamefont {Giuliani}, \citenamefont {Beyer}, \citenamefont {Catacora-Rios}, \citenamefont {Chan}, \citenamefont {Bonilla}, \citenamefont {Furnstahl}, \citenamefont {Godbey},\ and\ \citenamefont {Nunes}}]{Odell:2023cun}%
  \BibitemOpen
  \bibfield  {author} {\bibinfo {author} {\bibfnamefont {D.}~\bibnamefont {Odell}}, \bibinfo {author} {\bibfnamefont {P.}~\bibnamefont {Giuliani}}, \bibinfo {author} {\bibfnamefont {K.}~\bibnamefont {Beyer}}, \bibinfo {author} {\bibfnamefont {M.}~\bibnamefont {Catacora-Rios}}, \bibinfo {author} {\bibfnamefont {M.~Y.~H.}\ \bibnamefont {Chan}}, \bibinfo {author} {\bibfnamefont {E.}~\bibnamefont {Bonilla}}, \bibinfo {author} {\bibfnamefont {R.~J.}\ \bibnamefont {Furnstahl}}, \bibinfo {author} {\bibfnamefont {K.}~\bibnamefont {Godbey}}, \ and\ \bibinfo {author} {\bibfnamefont {F.~M.}\ \bibnamefont {Nunes}},\ }\href {\doibase 10.1103/PhysRevC.109.044612} {\bibfield  {journal} {\bibinfo  {journal} {Phys. Rev. C}\ }\textbf {\bibinfo {volume} {109}},\ \bibinfo {pages} {044612} (\bibinfo {year} {2024})},\ \Eprint {http://arxiv.org/abs/2312.12426} {arXiv:2312.12426 [physics.comp-ph]} \BibitemShut {NoStop}%
\bibitem [{\citenamefont {Cook}\ \emph {et~al.}(2025)\citenamefont {Cook}, \citenamefont {Jammooa}, \citenamefont {Hjorth-Jensen}, \citenamefont {Lee},\ and\ \citenamefont {Lee}}]{Cook:2024toj}%
  \BibitemOpen
  \bibfield  {author} {\bibinfo {author} {\bibfnamefont {P.}~\bibnamefont {Cook}}, \bibinfo {author} {\bibfnamefont {D.}~\bibnamefont {Jammooa}}, \bibinfo {author} {\bibfnamefont {M.}~\bibnamefont {Hjorth-Jensen}}, \bibinfo {author} {\bibfnamefont {D.~D.}\ \bibnamefont {Lee}}, \ and\ \bibinfo {author} {\bibfnamefont {D.}~\bibnamefont {Lee}},\ }\href {\doibase 10.1038/s41467-025-61362-4} {\bibfield  {journal} {\bibinfo  {journal} {Nature Commun.}\ }\textbf {\bibinfo {volume} {16}},\ \bibinfo {pages} {5929} (\bibinfo {year} {2025})},\ \Eprint {http://arxiv.org/abs/2401.11694} {arXiv:2401.11694 [cs.LG]} \BibitemShut {NoStop}%
\bibitem [{\citenamefont {Bonilla}\ \emph {et~al.}(2022)\citenamefont {Bonilla}, \citenamefont {Giuliani}, \citenamefont {Godbey},\ and\ \citenamefont {Lee}}]{Bonilla:2022rph}%
  \BibitemOpen
  \bibfield  {author} {\bibinfo {author} {\bibfnamefont {E.}~\bibnamefont {Bonilla}}, \bibinfo {author} {\bibfnamefont {P.}~\bibnamefont {Giuliani}}, \bibinfo {author} {\bibfnamefont {K.}~\bibnamefont {Godbey}}, \ and\ \bibinfo {author} {\bibfnamefont {D.}~\bibnamefont {Lee}},\ }\href {\doibase 10.1103/PhysRevC.106.054322} {\bibfield  {journal} {\bibinfo  {journal} {Phys. Rev. C}\ }\textbf {\bibinfo {volume} {106}},\ \bibinfo {pages} {054322} (\bibinfo {year} {2022})},\ \Eprint {http://arxiv.org/abs/2203.05284} {arXiv:2203.05284 [nucl-th]} \BibitemShut {NoStop}%
\bibitem [{\citenamefont {Gl{\"o}ckle}(1983)}]{Glockle:1983}%
  \BibitemOpen
  \bibfield  {author} {\bibinfo {author} {\bibfnamefont {W.}~\bibnamefont {Gl{\"o}ckle}},\ }\href {\doibase https://doi.org/10.1007/978-3-642-82081-6} {\emph {\bibinfo {title} {The Quantum Mechanical Few-Body Problem}}}\ (\bibinfo  {publisher} {Springer-Verlag},\ \bibinfo {address} {Berlin},\ \bibinfo {year} {1983})\BibitemShut {NoStop}%
\bibitem [{\citenamefont {Nielsen}\ \emph {et~al.}(2001)\citenamefont {Nielsen}, \citenamefont {Fedorov}, \citenamefont {Jensen},\ and\ \citenamefont {Garrido}}]{Nielsen:2001hbm}%
  \BibitemOpen
  \bibfield  {author} {\bibinfo {author} {\bibfnamefont {E.}~\bibnamefont {Nielsen}}, \bibinfo {author} {\bibfnamefont {D.~V.}\ \bibnamefont {Fedorov}}, \bibinfo {author} {\bibfnamefont {A.~S.}\ \bibnamefont {Jensen}}, \ and\ \bibinfo {author} {\bibfnamefont {E.}~\bibnamefont {Garrido}},\ }\href {\doibase 10.1016/s0370-1573(00)00107-1} {\bibfield  {journal} {\bibinfo  {journal} {Phys. Rept.}\ }\textbf {\bibinfo {volume} {347}},\ \bibinfo {pages} {373} (\bibinfo {year} {2001})}\BibitemShut {NoStop}%
\bibitem [{\citenamefont {Lazauskas}\ and\ \citenamefont {Carbonell}(2020)}]{Lazauskas:2019hil}%
  \BibitemOpen
  \bibfield  {author} {\bibinfo {author} {\bibfnamefont {R.}~\bibnamefont {Lazauskas}}\ and\ \bibinfo {author} {\bibfnamefont {J.}~\bibnamefont {Carbonell}},\ }\href {\doibase 10.3389/fphy.2019.00251} {\bibfield  {journal} {\bibinfo  {journal} {Front. in Phys.}\ }\textbf {\bibinfo {volume} {7}},\ \bibinfo {pages} {251} (\bibinfo {year} {2020})},\ \Eprint {http://arxiv.org/abs/2002.05876} {arXiv:2002.05876 [nucl-th]} \BibitemShut {NoStop}%
\bibitem [{\citenamefont {Reinhardt}(1982)}]{reinhardt1982complex}%
  \BibitemOpen
  \bibfield  {author} {\bibinfo {author} {\bibfnamefont {W.~P.}\ \bibnamefont {Reinhardt}},\ }\href@noop {} {\bibfield  {journal} {\bibinfo  {journal} {Annual Review of Physical Chemistry}\ }\textbf {\bibinfo {volume} {33}},\ \bibinfo {pages} {223} (\bibinfo {year} {1982})}\BibitemShut {NoStop}%
\bibitem [{\citenamefont {Moiseyev}(2011)}]{Moiseyev_2011}%
  \BibitemOpen
  \bibfield  {author} {\bibinfo {author} {\bibfnamefont {N.}~\bibnamefont {Moiseyev}},\ }\href {\doibase https://doi.org/10.1017/CBO9780511976186} {\emph {\bibinfo {title} {Non-Hermitian Quantum Mechanics}}}\ (\bibinfo  {publisher} {Cambridge University Press},\ \bibinfo {year} {2011})\BibitemShut {NoStop}%
\bibitem [{\citenamefont {Myo}\ \emph {et~al.}(2014)\citenamefont {Myo}, \citenamefont {Kikuchi}, \citenamefont {Masui},\ and\ \citenamefont {Kat\={o}}}]{Myo:2014ypa}%
  \BibitemOpen
  \bibfield  {author} {\bibinfo {author} {\bibfnamefont {T.}~\bibnamefont {Myo}}, \bibinfo {author} {\bibfnamefont {Y.}~\bibnamefont {Kikuchi}}, \bibinfo {author} {\bibfnamefont {H.}~\bibnamefont {Masui}}, \ and\ \bibinfo {author} {\bibfnamefont {K.}~\bibnamefont {Kat\={o}}},\ }\href {\doibase 10.1016/j.ppnp.2014.08.001} {\bibfield  {journal} {\bibinfo  {journal} {Prog. Part. Nucl. Phys.}\ }\textbf {\bibinfo {volume} {79}},\ \bibinfo {pages} {1} (\bibinfo {year} {2014})},\ \Eprint {http://arxiv.org/abs/1410.4356} {arXiv:1410.4356 [nucl-th]} \BibitemShut {NoStop}%
\bibitem [{\citenamefont {Michel}\ and\ \citenamefont {P\l{}oszajczak}(2021)}]{Michel:2021jkx}%
  \BibitemOpen
  \bibfield  {author} {\bibinfo {author} {\bibfnamefont {N.}~\bibnamefont {Michel}}\ and\ \bibinfo {author} {\bibfnamefont {M.}~\bibnamefont {P\l{}oszajczak}},\ }\href {\doibase 10.1007/978-3-030-69356-5} {\emph {\bibinfo {title} {{Gamow Shell Model: The Unified Theory of Nuclear Structure and Reactions}}}},\ Vol.\ \bibinfo {volume} {983}\ (\bibinfo  {publisher} {Springer Cham},\ \bibinfo {year} {2021})\BibitemShut {NoStop}%
\bibitem [{\citenamefont {Zhang}(2025{\natexlab{a}})}]{Zhang:2024ril}%
  \BibitemOpen
  \bibfield  {author} {\bibinfo {author} {\bibfnamefont {X.}~\bibnamefont {Zhang}},\ }\href {\doibase 10.1103/5frj-w5xh} {\bibfield  {journal} {\bibinfo  {journal} {Phys. Rev. Lett.}\ ,\ } (\bibinfo {year} {2025}{\natexlab{a}})},\ \Eprint {http://arxiv.org/abs/2408.03309} {arXiv:2408.03309 [nucl-th]} \BibitemShut {NoStop}%
\bibitem [{\citenamefont {Zhang}(2025{\natexlab{b}})}]{Zhang:2024gac}%
  \BibitemOpen
  \bibfield  {author} {\bibinfo {author} {\bibfnamefont {X.}~\bibnamefont {Zhang}},\ }\href {\doibase 10.1103/4wbf-gzk5} {\bibfield  {journal} {\bibinfo  {journal} {Phys. Rev. C}\ ,\ } (\bibinfo {year} {2025}{\natexlab{b}})},\ \Eprint {http://arxiv.org/abs/2411.06712} {arXiv:2411.06712 [nucl-th]} \BibitemShut {NoStop}%
\bibitem [{\citenamefont {Schlessinger}\ and\ \citenamefont {Schwartz}(1966)}]{Schlessinger:1966zz}%
  \BibitemOpen
  \bibfield  {author} {\bibinfo {author} {\bibfnamefont {L.}~\bibnamefont {Schlessinger}}\ and\ \bibinfo {author} {\bibfnamefont {C.}~\bibnamefont {Schwartz}},\ }\href {\doibase 10.1103/PhysRevLett.16.1173} {\bibfield  {journal} {\bibinfo  {journal} {Phys. Rev. Lett.}\ }\textbf {\bibinfo {volume} {16}},\ \bibinfo {pages} {1173} (\bibinfo {year} {1966})}\BibitemShut {NoStop}%
\bibitem [{\citenamefont {Carbonell}\ \emph {et~al.}(2014)\citenamefont {Carbonell}, \citenamefont {Deltuva}, \citenamefont {Fonseca},\ and\ \citenamefont {Lazauskas}}]{Carbonell:2013ywa}%
  \BibitemOpen
  \bibfield  {author} {\bibinfo {author} {\bibfnamefont {J.}~\bibnamefont {Carbonell}}, \bibinfo {author} {\bibfnamefont {A.}~\bibnamefont {Deltuva}}, \bibinfo {author} {\bibfnamefont {A.~C.}\ \bibnamefont {Fonseca}}, \ and\ \bibinfo {author} {\bibfnamefont {R.}~\bibnamefont {Lazauskas}},\ }\href {\doibase 10.1016/j.ppnp.2013.10.003} {\bibfield  {journal} {\bibinfo  {journal} {Prog. Part. Nucl. Phys.}\ }\textbf {\bibinfo {volume} {74}},\ \bibinfo {pages} {55} (\bibinfo {year} {2014})},\ \Eprint {http://arxiv.org/abs/1310.6631} {arXiv:1310.6631 [nucl-th]} \BibitemShut {NoStop}%
\bibitem [{\citenamefont {Efros}\ \emph {et~al.}(2007)\citenamefont {Efros}, \citenamefont {Leidemann}, \citenamefont {Orlandini},\ and\ \citenamefont {Barnea}}]{Efros:2007nq}%
  \BibitemOpen
  \bibfield  {author} {\bibinfo {author} {\bibfnamefont {V.~D.}\ \bibnamefont {Efros}}, \bibinfo {author} {\bibfnamefont {W.}~\bibnamefont {Leidemann}}, \bibinfo {author} {\bibfnamefont {G.}~\bibnamefont {Orlandini}}, \ and\ \bibinfo {author} {\bibfnamefont {N.}~\bibnamefont {Barnea}},\ }\href {\doibase 10.1088/0954-3899/34/12/R02} {\bibfield  {journal} {\bibinfo  {journal} {J. Phys. G}\ }\textbf {\bibinfo {volume} {34}},\ \bibinfo {pages} {R459} (\bibinfo {year} {2007})},\ \Eprint {http://arxiv.org/abs/0708.2803} {arXiv:0708.2803 [nucl-th]} \BibitemShut {NoStop}%
\bibitem [{\citenamefont {Liu}\ \emph {et~al.}(2024)\citenamefont {Liu}, \citenamefont {Lei},\ and\ \citenamefont {Ren}}]{Liu:2024pqp}%
  \BibitemOpen
  \bibfield  {author} {\bibinfo {author} {\bibfnamefont {J.}~\bibnamefont {Liu}}, \bibinfo {author} {\bibfnamefont {J.}~\bibnamefont {Lei}}, \ and\ \bibinfo {author} {\bibfnamefont {Z.}~\bibnamefont {Ren}},\ }\href {\doibase 10.1016/j.physletb.2024.139070} {\bibfield  {journal} {\bibinfo  {journal} {Phys. Lett. B}\ }\textbf {\bibinfo {volume} {858}},\ \bibinfo {pages} {139070} (\bibinfo {year} {2024})},\ \Eprint {http://arxiv.org/abs/2408.07954} {arXiv:2408.07954 [nucl-th]} \BibitemShut {NoStop}%
\bibitem [{buq()}]{buqeye}%
  \BibitemOpen
  \href@noop {} {\enquote {\bibinfo {title} {Buqeye collaboration},}\ }\bibinfo {howpublished} {\url{https://buqeye.github.io/publications/}}\BibitemShut {NoStop}%
\bibitem [{\citenamefont {Piarulli}\ \emph {et~al.}(2015)\citenamefont {Piarulli}, \citenamefont {Girlanda}, \citenamefont {Schiavilla}, \citenamefont {P\'erez}, \citenamefont {Amaro},\ and\ \citenamefont {Arriola}}]{Piarulli2015}%
  \BibitemOpen
  \bibfield  {author} {\bibinfo {author} {\bibfnamefont {M.}~\bibnamefont {Piarulli}}, \bibinfo {author} {\bibfnamefont {L.}~\bibnamefont {Girlanda}}, \bibinfo {author} {\bibfnamefont {R.}~\bibnamefont {Schiavilla}}, \bibinfo {author} {\bibfnamefont {R.~N.}\ \bibnamefont {P\'erez}}, \bibinfo {author} {\bibfnamefont {J.~E.}\ \bibnamefont {Amaro}}, \ and\ \bibinfo {author} {\bibfnamefont {E.~R.}\ \bibnamefont {Arriola}},\ }\href {\doibase 10.1103/PhysRevC.91.024003} {\bibfield  {journal} {\bibinfo  {journal} {Phys. Rev. C}\ }\textbf {\bibinfo {volume} {91}},\ \bibinfo {pages} {024003} (\bibinfo {year} {2015})}\BibitemShut {NoStop}%
\bibitem [{\citenamefont {Piarulli}\ \emph {et~al.}(2016)\citenamefont {Piarulli}, \citenamefont {Girlanda}, \citenamefont {Schiavilla}, \citenamefont {Kievsky}, \citenamefont {Lovato}, \citenamefont {Marcucci}, \citenamefont {Pieper}, \citenamefont {Viviani},\ and\ \citenamefont {Wiringa}}]{Piarulli2016}%
  \BibitemOpen
  \bibfield  {author} {\bibinfo {author} {\bibfnamefont {M.}~\bibnamefont {Piarulli}}, \bibinfo {author} {\bibfnamefont {L.}~\bibnamefont {Girlanda}}, \bibinfo {author} {\bibfnamefont {R.}~\bibnamefont {Schiavilla}}, \bibinfo {author} {\bibfnamefont {A.}~\bibnamefont {Kievsky}}, \bibinfo {author} {\bibfnamefont {A.}~\bibnamefont {Lovato}}, \bibinfo {author} {\bibfnamefont {L.~E.}\ \bibnamefont {Marcucci}}, \bibinfo {author} {\bibfnamefont {S.~C.}\ \bibnamefont {Pieper}}, \bibinfo {author} {\bibfnamefont {M.}~\bibnamefont {Viviani}}, \ and\ \bibinfo {author} {\bibfnamefont {R.~B.}\ \bibnamefont {Wiringa}},\ }\href {\doibase 10.1103/PhysRevC.94.054007} {\bibfield  {journal} {\bibinfo  {journal} {Phys. Rev. C}\ }\textbf {\bibinfo {volume} {94}},\ \bibinfo {pages} {054007} (\bibinfo {year} {2016})}\BibitemShut {NoStop}%
\bibitem [{\citenamefont {Piarulli}\ \emph {et~al.}(2018)\citenamefont {Piarulli} \emph {et~al.}}]{Piarulli:2017dwd}%
  \BibitemOpen
  \bibfield  {author} {\bibinfo {author} {\bibfnamefont {M.}~\bibnamefont {Piarulli}} \emph {et~al.},\ }\href {\doibase 10.1103/PhysRevLett.120.052503} {\bibfield  {journal} {\bibinfo  {journal} {Phys. Rev. Lett.}\ }\textbf {\bibinfo {volume} {120}},\ \bibinfo {pages} {052503} (\bibinfo {year} {2018})},\ \Eprint {http://arxiv.org/abs/1707.02883} {arXiv:1707.02883 [nucl-th]} \BibitemShut {NoStop}%
\bibitem [{\citenamefont {Kohn}(1948)}]{Kohn:1948col}%
  \BibitemOpen
  \bibfield  {author} {\bibinfo {author} {\bibfnamefont {W.}~\bibnamefont {Kohn}},\ }\href {\doibase 10.1103/PhysRev.74.1763} {\bibfield  {journal} {\bibinfo  {journal} {Phys. Rev.}\ }\textbf {\bibinfo {volume} {74}},\ \bibinfo {pages} {1763} (\bibinfo {year} {1948})}\BibitemShut {NoStop}%
\bibitem [{\citenamefont {Zakharev}\ \emph {et~al.}(1969)\citenamefont {Zakharev}, \citenamefont {Pustovalov},\ and\ \citenamefont {Efros}}]{Efros1969HH}%
  \BibitemOpen
  \bibfield  {author} {\bibinfo {author} {\bibfnamefont {B.}~\bibnamefont {Zakharev}}, \bibinfo {author} {\bibfnamefont {V.}~\bibnamefont {Pustovalov}}, \ and\ \bibinfo {author} {\bibfnamefont {V.}~\bibnamefont {Efros}},\ }\href@noop {} {\bibfield  {journal} {\bibinfo  {journal} {Sov. J. Nucl. Phys.(Engl. Transl.);(United States)}\ }\textbf {\bibinfo {volume} {8}},\ \bibinfo {pages} {234} (\bibinfo {year} {1969})}\BibitemShut {NoStop}%
\bibitem [{\citenamefont {Efros}\ and\ \citenamefont {Zhukov}(1971)}]{Efros:1971vff}%
  \BibitemOpen
  \bibfield  {author} {\bibinfo {author} {\bibfnamefont {V.~D.}\ \bibnamefont {Efros}}\ and\ \bibinfo {author} {\bibfnamefont {M.~V.}\ \bibnamefont {Zhukov}},\ }\href {\doibase 10.1016/0370-2693(71)90558-2} {\bibfield  {journal} {\bibinfo  {journal} {Phys. Lett. B}\ }\textbf {\bibinfo {volume} {37}},\ \bibinfo {pages} {18} (\bibinfo {year} {1971})}\BibitemShut {NoStop}%
\bibitem [{\citenamefont {Joachain}(1975)}]{JoachainQCT1975}%
  \BibitemOpen
  \bibfield  {author} {\bibinfo {author} {\bibfnamefont {C.~J.}\ \bibnamefont {Joachain}},\ }\href@noop {} {\emph {\bibinfo {title} {Quantum Collision Theory}}}\ (\bibinfo  {publisher} {North-Holland},\ \bibinfo {address} {Amsterdam},\ \bibinfo {year} {1975})\BibitemShut {NoStop}%
\bibitem [{\citenamefont {Newton}(2002)}]{newton2002scattering}%
  \BibitemOpen
  \bibfield  {author} {\bibinfo {author} {\bibfnamefont {R.~G.}\ \bibnamefont {Newton}},\ }\href@noop {} {\emph {\bibinfo {title} {Scattering theory of waves and particles}}}\ (\bibinfo  {publisher} {Dover},\ \bibinfo {address} {Mineola, New York},\ \bibinfo {year} {2002})\BibitemShut {NoStop}%
\bibitem [{\citenamefont {Schwartz}(1961)}]{PhysRev.124.1468}%
  \BibitemOpen
  \bibfield  {author} {\bibinfo {author} {\bibfnamefont {C.}~\bibnamefont {Schwartz}},\ }\href {\doibase 10.1103/PhysRev.124.1468} {\bibfield  {journal} {\bibinfo  {journal} {Phys. Rev.}\ }\textbf {\bibinfo {volume} {124}},\ \bibinfo {pages} {1468} (\bibinfo {year} {1961})}\BibitemShut {NoStop}%
\bibitem [{\citenamefont {Nesbet}(1980)}]{nesbet1980variational}%
  \BibitemOpen
  \bibfield  {author} {\bibinfo {author} {\bibfnamefont {R.}~\bibnamefont {Nesbet}},\ }\href {https://books.google.com/books?id=0mF5AAAAIAAJ} {\emph {\bibinfo {title} {Variational methods in electron-atom scattering theory}}},\ Physics of atoms and molecules\ (\bibinfo  {publisher} {Plenum Press},\ \bibinfo {year} {1980})\BibitemShut {NoStop}%
\bibitem [{\citenamefont {Hunter}(2007)}]{Hunter:2007}%
  \BibitemOpen
  \bibfield  {author} {\bibinfo {author} {\bibfnamefont {J.~D.}\ \bibnamefont {Hunter}},\ }\href {\doibase 10.1109/MCSE.2007.55} {\bibfield  {journal} {\bibinfo  {journal} {Computing in Science \& Engineering}\ }\textbf {\bibinfo {volume} {9}},\ \bibinfo {pages} {90} (\bibinfo {year} {2007})}\BibitemShut {NoStop}%
\bibitem [{\citenamefont {Harris}\ \emph {et~al.}(2020)\citenamefont {Harris}, \citenamefont {Millman}, \citenamefont {van~der Walt}, \citenamefont {Gommers}, \citenamefont {Virtanen}, \citenamefont {Cournapeau}, \citenamefont {Wieser}, \citenamefont {Taylor}, \citenamefont {Berg}, \citenamefont {Smith}, \citenamefont {Kern}, \citenamefont {Picus}, \citenamefont {Hoyer}, \citenamefont {van Kerkwijk}, \citenamefont {Brett}, \citenamefont {Haldane}, \citenamefont {del R{\'{i}}o}, \citenamefont {Wiebe}, \citenamefont {Peterson}, \citenamefont {G{\'{e}}rard-Marchant}, \citenamefont {Sheppard}, \citenamefont {Reddy}, \citenamefont {Weckesser}, \citenamefont {Abbasi}, \citenamefont {Gohlke},\ and\ \citenamefont {Oliphant}}]{harris2020array}%
  \BibitemOpen
  \bibfield  {author} {\bibinfo {author} {\bibfnamefont {C.~R.}\ \bibnamefont {Harris}}, \bibinfo {author} {\bibfnamefont {K.~J.}\ \bibnamefont {Millman}}, \bibinfo {author} {\bibfnamefont {S.~J.}\ \bibnamefont {van~der Walt}}, \bibinfo {author} {\bibfnamefont {R.}~\bibnamefont {Gommers}}, \bibinfo {author} {\bibfnamefont {P.}~\bibnamefont {Virtanen}}, \bibinfo {author} {\bibfnamefont {D.}~\bibnamefont {Cournapeau}}, \bibinfo {author} {\bibfnamefont {E.}~\bibnamefont {Wieser}}, \bibinfo {author} {\bibfnamefont {J.}~\bibnamefont {Taylor}}, \bibinfo {author} {\bibfnamefont {S.}~\bibnamefont {Berg}}, \bibinfo {author} {\bibfnamefont {N.~J.}\ \bibnamefont {Smith}}, \bibinfo {author} {\bibfnamefont {R.}~\bibnamefont {Kern}}, \bibinfo {author} {\bibfnamefont {M.}~\bibnamefont {Picus}}, \bibinfo {author} {\bibfnamefont {S.}~\bibnamefont {Hoyer}}, \bibinfo {author} {\bibfnamefont {M.~H.}\ \bibnamefont {van Kerkwijk}}, \bibinfo {author} {\bibfnamefont {M.}~\bibnamefont {Brett}}, \bibinfo {author} {\bibfnamefont
  {A.}~\bibnamefont {Haldane}}, \bibinfo {author} {\bibfnamefont {J.~F.}\ \bibnamefont {del R{\'{i}}o}}, \bibinfo {author} {\bibfnamefont {M.}~\bibnamefont {Wiebe}}, \bibinfo {author} {\bibfnamefont {P.}~\bibnamefont {Peterson}}, \bibinfo {author} {\bibfnamefont {P.}~\bibnamefont {G{\'{e}}rard-Marchant}}, \bibinfo {author} {\bibfnamefont {K.}~\bibnamefont {Sheppard}}, \bibinfo {author} {\bibfnamefont {T.}~\bibnamefont {Reddy}}, \bibinfo {author} {\bibfnamefont {W.}~\bibnamefont {Weckesser}}, \bibinfo {author} {\bibfnamefont {H.}~\bibnamefont {Abbasi}}, \bibinfo {author} {\bibfnamefont {C.}~\bibnamefont {Gohlke}}, \ and\ \bibinfo {author} {\bibfnamefont {T.~E.}\ \bibnamefont {Oliphant}},\ }\href {\doibase 10.1038/s41586-020-2649-2} {\bibfield  {journal} {\bibinfo  {journal} {Nature}\ }\textbf {\bibinfo {volume} {585}},\ \bibinfo {pages} {357} (\bibinfo {year} {2020})}\BibitemShut {NoStop}%
\bibitem [{\citenamefont {Virtanen}\ \emph {et~al.}(2020)\citenamefont {Virtanen}, \citenamefont {Gommers}, \citenamefont {Oliphant}, \citenamefont {Haberland}, \citenamefont {Reddy}, \citenamefont {Cournapeau}, \citenamefont {Burovski}, \citenamefont {Peterson}, \citenamefont {Weckesser}, \citenamefont {Bright}, \citenamefont {{van der Walt}}, \citenamefont {Brett}, \citenamefont {Wilson}, \citenamefont {Millman}, \citenamefont {Mayorov}, \citenamefont {Nelson}, \citenamefont {Jones}, \citenamefont {Kern}, \citenamefont {Larson}, \citenamefont {Carey}, \citenamefont {Polat}, \citenamefont {Feng}, \citenamefont {Moore}, \citenamefont {{VanderPlas}}, \citenamefont {Laxalde}, \citenamefont {Perktold}, \citenamefont {Cimrman}, \citenamefont {Henriksen}, \citenamefont {Quintero}, \citenamefont {Harris}, \citenamefont {Archibald}, \citenamefont {Ribeiro}, \citenamefont {Pedregosa}, \citenamefont {{van Mulbregt}},\ and\ \citenamefont {{SciPy 1.0 Contributors}}}]{2020SciPy-NMeth}%
  \BibitemOpen
  \bibfield  {author} {\bibinfo {author} {\bibfnamefont {P.}~\bibnamefont {Virtanen}}, \bibinfo {author} {\bibfnamefont {R.}~\bibnamefont {Gommers}}, \bibinfo {author} {\bibfnamefont {T.~E.}\ \bibnamefont {Oliphant}}, \bibinfo {author} {\bibfnamefont {M.}~\bibnamefont {Haberland}}, \bibinfo {author} {\bibfnamefont {T.}~\bibnamefont {Reddy}}, \bibinfo {author} {\bibfnamefont {D.}~\bibnamefont {Cournapeau}}, \bibinfo {author} {\bibfnamefont {E.}~\bibnamefont {Burovski}}, \bibinfo {author} {\bibfnamefont {P.}~\bibnamefont {Peterson}}, \bibinfo {author} {\bibfnamefont {W.}~\bibnamefont {Weckesser}}, \bibinfo {author} {\bibfnamefont {J.}~\bibnamefont {Bright}}, \bibinfo {author} {\bibfnamefont {S.~J.}\ \bibnamefont {{van der Walt}}}, \bibinfo {author} {\bibfnamefont {M.}~\bibnamefont {Brett}}, \bibinfo {author} {\bibfnamefont {J.}~\bibnamefont {Wilson}}, \bibinfo {author} {\bibfnamefont {K.~J.}\ \bibnamefont {Millman}}, \bibinfo {author} {\bibfnamefont {N.}~\bibnamefont {Mayorov}}, \bibinfo {author} {\bibfnamefont
  {A.~R.~J.}\ \bibnamefont {Nelson}}, \bibinfo {author} {\bibfnamefont {E.}~\bibnamefont {Jones}}, \bibinfo {author} {\bibfnamefont {R.}~\bibnamefont {Kern}}, \bibinfo {author} {\bibfnamefont {E.}~\bibnamefont {Larson}}, \bibinfo {author} {\bibfnamefont {C.~J.}\ \bibnamefont {Carey}}, \bibinfo {author} {\bibfnamefont {{\.I}.}~\bibnamefont {Polat}}, \bibinfo {author} {\bibfnamefont {Y.}~\bibnamefont {Feng}}, \bibinfo {author} {\bibfnamefont {E.~W.}\ \bibnamefont {Moore}}, \bibinfo {author} {\bibfnamefont {J.}~\bibnamefont {{VanderPlas}}}, \bibinfo {author} {\bibfnamefont {D.}~\bibnamefont {Laxalde}}, \bibinfo {author} {\bibfnamefont {J.}~\bibnamefont {Perktold}}, \bibinfo {author} {\bibfnamefont {R.}~\bibnamefont {Cimrman}}, \bibinfo {author} {\bibfnamefont {I.}~\bibnamefont {Henriksen}}, \bibinfo {author} {\bibfnamefont {E.~A.}\ \bibnamefont {Quintero}}, \bibinfo {author} {\bibfnamefont {C.~R.}\ \bibnamefont {Harris}}, \bibinfo {author} {\bibfnamefont {A.~M.}\ \bibnamefont {Archibald}}, \bibinfo {author}
  {\bibfnamefont {A.~H.}\ \bibnamefont {Ribeiro}}, \bibinfo {author} {\bibfnamefont {F.}~\bibnamefont {Pedregosa}}, \bibinfo {author} {\bibfnamefont {P.}~\bibnamefont {{van Mulbregt}}}, \ and\ \bibinfo {author} {\bibnamefont {{SciPy 1.0 Contributors}}},\ }\href {\doibase 10.1038/s41592-019-0686-2} {\bibfield  {journal} {\bibinfo  {journal} {Nature Methods}\ }\textbf {\bibinfo {volume} {17}},\ \bibinfo {pages} {261} (\bibinfo {year} {2020})}\BibitemShut {NoStop}%
\end{thebibliography}

%

\end{document}